\documentclass[article,shortnames,noheader,nojss]{jss}

\usepackage{xcolor}
\usepackage{utfsym}
\usepackage{orcidlink,thumbpdf,lmodern,amsmath,amssymb}
\usepackage{enumitem}
\usepackage{framed}
\usepackage{hyperref}
\hypersetup{
    colorlinks=true,
    linkcolor=blue,
    filecolor=magenta,      
    urlcolor=cyan,
    pdftitle={StochTree},
    pdfpagemode=FullScreen,
}
\usepackage{multirow}
\usepackage{graphicx}
\usepackage{tikz}
\usetikzlibrary{positioning}

\definecolor{darkgreen}{RGB}{90,192,66}
\DeclareMathOperator{\iid}{\stackrel{\mbox{\tiny iid} }{\sim}}

\usepackage{tikz,pgfplots,graphicx}
\usetikzlibrary{shapes,decorations,arrows,calc,arrows.meta,fit,positioning}
\pgfplotsset{
	compat=1.11,
}
\pgfkeys{
	/pgf/declare function={f(\x,\a,\b) = \a*\x+\b;}
}
\pgfkeys{
	/pgf/declare function={g(\x,\a,\b) = \a*\x+\b;}
}
\newsavebox{\partitionA}
\newsavebox{\partitionB}
\newsavebox{\partitionC}

\usetikzlibrary{arrows}
\usetikzlibrary{positioning}
\usetikzlibrary{shapes.arrows}
\newdimen\nodeDist
\tikzset{
    myarrow/.style={
        draw,
        fill=orange,
        single arrow,
        minimum height=3.5ex,
        single arrow head extend=1ex
    }
}

\tikzset{dist1/.style={path picture= {
    \begin{scope}[x=1pt,y=10pt]
      \draw plot[domain=-6:6] (\x,{\x/24});
    \end{scope}
    }
  }
}
\tikzset{dist2/.style={path picture= {
    \begin{scope}[x=1pt,y=10pt]
      \draw plot[domain=-6:6] (\x,{0.01*(\x-2)^2-0.4});
    \end{scope}
    }
  }
}
\tikzset{dist3/.style={path picture= {
    \begin{scope}[x=1pt,y=10pt]
      \draw plot[domain=-6:6] (\x,{0.85/(1 + exp(\x))-0.5});
    \end{scope}
    }
  }
}
\tikzstyle{f1}=[draw,circle,minimum size=25pt,inner sep=0pt,dist1]
\tikzstyle{f2}=[draw,circle,minimum size=25pt,inner sep=0pt,dist2]
\tikzstyle{f3}=[draw,circle,minimum size=25pt,inner sep=0pt,dist3]

\usepackage{subcaption}

\author{Andrew Herren~\orcidlink{0000-0003-4109-6611}\\University of Texas at Austin
   \And P. Richard Hahn\\Arizona State University
   \And Jared Murray\\University of Texas at Austin
   \AND Carlos Carvalho\\University of Austin}
\Plainauthor{Andrew Herren, Richard Hahn, Jared Murray, Carlos Carvalho}

\title{StochTree: BART-based modeling in \proglang{R} and \proglang{Python}}
\Plaintitle{StochTree: BART-based modeling in R and Python}
\Shorttitle{StochTree: BART-based modeling in \proglang{R} and \proglang{Python}}

\Abstract{
   \pkg{stochtree} is a \proglang{C++} library for Bayesian tree ensemble models such as BART and Bayesian Causal Forests (BCF), as well as user-specified variations. Unlike previous BART packages, \pkg{stochtree} provides bindings to both \proglang{R} and \proglang{Python} for full interoperability.  \pkg{stochtree} boasts a more comprehensive range of models relative to previous packages, including heteroskedastic forests, random effects, and treed linear models. Additionally, \pkg{stochtree} offers flexible handling of model fits: the ability to save model fits, reinitialize models from existing fits (facilitating improved model initialization heuristics), and pass fits between \proglang{R} and \proglang{Python}. On both platforms, \pkg{stochtree} exposes lower-level functionality, allowing users to specify models incorporating Bayesian tree ensembles without needing to modify \proglang{C++} code. We illustrate the use of \pkg{stochtree} in three settings: i) straightfoward applications of existing models such as BART and BCF, ii) models that include more sophisticated components like heteroskedasticity and leaf-wise regression models, and iii) as a component of custom MCMC routines to fit nonstandard tree ensemble models.  
 }

\Plainkeywords{BART, causal inference, decision trees, tree ensembles, R, Python, C++}

\Address{
  Andrew Herren\\
  University of Texas at Austin\\
  Robert A. Welch Hall 5.216, 105 E 24th St\\
  Austin, TX 78712\\
  E-mail: \email{andrew.herren@mccombs.utexas.edu}\\
  URL: \url{https://andrewherren.github.io}
}

\begin{document}


\section[Introduction]{Preliminaries} \label{sec:intro}

\subsection{Introduction}

In the 15 years since Bayesian additive regression trees were introduced \citep{chipman2010bart}, BART has become a go-to method for supervised learning and causal inference, particularly among users interested in Bayesian uncertainty quantification in these contexts. In that time, BART-based methods have seen extensive development in terms of both modeling and computation. However, such innovations are often slow to trickle down to BART users, being introduced in researcher-specific packages that differ from each other in terms of both basic functionality and user-interface. In particular, some packages are in \proglang{Python} only and others are in \proglang{R} only. 

Against this backdrop, \pkg{stochtree} was built with three complementary goals in mind:
\begin{enumerate}
\item To have a common \proglang{C++} code library that serves inter-operable \proglang{R} and \proglang{Python} packages.
\item To incorporate, in one place, many of the recent advances to BART computation and modeling that do not currently exist in user-facing packages.
\item To expose the \proglang{C++} model fitting objects and functions so that novel BART models may be prototyped directly in \proglang{R} or \proglang{Python} while still realizing the substantial performance benefits of core operations being executed from compiled code with BART specific sampling algorithms. 
\end{enumerate}

\subsection{Comparison to Existing Packages}

Here, we list for reference the most prominent currently-available BART packages and briefly compare their functionality to \pkg{stochtree}.
\begin{itemize}
\item \pkg{BART} \citep{bart2021pkg} is the current reference implementation \proglang{R} package of \cite{chipman2010bart}, providing the classic BART model along with a number of extensions for classification and survival models \citep{bart2021pkg}. 

\item \pkg{dbarts} \citep{dbarts2024pkg} offers a limited interface for interchanging a forest MCMC step with other samplers.

\item \pkg{bartMachine} \citep{bartMachine2016pkg} was developed to address many of the drawbacks of \pkg{BayesTree} \citep{bayestree2024pkg}, the first-generation reference implementation of BART.

\item \pkg{pymc-bart} \citep{quiroga2022bart}  is a \proglang{Python} package employing a ``particle Gibbs'' BART sampler \citep{lakshminarayanan2015particle} rather than the traditional MCMC sampler of \cite{chipman2010bart}.

\item \pkg{bartz} \citep{petrillo2024very}  offers a GPU-accelerated \pkg{bartz} sampler for a stripped-down BART model which has fixed-depth trees and only accommodates continuous predictor variables.

\item \pkg{flexBART} \citep{deshpande2024flexbart} provides a flexible, GLM-like interface for BART models in \proglang{R} with careful handling of unordered categorical variables and graph-structured categorical variables. It supports univariate leaf regression through a formula interface.

\item \pkg{softBART} \citep{linero2022softbart} implements the ``soft BART'' model of \cite{linero2018bayesian} in \proglang{R} with a flexible, GLM-like interface and an interface for interchanging forest MCMC with other samplers. The soft BART model differs from the classic BART model in that the implied bases of the model are logistic functions that determine each observation's probability of being routed to a given leaf, not mutually-exclusive, deterministic leaf membership indicators.
\end{itemize}

New to \pkg{stochtree} are the following features:
\begin{itemize}
\item  The availability of ``warm-start'' sampler acceleration \citep{he2023stochastic}, introduced in Section \ref{sec:friedman-bart-supervised}.

\item  The Bayesian causal forests (BCF) model of \cite{hahn2020bayesian} and the ability to warm-start it using the accelerated algorithm of \cite{krantsevich2023stochastic}, introduced in Section \ref{sec:acic-data}.

\item Estimation of additive group random effects (\cite{gelman2008redundant}, \cite{yeager2022teacher}), introduced in Section \ref{sec:acic-bcf-rfx}.

\item Extensions to allow for linear leaf regression models with arbitrary, user-defined bases (a generalization of \cite{chipman2002bayesian} which is articulated in part in \cite{starling2020bart}), introduced in Section \ref{sec:bart-rdd}.

\item Heteroskedastic BART forests as implemented in \cite{murray2021log} (\pkg{BART} offers a similar model based on \cite{pratola2020heteroscedastic} and \pkg{pymc-bart} also makes it possible to specify a similar model), introduced in Section \ref{sec:bart-motorcycle}. 

\end{itemize}
 
Table \ref{tab:overview} presents a visual overview of these comparisons. 

\begin{table}[t!]
\centering
\begin{tabular}{lp{16em}p{0.25cm}p{0.25cm}p{0.25cm}p{0.25cm}p{0.25cm}p{0.25cm}p{0.25cm}p{0.25cm}p{0.25cm}}
 & Feature                                                   & \rotatebox{45}{stochtree}  & \rotatebox{45}{BART} & \rotatebox{45}{dbarts} & \rotatebox{45}{bartMachine} & \rotatebox{45}{pymc-bart} & \rotatebox{45}{bartz}  & \rotatebox{45}{flexBART} & \rotatebox{45}{softBART} \\ \hline\hline
\multirow{7}{6em}{Computational Features} & \proglang{R} API                                                     & \leavevmode\color{darkgreen}\checkmark  & \leavevmode\color{darkgreen}\checkmark  & \leavevmode\color{darkgreen}\checkmark  & \leavevmode\color{darkgreen}\checkmark  & \leavevmode\color{red}\usym{2717} & \leavevmode\color{red}\usym{2717} & \leavevmode\color{darkgreen}\checkmark & \leavevmode\color{darkgreen}\checkmark \\ 
 & \proglang{Python} API                                                & \leavevmode\color{darkgreen}\checkmark  & \leavevmode\color{red}\usym{2717} & \leavevmode\color{red}\usym{2717} & \leavevmode\color{red}\usym{2717} & \leavevmode\color{darkgreen}\checkmark  & \leavevmode\color{darkgreen}\checkmark & \leavevmode\color{red}\usym{2717} & \leavevmode\color{red}\usym{2717}  \\ 
 & \proglang{C}/\proglang{C++} API                                                 & \leavevmode\color{darkgreen}\checkmark  & \leavevmode\color{red}\usym{2717} & \leavevmode\color{red}\usym{2717} & \leavevmode\color{red}\usym{2717} & \leavevmode\color{red}\usym{2717} & \leavevmode\color{red}\usym{2717} & \leavevmode\color{red}\usym{2717} & \leavevmode\color{red}\usym{2717} \\ 
 & Model serialization                                       & \leavevmode\color{darkgreen}\checkmark  & \leavevmode\color{darkgreen}\checkmark  & \leavevmode\color{darkgreen}\checkmark  & \leavevmode\color{darkgreen}\checkmark  & \leavevmode\color{darkgreen}\checkmark  & \leavevmode\color{darkgreen}\checkmark & \leavevmode\color{darkgreen}\checkmark & \leavevmode\color{red}\usym{2717}  \\ 
 & Parallelism/multiple chains                               & \leavevmode\color{darkgreen}\checkmark  & \leavevmode\color{darkgreen}\checkmark  & \leavevmode\color{darkgreen}\checkmark  & \leavevmode\color{darkgreen}\checkmark  & \leavevmode\color{darkgreen}\checkmark & \leavevmode\color{darkgreen}\checkmark  & \leavevmode\color{darkgreen}\checkmark & \leavevmode\color{red}\usym{2717}  \\ 
 & Custom sampler interface                 & \leavevmode\color{darkgreen}\checkmark  & \leavevmode\color{red}\usym{2717} & \leavevmode\color{darkgreen}\checkmark  & \leavevmode\color{red}\usym{2717} & \leavevmode\color{darkgreen}\checkmark & \leavevmode\color{red}\usym{2717}  & \leavevmode\color{red}\usym{2717} & \leavevmode\color{darkgreen}\checkmark \\ \hline
\multirow{6}{6em}{Modeling Features} & Continuous outcomes                                       & \leavevmode\color{darkgreen}\checkmark  & \leavevmode\color{darkgreen}\checkmark  & \leavevmode\color{darkgreen}\checkmark  & \leavevmode\color{darkgreen}\checkmark  & \leavevmode\color{darkgreen}\checkmark & \leavevmode\color{darkgreen}\checkmark  & \leavevmode\color{darkgreen}\checkmark & \leavevmode\color{darkgreen}\checkmark  \\ 
 & Binary outcomes                                       & \leavevmode\color{darkgreen}\checkmark  & \leavevmode\color{darkgreen}\checkmark  & \leavevmode\color{darkgreen}\checkmark  & \leavevmode\color{darkgreen}\checkmark  & \leavevmode\color{darkgreen}\checkmark & \leavevmode\color{darkgreen}\checkmark  & \leavevmode\color{darkgreen}\checkmark & \leavevmode\color{darkgreen}\checkmark \\ 
 & Random effects                                       & \leavevmode\color{darkgreen}\checkmark  & \leavevmode\color{red}\usym{2717} & \leavevmode\color{red}\usym{2717} & \leavevmode\color{red}\usym{2717} & \leavevmode\color{red}\usym{2717} & \leavevmode\color{red}\usym{2717} & \leavevmode\color{red}\usym{2717} & \leavevmode\color{red}\usym{2717} \\  
 & Multivariate linear leaf model                                       & \leavevmode\color{darkgreen}\checkmark  & \leavevmode\color{red}\usym{2717} & \leavevmode\color{red}\usym{2717} & \leavevmode\color{red}\usym{2717} & \leavevmode\color{red}\usym{2717} & \leavevmode\color{red}\usym{2717} & \leavevmode\color{red}\usym{2717} & \leavevmode\color{red}\usym{2717} \\ 
 & Forest-based heteroskedasticity                                       & \leavevmode\color{darkgreen}\checkmark  & \leavevmode\color{darkgreen}\checkmark  & \leavevmode\color{red}\usym{2717} & \leavevmode\color{red}\usym{2717} & \leavevmode\color{darkgreen}\checkmark & \leavevmode\color{red}\usym{2717} & \leavevmode\color{red}\usym{2717}  & \leavevmode\color{red}\usym{2717}\\ 
 & Causal effect estimation                             & \leavevmode\color{darkgreen}\checkmark  & \leavevmode\color{red}\usym{2717} & \leavevmode\color{red}\usym{2717} & \leavevmode\color{red}\usym{2717} & \leavevmode\color{red}\usym{2717} & \leavevmode\color{red}\usym{2717} & \leavevmode\color{darkgreen}\checkmark & \leavevmode\color{darkgreen}\checkmark \\  
 & Monotone regression                             & \leavevmode\color{red}\usym{2717}  & \leavevmode\color{darkgreen}\checkmark & \leavevmode\color{red}\usym{2717} & \leavevmode\color{red}\usym{2717} & \leavevmode\color{red}\usym{2717} & \leavevmode\color{red}\usym{2717} & \leavevmode\color{red}\usym{2717} & \leavevmode\color{red}\usym{2717} \\  
 & Survival analysis                             & \leavevmode\color{red}\usym{2717}  & \leavevmode\color{darkgreen}\checkmark & \leavevmode\color{red}\usym{2717} & \leavevmode\color{red}\usym{2717} & \leavevmode\color{darkgreen}\checkmark & \leavevmode\color{red}\usym{2717} & \leavevmode\color{red}\usym{2717} & \leavevmode\color{red}\usym{2717} \\  
 & Dirichlet process errors                             & \leavevmode\color{red}\usym{2717}  & \leavevmode\color{darkgreen}\checkmark & \leavevmode\color{red}\usym{2717} & \leavevmode\color{red}\usym{2717} & \leavevmode\color{darkgreen}\checkmark & \leavevmode\color{red}\usym{2717} & \leavevmode\color{red}\usym{2717} & \leavevmode\color{red}\usym{2717} \\  \hline
\end{tabular}
\caption{\label{tab:overview} Comparison of BART packages}
\end{table}

\subsection{The BART prior}

The classic supervised learning BART model (\cite{chipman2010bart}) with $m$ trees is given by 
\begin{equation} \label{eq:bart-model}
\begin{aligned}
Y_i \mid X_i = x_i &\iid \mathrm{N}\left(f(x_i), \sigma^2\right),\\
f(x_i) &= \sum_{s=1}^m g_s(x_i),\\
\sigma^2 &\sim \mathrm{IG}\left(a, b \right),
\end{aligned}
\end{equation}
where $g_s(x)$ denotes a decision tree, which partitions $\mathcal{X}$ into $k_s$ disjoint regions (denoted $\mathcal{A}_{s,j})$ each of which is associated with a constant scalar parameter $\mu_{s,j}$. Thus, for the $s$th tree 
\begin{equation}\label{eq:g-function}
\begin{aligned}
g_s(x) &= \sum_{j = 1}^{k_s} \mu_{s,j} \mathbb{I}\left(x \in \mathcal{A}_{s,j}\right).
\end{aligned}
\end{equation}
The leaf means, $\mu_{s,j}$, are assigned independent normal priors:
\begin{equation}
\begin{aligned}
\mu_{s,j} &\iid \mathrm{N}\left(0, \sigma^2_{\mu}\right).
\end{aligned}
\end{equation}
In \pkg{stochtree}, the prior variance over the leaf means, $\sigma^2_{\mu}$, can be either user-assigned or estimated, in which case it is given an inverse gamma prior. 

\cite{chipman1998bayesian} (CGM98) introduces a ``process prior'' over decision trees and their associated partitions, $\mathcal{A}_{s,1} \dots \mathcal{A}_{s, k_s}$, describing how one would sample from the tree prior. Essentially, trees are grown sequentially, with both the splitting variable and the cut point drawn uniformly from the available features and cut point values. What remains is to specify how this process (stochastically) terminates in a way that yields valid prior probabilities for any given decision tree. This is achieved by defining the split probability at depth $d$ to be 

\begin{equation}
P\left(\text{split at depth }d\right) = \frac{\alpha}{(1+d)^{\beta}}\;\;;\;\; 0 < \alpha \leq 1 ; \; \beta > 0.
\end{equation}

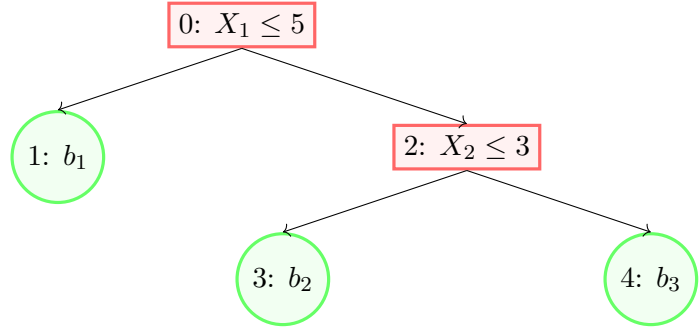
\begin{figure}
\centering
\begin{tikzpicture}[
roundnode/.style={circle, draw=green!60, fill=green!5, very thick, minimum size=7mm},
squarednode/.style={rectangle, draw=red!60, fill=red!5, very thick, minimum size=5mm},
]

\node[squarednode]      (split1)       {0: $X_1 \leq 5$};
\node[roundnode]          (leaf1)       [below left=of split1] {1: $b_1$};
\node[squarednode]      (split2)       [below right=of split1] {2: $X_2 \leq 3$};
\node[roundnode]          (leaf2)       [below left=of split2] {3: $b_2$};
\node[roundnode]          (leaf3)       [below right=of split2] {4: $b_3$};

\draw[->] (split1.south) -- (leaf1.north);
\draw[->] (split1.south) -- (split2.north);
\draw[->] (split2.south) -- (leaf2.north);
\draw[->] (split2.south) -- (leaf3.north);

\end{tikzpicture}
\caption{Example of a decision tree with five nodes and two splits.}
\label{fig:tree_diagram}
\end{figure}

Thus, for example, the tree depicted in Figure \ref{fig:tree_diagram} would have prior probability of 
$$\alpha\left(\frac{\alpha}{2^{\beta}}\right)\left(1-\frac{\alpha}{2^{\beta}}\right)\left(1-\frac{\alpha}{3^{\beta}}\right)^2.$$

The \pkg{stochtree} package contains a number of built-in extensions of the original BART model, as well as the ability to specify custom variants. For this reason, it will be useful to introduce the following notation: any function $f(x) = \sum_{s=1}^m g_s(x)$ composed of $m$ trees given the CGM98 prior with parameters $\alpha$ and $\beta$ will be said to have a ``BART prior'', denoted in modeling notation as
\begin{equation}
f \sim \mathrm{BART}(\alpha, \beta, m).
\end{equation}

\subsection{Outline}

We illustrate the use of \pkg{stochtree} in three settings:
\begin{enumerate}
 \item straightfoward applications of existing models such as BART (Section \ref{sec:friedman-bart-supervised}) and BCF (Section \ref{sec:acic-data}), 
 \item models that include more sophisticated components like random effects (Section \ref{sec:random-effects}), heteroskedasticity (Section \ref{sec:user-guide-heteroskedasticity}) and linear leaf models (Section \ref{sec:user-guide-linear-leaf-model}), and 
 \item as a component of custom MCMC routines to fit nonstandard tree ensemble models (Section \ref{sec:user-guide-customization}).  
\end{enumerate}

Below, we illustrate the key functions for supervised learning and causal inference (i.e. treatment effect estimation) using BART and BCF (respectively) in \pkg{stochtree}. We review the \proglang{R} interface in the body of this document and point readers to the appendix to see the same workflow in \proglang{Python}.


\subsection{Datasets} \label{sec:demo-datasets}

We will use the following four datasets to demonstrate \pkg{stochtree}'s functionality. We briefly describe these datasets here, with detailed code for loading and preprocessing them, along with dictionaries giving feature names and descriptions, in Appendix \ref{app:dataset-details}. 

\begin{itemize}
    \item {\bf Friedman dataset}. Our first dataset is the simulated data generating process (DGP) presented in \cite{friedman1991multivariate}. This function uses the five features of a dataset to define a nonlinear conditional mean model, so it is a useful demo dataset for nonlinearity and sparsity in supervised learning settings. 
    \item {\bf Causal Friedman dataset}. The second demo dataset is a modified version of the Friedman dataset which includes an endogeneous binary treatment which has an additive, heterogeneous effect on the outcome. We use this dataset to demonstrate \pkg{stochtree}'s basic causal inference functionality.
    \item {\bf ACIC dataset}. The third demo dataset we use is a semi-synthetic dataset introduced by \cite{carvalho2019assessing}. The covariates are based on the randomized controlled trial of \cite{yeager2019national}, while the treatment assignment and outcome are generated in order to simulate confounding. We use this dataset to demonstrate some advanced causal inference features of \pkg{stochtree}, including the incorporation of group-level random effects (as the original study is clustered by school).
    \item {\bf Academic probation dataset}. We use the dataset from \cite{lindo2010ability} to demonstrate how \pkg{stochtree}'s ``leaf linear model'' interface can be used to fit a regression discontinuity design (RDD) model.
    \item {\bf Motorcycle dataset}. The motorcycle accident dataset was originally used in \cite{silverman1985some} and has since been used in many papers and demos (cf. \cite{rasmussen2001infinite} and \cite{gramacy2007tgp}). The dataset records a simulated motorcycle accident, measuring head acceleration at various times since the initial impact. 
\end{itemize}

\section{Basic stochtree Workflow} \label{sec:user-guide-basic-function-call}

\subsection{Overview}
The \code{stochtree::bart()} function in \pkg{stochtree} takes as input response and predictor data and returns posterior samples from a BART model. BART models reside in-memory as a \code{bartmodel} object, which includes model metadata, parameter draws, and pointers to \proglang{C++} representations of tree ensemble samples. Out-of-sample prediction is accomplished through a \code{predict()} function which accepts a \code{bartmodel} object and an array of predictors dictating the points at which predictions are desired.

In its simplest form (using default priors and default MCMC settings) this process consists of one call to \code{stochtree::bart()}:
\begin{CodeChunk}
\begin{CodeInput}
R> bart_model <- stochtree::bart(X_train = X_train, y_train = y_train)
\end{CodeInput}
\end{CodeChunk}
and one call to \code{predict()}:
\begin{CodeChunk}
\begin{CodeInput}
R> bart_preds <- predict(bart_model, X = X_test)
\end{CodeInput}
\end{CodeChunk}

However, the \pkg{stochtree} workflow consists of a number of choices prior to calling these core functions and a number of choices after as well. In the following subsections we will examine three preparatory steps: data preprocessing (Section \ref{sec:preprocessing}), prior specification (Section \ref{sec:prior-parameters}), and algorithm settings (Section \ref{sec:mcmc-settings}); as well as
 three post-fitting steps: convergence diagnostic plots, extracting model fit information, and extracting (point and interval) predictions.

We present this basic workflow in terms of the \code{stochtree::bart()} function, but the other core function \code{stochtree::bcf()} for causal inference, is closely analagous and will be covered in one of the illustrations. 

The Python API is discussed in depth in Appendix \ref{sec:python-user-guide-basic-function-call}, but we note here that it follows a very similar pattern to the R interface, with sampling and prediction achieved by running:
\begin{CodeChunk}
\begin{CodeInput}
>>> bart_model = BARTModel()
>>> bart_model.sample(X_train = X_train, y_train = y_train)
>>> bart_preds = bart_model.predict(X = X_test)
\end{CodeInput}
\end{CodeChunk}

\subsubsection{Data Preprocessing} \label{sec:preprocessing}

\pkg{stochtree} can accommodate both categorical and numerical predictor variables, but some preprocessing is necessary for the algorithm to utilize them properly. We demonstrate this below on the ACIC dataset, assuming that it has been loaded into a dataframe called \code{df} (see Appendix \ref{app:acic-dataset} for details on loading this dataset).

We begin by extracting the outcome (and treatment variable, in the case of BCF) from our data frame and storing them as vectors:
\begin{CodeChunk}
\begin{CodeInput}
R> y <- df$Y
R> Z <- df$Z
\end{CodeInput}
\end{CodeChunk}

Categorical variables should be coded as ordered or unordered factors; by default the response variable is standardized internally but results are reported on the original scale.
\begin{CodeChunk}
\begin{CodeInput}
R> unordered_categorical_cols <- c("C1","XC")
R> ordered_categorical_cols <- c("S3","C2","C3")
R> for (col in unordered_categorical_cols) {
R>     df[,col] <- factor(df[,col], ordered = F)
R> }
R> for (col in ordered_categorical_cols) {
R>     df[,col] <- factor(df[,col], ordered = T)
R> }
\end{CodeInput}
\end{CodeChunk}

\subsubsection{Specifying Prior Parameters} \label{sec:prior-parameters}

While BART models are known to work remarkably well using default prior distributions, users can depart freely from these defaults in various ways. \pkg{stochtree}'s prior parameters are organized according to user-passed lists. In the basic case of a BART fit, the relevant lists are  \code{mean_forest_params}, which governs the prior over the trees, and \code{general_params}, which governs global model parameters as well as MCMC parameters, including model initialization. 

If we wish to change model parameters, such as the number of trees or node split hyperparameters $\alpha$ and $\beta$, we do so via parameter lists
\begin{CodeChunk}
\begin{CodeInput}
R> mean_forest_params <- list(alpha = 0.25, beta = 3, num_trees = 20)
R> bart_model <- stochtree::bart(X_train = X_train, y_train = y_train, 
R>                    mean_forest_params = mean_forest_params, 
R>                    num_gfr = 0, num_burnin = 1000, num_mcmc = 1000)
\end{CodeInput}
\end{CodeChunk}

\subsubsection{Algorithm Settings} \label{sec:mcmc-settings}

Users may specify the length of the desired Markov chain, number of burn-in steps, and certain Markov chain initialization choices. 

In particular, initialization can be determined by the XBART algorithm, a greedy approximation to the BART MCMC procedure that rapidly converges to high likelihood tree ensembles \citep{he2023stochastic}, by regrowing each tree from root at each iteration of the algorithm. Because XBART involves a number of additional innovations, here we will refer to its core algorithm separately as grow-from-root (GFR). On its own, grow-from-root represents a fast approximate BART fit for prediction problems, particularly those with large training sample sizes; but grow-from-root can also be used to initialize the BART MCMC algorithm at favorable parameter values, thereby substantially reducing the necessary burn-in period. Following \cite{he2023stochastic} we refer to this strategy as ``warm start''.

By default, \code{stochtree::bart()} runs 10 iterations of the grow-from-root (GFR) algorithm \citep{he2023stochastic} and 100 MCMC iterations initialized by the final GFR forest. We can modify these values directly in the \code{stochtree::bart()} function call
\begin{CodeChunk}
\begin{CodeInput}
R> bart_model <- stochtree::bart(X_train = X_train, y_train = y_train, 
R>                    num_gfr = 0, num_burnin = 1000, num_mcmc = 1000)
\end{CodeInput}
\end{CodeChunk}

\subsubsection{Prediction}
Finally, predictions are obtained as pointwise posterior means via the \code{predict()} function. 
\begin{CodeChunk}
\begin{CodeInput}
R> y_hat_test <- predict(bart_model, X = X_test, 
R>                       terms = "y_hat", type = "mean")
\end{CodeInput}
\end{CodeChunk}

\subsubsection{Serialization and Advanced Sampling Options} \label{sec:user-guide-mcmc-niceties}

Should one need to revisit and modify a previously-fit BART model, \pkg{stochtree} model objects can be saved to JSON via the \code{saveBARTModelToJsonString()} function and reloaded via the \code{createBARTModelFromJsonString()} function. This can be useful when further MCMC iterations are needed or when a BART model is being used as one step of a more elaborate model. This functionality will be demonstrated in Section \ref{sec:friedman-bart-supervised}. Note that these objects can be ported back and forth between \proglang{R} and \proglang{Python}, facilitating convenient cross-platform interoperability. 

\subsection{Example: Supervised Learning with BART (Friedman dataset)} \label{sec:friedman-bart-supervised}

The Friedman dataset is generated by a nonlinear function of 5 variables and additive Gaussian noise.
\begin{equation*}
\begin{aligned}
Y_i \mid X_i = x_i &\iid \mathrm{N}\left(f(x_i), \sigma^2\right)\\
f(x) &= 10 \sin \left(\pi x_1 x_2\right) + 20 (x_3 - 1/2)^2 + 10 x_4 + 5 x_5\\
X_{i,j} &\iid \text{U}\left(0,1\right), \;\;\;\; j = 1 \dots p.
\end{aligned}
\end{equation*}
Note that $p = \mbox{dim}(X)$ must be at least five but can potentially be much larger. This makes the Friedman dataset appropriate for evaluating models' ability to handle nonlinearity and sparsity (for $p \gg 5$).

Here we have $p = 100$ and $n = 500$. Our Bayesian model will simply be $f \sim \mbox{BART}(\alpha = 0.25,\beta = 2, m = 200)$. We initialize the model by running 20 GFR iterations using default settings.

\begin{CodeChunk}
\begin{CodeInput}
R> xbart_model <- stochtree::bart(X_train = X_train, y_train = y_train, X_test = X_test, 
R>                     num_gfr = 20, num_mcmc = 0)
\end{CodeInput}
\end{CodeChunk}

We then save the model and restart sampling according to the Markov chain, obtaining 10,000 samples under a more conservative prior ($\alpha = 0.25$ and a maximum tree depth of eight). 

\begin{CodeChunk}
\begin{CodeInput}
R> xbart_json <- saveBARTModelToJsonString(xbart_model)
R> mean_forest_params = list(
R>     alpha = 0.25, beta = 2, min_samples_leaf = 10, max_depth = 8
R> )
R> bart_model <- stochtree::bart(X_train = X_train, y_train = y_train, X_test = X_test, 
R>                    num_gfr = 0, num_burnin = 0, num_mcmc = 10000,
R>                    mean_forest_params = mean_forest_params, 
R>                    previous_model_json = xbart_json, 
R>                    previous_model_warmstart_sample_num = 20)
\end{CodeInput}
\end{CodeChunk}

\begin{figure}[t!]
\centering
\includegraphics{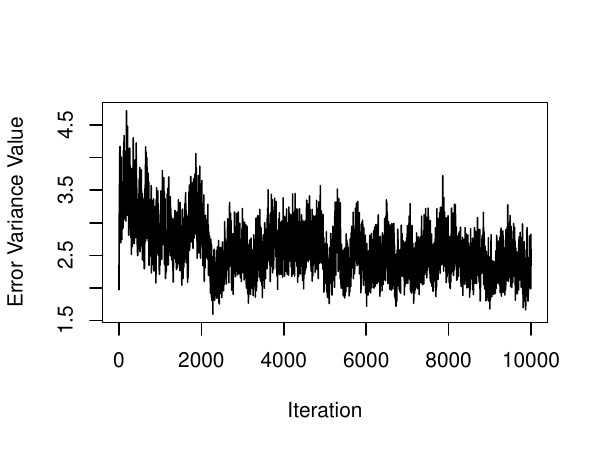}
\caption{\label{fig:friedman-bart-r-traceplot-warm-start} Traceplot of the global error variance parameter, $\sigma^2$, sampled as part of a homoskedastic BART model fit with the \pkg{stochtree} \proglang{R} package.}
\end{figure}

Figure \ref{fig:friedman-bart-r-traceplot-warm-start} shows the traceplot for the residual variance paramter, $\sigma^2$, produced with the following code:
\begin{CodeChunk}
\begin{CodeInput}
R> plot(bart_model$sigma2_global_samples, type = "l"
R>      xlab = "Iteration", ylab = "Error Variance Value")
\end{CodeInput}
\end{CodeChunk}

\begin{figure}[t!]
\centering
\includegraphics{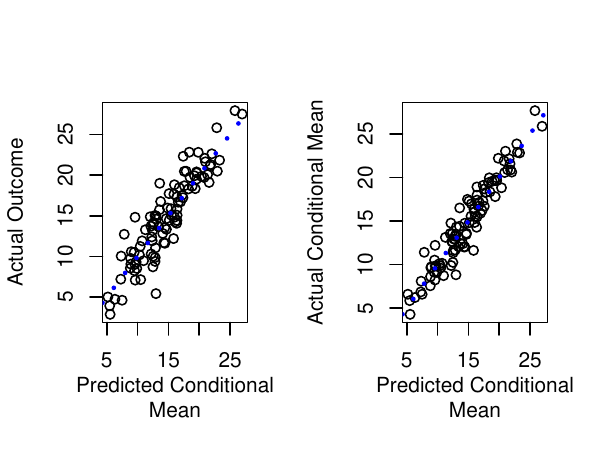}
\caption{\label{fig:friedman-bart-r-pred-actual-warm-start} Predictions (posterior means) plotted against actual outcomes on a hold-out set, for a homoskedastic BART model fit with the \pkg{stochtree} \proglang{R} package.}
\end{figure}

Figure \ref{fig:friedman-bart-r-pred-actual-warm-start} plots the predicted conditional mean estimates to observed outcomes for a holdout set, produced with the following code:
\begin{CodeChunk}
\begin{CodeInput}
R> y_hat_test <- predict(bart_model, X = X_test, 
R>                    terms = "y_hat", type = "mean")
R> par(mfrow = c(1, 2))
R> plot(y_hat_test, y_test, xlab = "Predicted Conditional\nMean", 
R>        ylab = "Actual Outcome")
R> abline(0,1,col="blue", lty=3, lwd=3)
R> plot(y_hat_test, m_x_test, xlab = "Predicted Conditional\nMean", 
R>        ylab = "Actual Conditional Mean")
R> abline(0,1,col="blue", lty=3, lwd=3)
\end{CodeInput}
\end{CodeChunk}

\subsection{Example: Causal Inference with BCF (Causal Friedman dataset)}
\label{sec:acic-data}

\cite{hahn2020bayesian} introduce a model for the effect of a binary treatment $Z$ on an outcome $Y$ given covariates $x$
\begin{equation} \label{eq:bcf-model}
\begin{aligned}
Y_i \mid x_i, z_i &\sim \mathrm{N}(f_0(x_i) + \tau(x_i) z_i, \sigma^2)\\
f_0 &\sim \mathrm{BART}(\alpha_0, \beta_0, m_0)\\
\tau &\sim \mathrm{BART}(\alpha_{\tau}, \beta_{\tau}, m_{\tau}).
\end{aligned}
\end{equation}
Provided several standard assumptions are met, $\tau(x)$ may be interpreted as the {\it causal} effect of $Z$ on $Y$, conditional on $X$. In potential outcomes notation,
\begin{equation}
\E[Y_i(1) - Y_i(0) \mid X=x_i] = \tau(x_{i})
\end{equation}
\cite{krantsevich2023stochastic} develop the ``grow-from-root'' analog for BCF and refer to the resulting algorithm as XBCF. \pkg{stochtree} supports BCF and XBCF, with extensions to allow for continuous and multivariate treatments and forest-based heteroskedasticity.

Basic function calls to BCF mirror those of BART. Models are fitted with a call to \code{stochtree::bcf()}, which returns a \code{bcfmodel} object, and \code{predict()} computes response predictions, as well as prognostic ($f_0(x))$ and treatment effect ($\tau(x)$) functions. Likewise, models are serialized via \code{saveBCFModelToJsonString()} and reloaded via \code{createBCFModelFromJsonString()}.

A number of features make BCF more suitable for causal inference problems than regular BART. One, the BART priors on $f_0$ and $\tau$ can be specified separately, including involving distinct subsets of features. Two, a propensity function is automatically fit and used as a control covariate (or one may be supplied by the user), mitigating ``regularization-induced confounding", cf. \cite{hahn2020bayesian}. 

The following code fits a BCF model to the causal Friedman data set. We begin by fitting a propensity score model, also using \pkg{stochtree}, and specifying a probit link, as described in \cite{chipman2010bart}. That is, for binary treatment variable $Z_i$ and covariate vector $x_i$,
\begin{equation} \label{eq:bart-model-probit}
\begin{aligned}
\Pr(Z_i = 1 \mid x_i) &\sim \Phi(\pi(x_i)),\\
\pi &\sim \mathrm{BART}(\alpha, \beta, m),
\end{aligned}
\end{equation}
where $\Phi(\cdot)$ denotes the standard normal CDF.

This model can be fit in \pkg{stochtree} simply by specifying a probit model in the general parameters list and suppressing sampling of the residual error variance ($\sigma^2$), which is unidentified for probit models.

\begin{CodeChunk}
\begin{CodeInput}
R> general_params_propensity = list(probit_outcome_model = T, 
R>                                  sample_sigma2_global = F)
R> propensity_model <- stochtree::bart(
R>     X_train = X, y_train = Z, 
R>     general_params = general_params_propensity)
R> propensity <- predict(propensity_model, X = X, 
R>                       terms = "y_hat", type = "mean")
\end{CodeInput}
\end{CodeChunk}

We then pass this estimated propensity score vector to the \code{stochtree::bcf()} function as an argument.
\begin{CodeChunk}
\begin{CodeInput}
R> bcf_model <- stochtree::bcf(
R>     X_train = X, Z_train = Z, y_train = y, 
R>     propensity_train = propensity_train
R> )
\end{CodeInput}
\end{CodeChunk}

If propensities are not provided to the \code{stochtree::bcf()} function, \pkg{stochtree} will fit a propensity model automatically unless the user specifies otherwise by argument \code{propensity_covariate = "none"} in the \code{general_params} list.
Users retain more control\footnote{It could be fit with a method other than \pkg{stochtree}, as \code{stochtree::bcf()} merely requires a vector of estimated propensity scores.} over the propensity model specification by pre-fitting it, as shown here. 
By default, this option is set to \code{propensity_covariate = "mu"}, indicating that the propensity score is included in the $f_0$ function.

\begin{CodeChunk}
\begin{CodeInput}
R> bcf_model <- stochtree::bcf(
R>     X_train = X, Z_train = Z,
R>     y_train = y, propensity_train = propensity
R> )
\end{CodeInput}
\end{CodeChunk}

\begin{figure}[t!]
\centering
\includegraphics[width=0.7\linewidth]{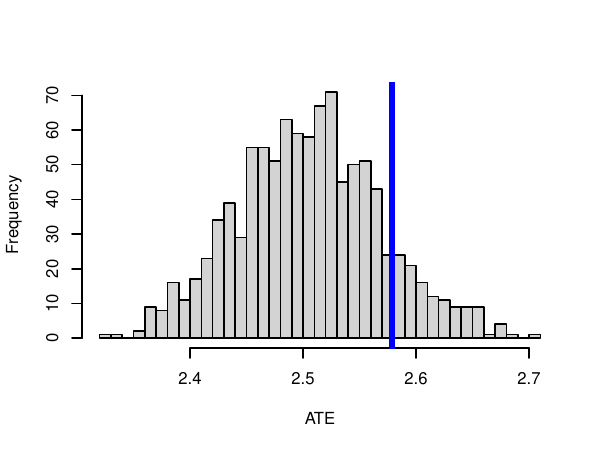}
\caption{\label{fig:simulated-ate-r} Posterior distribution of average treatment effect (ATE) estimated by homoskedastic BCF model using the \pkg{stochtree} \proglang{R} package.}
\end{figure}

Figure \ref{fig:simulated-ate-r} plots a histogram of posterior samples of the average treatment effect (ATE) as estimated by BCF: 
\begin{CodeChunk}
\begin{CodeInput}
R> tau_hat_posterior <- predict(bcf_model, X = covariate_df,
R>     Z = Z, propensity = propensity,
R>     type = "posterior", terms = "cate"
R> )
R> ate_posterior <- colMeans(tau_hat_posterior)
R> hist(ate_posterior, xlab = "ATE", main = NULL)
\end{CodeInput}
\end{CodeChunk}

\begin{figure}[t!]
\centering
\includegraphics[width=0.6\linewidth]{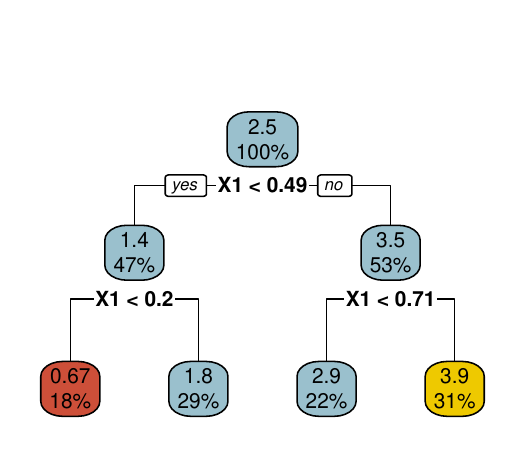}
\caption{\label{fig:simulated-cate-tree} Decision tree summary of the CATE posterior mean, as estimated by a homoskedastic BCF model using the \pkg{stochtree} \proglang{R} package.}
\end{figure}

Now, we investigate the nature of the observed treatment effect moderation by fitting a CART tree to the posterior mean of $\tau(X)$. The posterior summary tree presented in Figure \ref{fig:simulated-cate-tree} flags $X_1$ as an important moderating variable, successfully recovering the true DGP, which has $\tau(X) = 5 X_1$. 

\begin{figure}[t!]
\centering
\includegraphics{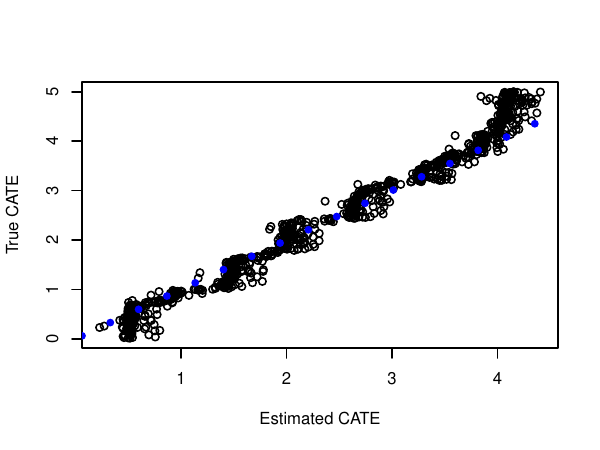}
\caption{\label{fig:simulated-cate-true-fitted} True conditional average treatment effect (CATE) compared to the CATE estimated by a homoskedastic BCF model using the \pkg{stochtree} \proglang{R} package.}
\end{figure}

Figure \ref{fig:simulated-cate-true-fitted} plots the estimated conditional average treatment effect (CATE) against the true conditional average treatment effect function.

Note that, in addition to a continuous outcome and a binary treatment (as shown in this example), \code{stochtree::bcf()} can accommodate binary outcomes and/or a continuous treatment. We do not demonstrate these variants of the model here due to space constraints; see package documentation for details.

\section{Advanced Models} \label{sec:user-guide-advanced}

\subsection{Additive Random Effects Model}\label{sec:random-effects}

\subsubsection{A Random Effects Model} \label{sec:rfx-model}

BART and BCF models in \pkg{stochtree} include specialized algorithms for fitting one-way random intercepts and random slopes with independent variance components (using the redundant parameterizations of \citep{gelman2008redundant}). These models are appropriate for observations with group-wise structure and dependence, such as patients nested within providers or students within schools. 

These random effects essentially serve two purposes. Thinking about the model after marginalizing the random effects, they induce a correlated error structure that (heuristically) captures the effect of omitted group-level variables or non-random selection into groups. Thinking about the model conditionally on the random effects they capture the predictive impact of group-level membership -- for example, in a  model with random intercepts we can account for the fact that some groups have responses that are higher or lower than others, even after adjusting for the effect of group and individual-level covariates included in the model. 

The most common use-case in our applied work is augmenting the BCF model with a random intercept and slope (of treatment assignment):
\begin{equation} \label{eq:bcf-model-rfx}
\begin{aligned}
Y_{ij} \mid x_{ij}, z_{ij} &\sim \mathrm{N}(\gamma_{j} + f_0(x_{ij}) + [\xi_{j} + \tau(x_{ij})]z_{ij}, \sigma^2)\\
f_0 &\sim \mathrm{BART}(\alpha_0, \beta_0, m_0)\\
\tau &\sim \mathrm{BART}(\alpha_{\tau}, \beta_{\tau}, m_{\tau})\\
\gamma_j&\iid \mathrm{N}(0, \sigma^2_\gamma)\\
\xi_j&\iid \mathrm{N}(0, \sigma^2_\xi)
\end{aligned}
\end{equation}
In this model $\xi_j$ in particular captures variability in treatment effects across groups that isn't captured by the $\tau$ function. In a prediction context, a BART model with varying intercepts is recovered on fixing $\xi$ and $\tau$ to be zero.
The prior on the variance components $\sigma^2_\gamma$ and $\sigma^2_\xi$ is induced through parameter expansion following \cite{gelman2008redundant}. 

Generic one-way varying intercepts/slopes are possible in both \code{stochtree::bart()} and \code{stochtree::bcf()}. In this case, the conditional mean function would be
\begin{equation}\label{eq:bcf-model-rfx-gen}
\beta_j'w_{ij} + f_0(x_{ij}) + \tau(x_{ij})z_{ij},
\end{equation}
where $w_i$ includes an intercept (for group-wise intercepts) and other terms constructed from $x_{ij}$ and $z_{ij}$. The model in \eqref{eq:bcf-model-rfx} is recovered by setting $w_{ij} = (1,z_{ij})'$, in which case $\beta_j = (\gamma_j, \xi_j)'$.

The one-way grouping structure and {\em a priori} uncorrelated pairs $(\gamma_j, \xi_j)$ (or more generically $\beta_j$ with a diagonal covariance matrix in its prior) is a special (but very common) case of embedding BART functions into larger multilevel models. The \code{stochtree::bart()} and \code{stochtree::bcf()} functions do not handle these out-of-the-box, but in Section~\ref{sec:user-guide-additive-linear-model} we show how to add linear terms with arbitrary coefficient priors using \pkg{stochtree} to build a custom sampler\footnote{For at least some models the package \code{stan4bart} would be a viable alternative. It extends \code{rstanarm} to incorporate BART functions in generic multilevel models, but does not accommodate bcf-style specifications, heteroskedastic variance functions with BART priors, and other \pkg{stochtree} features.}.

\subsubsection{A random effects analysis on the ACIC 2019 Data} \label{sec:acic-bcf-rfx}

The data in \cite{carvalho2019assessing} (see Appendix \ref{app:acic-dataset}) has an individual student-level intervention (randomized at the student level) and students nested within 76 schools. Covariates are measured at the student and school level. This structure is naturally accommodated by the model in~\eqref{eq:bcf-model-rfx}. 
We fit school-level random intercept and treatment effects by specifying group IDs as \code{rfx_group_ids_train}. 
The model specified in \eqref{eq:bcf-model-rfx} is supported in \code{stochtree::bcf()} without users having to construct and specify \code{rfx_basis} manually. If \code{model_spec = "intercept_plus_treatment"} is specified in the \code{random_effects_params} parameter list, \pkg{stochtree} will handle basis construction at both sampling and prediction time.

\begin{itemize}
    \item \code{rfx_group_ids_train}: labels that define groups or clusters in a random effects model, and 
    \item \code{rfx_basis_train}: bases on which random slopes are fitted.
\end{itemize}

The random-intercepts-random-slopes model specified in \eqref{eq:bcf-model-rfx} is obtained by setting \code{model_spec = "intercept_plus_treatment"} in the \code{random_effects_params} parameter list; \pkg{stochtree} also supports a \code{model_spec = "intercept_only"} random effects specification for both BART and BCF. In this model the CATE function is 
\begin{equation}
\E[Y_i(1) - Y_i(0) \mid X=x_{ij}, i\in \text{group } j] = \xi_j + \tau(x_{ij}),    
\end{equation}
unlike in BCF where the CATE was simply given by $\tau$. In the random effects model we may be interested in the CATE function above or the covariate-dependent factor $\tau$.

\cite{hahn2020bayesian} note that, in many cases, researchers only consider a subset of their covariates to be plausible treatment effect moderators. 
\pkg{stochtree} offers a simple way to specify this expectation without providing separate covariate data for each forest. Parameter lists for the prognostic forest (\code{prognostic_forest_params}) and the treatment effect forest (\code{treatment_effect_forest_params}) both allow users to specify a list of variables to retain (\code{keep_vars}) or drop from a forest (\code{drop_vars}). In the analytical challenge that accompanied this dataset, \cite{carvalho2019assessing} asked researchers to assess whether $X_1$ and $X_2$ moderate the treatment effect, so we fit a BCF model with only these two variables included as moderators.

The BCF model with an additive random effects term and $\tau(X)$ specified as a function of $X_1$ and $X_2$ is fit as follows:
\begin{CodeChunk}
\begin{CodeInput}
R> treatment_forest_params <- list(
R>     keep_vars = c("X1", "X2")
R> )
R> rfx_params <- list(model_spec = "intercept_plus_treatment")
R> bcf_model_rfx <- stochtree::bcf(
R>     X_train = X, Z_train = Z, 
R>     y_train = y, propensity_train = propensity, 
R>     rfx_group_ids_train = group_ids, 
R>     treatment_effect_forest_params = treatment_forest_params, 
R>     random_effects_params = rfx_params
R> )
\end{CodeInput}
\end{CodeChunk}

\begin{figure}[t!]
\centering
\includegraphics{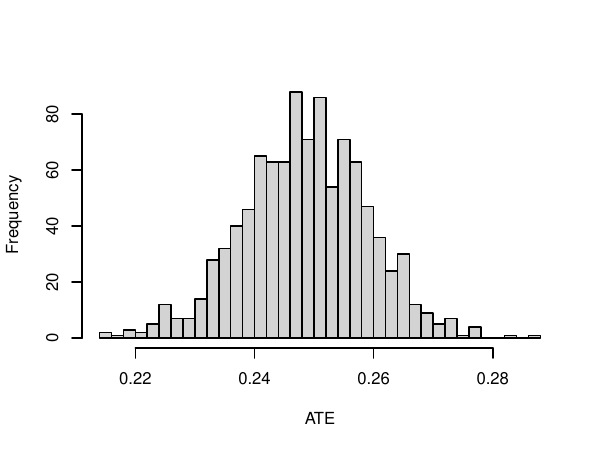}
\caption{\label{fig:acic-bcf-ate-posterior-rfx} Posterior distribution of average treatment effect (ATE) estimated by homoskedastic BCF model using the \pkg{stochtree} \proglang{R} package.}
\end{figure}

\begin{figure}[t!]
\centering
\includegraphics{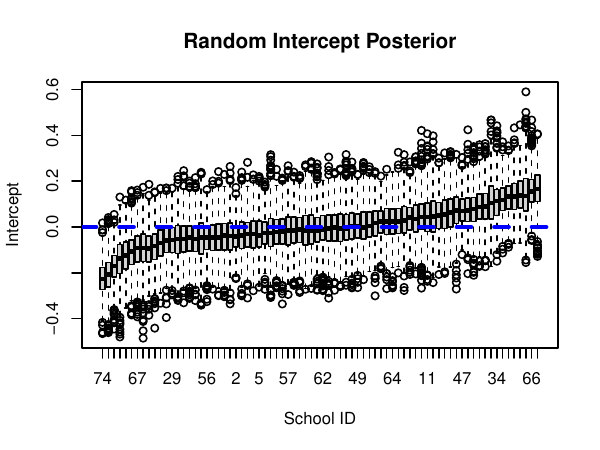}
\caption{\label{fig:acic-random-intercept-boxplot} Boxplot of random intercept posterior samples for all 76 schools, estimated using the \pkg{stochtree} \proglang{R} package.}
\end{figure}

Figure \ref{fig:acic-bcf-ate-posterior-rfx} reveals that the posterior ATE is largely unchanged by the more elaborate random effects specification, but that the school-level intercepts are in some cases quite pronounced (Figure \ref{fig:acic-random-intercept-boxplot}).

Next, we demonstrate how to extract posterior samples from the random effects model to produce Figure \ref{fig:acic-random-intercept-boxplot}. 
The \code{getRandomEffectSamples} function returns a list of arrays that correspond to terms in the \cite{gelman2008redundant} parameterization. The \code{beta_samples} entry of this list corresponds to $\beta_j$ in \eqref{eq:bcf-model-rfx-gen} and it is stored as an array of dimension $(k,l,m)$ where $k$ is the dimension of $\beta_j$, $l$ is the number of groups in the study, and $m$ is the number of posterior samples. As we are interested in the random intercepts, we restrict our attention to \code{rfx_betas[1, , ]}.
\begin{CodeChunk}
\begin{CodeInput}
R> rfx_samples <- getRandomEffectSamples(bcf_model_rfx)
R> rfx_betas <- rfx_samples$beta_samples
R> random_intercepts <- as.data.frame(rfx_betas[1, , ])
R> random_intercepts$schoolid <- 1:76
R> random_intercepts_long <- reshape(
R>     random_intercepts,
R>     idvar = "schoolid",
R>     varying = list(1:bcf_model_rfx$model_params$num_samples),
R>     v.names = "V",
R>     direction = "long"
R> )
R> box_out <- boxplot(
R>     V ~ schoolid,
R>     data = random_intercepts_long,
R>     coef = 0,
R>     xlab = "School ID",
R>     ylab = "Intercept",
R>     main = "Random Intercept Posterior"
R> )
R> abline(h = 0, lty = 2, lwd = 3, col = "blue")
\end{CodeInput}
\end{CodeChunk}

\begin{figure}[t!]
\centering
\includegraphics[width=0.8\linewidth]{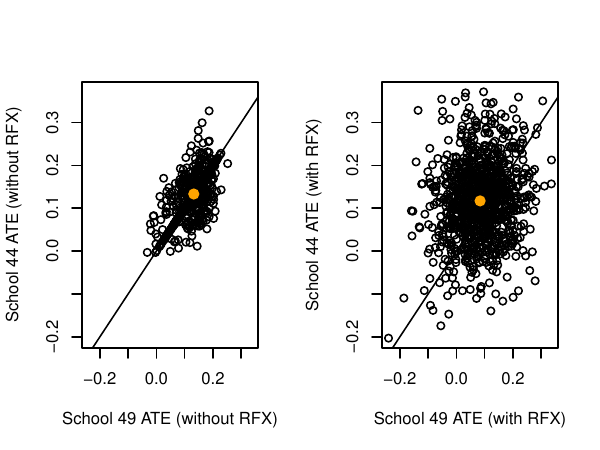}
\caption{\label{fig:acic-bcf-rfx-comparison} Comparison of subgroup ATE for two schools in BCF models fit with random effects and without random effects, both estimated using the \pkg{stochtree} \proglang{R} package.}
\end{figure}

To visualize the impact of fitting a random effects BCF model, compared to a BCF model with no random effects, Figure \ref{fig:acic-bcf-rfx-comparison} plots posterior samples of the school-average treatment effect for two schools, given by:
\begin{equation}
\frac{1}{n_j}\sum_{i\in \text{group } j}\E[Y_i(1) - Y_i(0) \mid X=x_{ij}, i\in \text{group } j] = \xi_j + \frac{1}{n_j}\sum_{i\in \text{group } j}\tau(x_{ij})
\end{equation}
Observe that the school-level treatment effects show stronger correlation in the no-random-effects model, due to shared group level covariates, while the random effects specification allows the observed differences in the data to be attributed to idiosyncratic random effects, leading to weaker posterior dependence. 

\subsection{Heteroskedastic errors} \label{sec:user-guide-heteroskedasticity}

\subsubsection{BART with Heteroskedastic Errors}

\cite{pratola2020heteroscedastic} and \cite{murray2021log} extend the classic BART model with a log-linear forest modeling the conditional variance function. 
\begin{equation*}
\begin{aligned}
Y_i \mid x_i &\sim \mathrm{N}\left(f(x_i), \sigma^2_0 \exp{(h(x_i)})\right)\\
f &\sim \mathrm{BART}(\alpha_f, \beta_f, m_f) \\
h &\sim \mathrm{logBART}(\alpha_h, \beta_h, m_h)
\end{aligned}
\end{equation*}
where $\mathrm{logBART}$ denotes the same tree prior as BART, but with leaf parameters $\lambda_{s,j}$ assigned independent (log) inverse-gamma priors: 
\begin{equation*}
\begin{aligned}
\exp(\lambda_{s,j}) &\iid \text{IG}\left(a, b\right).
\end{aligned}
\end{equation*}

Heteroskedastic forests are available in \pkg{stochtree} for both supervised learning (BART) and causal inference (BCF); the next example demonstrates a simple heteroskedastic supervised learning model. 

\cite{murray2021log} also applies the log-linear parameterization for multiclass classification and zero-inflated count regression, in addition to the conditional variance estimation example demonstrated here; \pkg{stochtree} will eventually support this broader set of log-linear models.

\subsubsection{Heteroskedastic Motorcycle Data} \label{sec:bart-motorcycle}

Conditional variance may be modeled in \pkg{stochtree} by specifying \code{num_trees > 0} in the \code{variance_forest_params} parameter list.
\begin{CodeChunk}
\begin{CodeInput}
R> num_gfr <- 10
R> num_burnin <- 0
R> num_mcmc <- 100
R> general_params <- list(sample_sigma2_global = F)
R> variance_forest_params <- list(num_trees = 20, alpha = 0.5, 
R>                                beta = 3.0, min_samples_leaf = 20)
R> bart_model_het <- stochtree::bart(
R>     X_train = as.matrix(mcycle$times), y_train = mcycle$accel, 
R>     num_gfr = num_gfr, num_burnin = num_burnin, num_mcmc = num_mcmc, 
R>     general_params = general_params, 
R>     variance_forest_params = variance_forest_params
R> )
\end{CodeInput}
\end{CodeChunk}

We also construct a homoskedastic model for comparison.
\begin{CodeChunk}
\begin{CodeInput}
R> general_params <- list(sample_sigma2_global = T)
R> bart_model <- stochtree::bart(
R>     X_train = as.matrix(mcycle$times), y_train = mcycle$accel, 
R>     num_gfr = num_gfr, num_burnin = num_burnin, num_mcmc = num_mcmc, 
R>     general_params = general_params, 
R> )
\end{CodeInput}
\end{CodeChunk}

\begin{figure}[t!]
\centering
\includegraphics{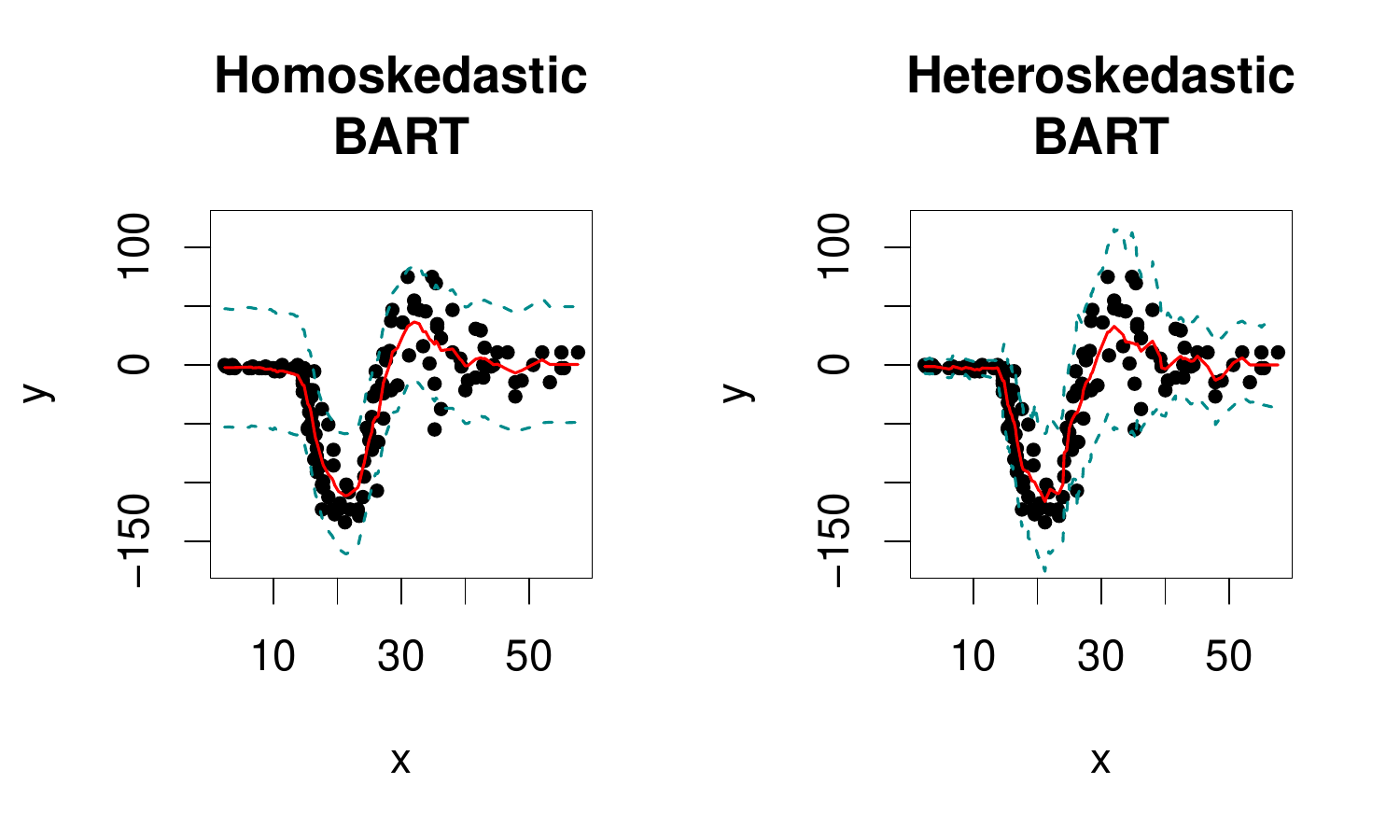}
\caption{\label{fig:motorcycle-model-comparison} Comparison of mean predictions and intervals on the motorcycle dataset from heteroskedastic and homoskedastic BART models using the \pkg{stochtree} \proglang{R} package.}
\end{figure}

Posterior mean and prediction interval from the two models are shown for comparison in Figure~\ref{fig:motorcycle-model-comparison}. The patent heteroskedasticity of the motorcycle data is well-captured by the heteroskedastic BART model (right), while the homoskedastic BART model (left) has too-wide prediction intervals for $x < 15$ or so.

\newpage
\subsection{Leafwise-Regression Models} \label{sec:user-guide-linear-leaf-model}

\begin{figure}[t!]
\centering
\begin{tikzpicture}[
    scale=0.7,
      node/.style={%
        draw,
        rectangle,
      },
    ]

      \node [node] (A) {$x_1<0.9$};
      \path (A) ++(-135:\nodeDist) node [label=below:$\left(\begin{smallmatrix}\beta_{11}\\\beta_{21}\\\dots\\\beta_{k1}\end{smallmatrix}\right)$] (B) {};
      \path (A) ++(-45:\nodeDist) node [node] (C) {$x_3<0.4$}; 
      \path (C) ++(-135:\nodeDist) node [label=below:$\left(\begin{smallmatrix}\beta_{12}\\\beta_{22}\\\dots\\\beta_{k2}\end{smallmatrix}\right)$] (D) {};
      \path (C) ++(-45:\nodeDist) node  [label=below:$\left(\begin{smallmatrix}\beta_{13}\\\beta_{23}\\\dots\\\beta_{k3}\end{smallmatrix}\right)$]  (E) {};

      \draw (A) -- (B) node [left,pos=0.25] {yes}(A);
      \draw (A) -- (C) node [right,pos=0.25] {no}(A);
      \draw (C) -- (D) node [left,pos=0.25] {yes}(A);
      \draw (C) -- (E) node [right,pos=0.25] {no}(A);
\end{tikzpicture} \begin{tikzpicture}[
    scale=0.7,
      node/.style={%
        draw,
        rectangle,
      },
    ]

      \node [node] (A) {$x_1<0.9$};
      \path (A) ++(-145:\nodeDist) node [label=below:{$\sum_{h=1}^k \beta_{h1}\psi_h(x,z)$}] (B) {};
      \path (A) ++(-35:\nodeDist) node [node] (C) {$x_3<0.4$}; 
      \path (C) ++(-145:\nodeDist) node [label=below:{$\sum_{h=1}^k \beta_{h2}\psi_h(x,z)$}] (D) {};
      \path (C) ++(-35:\nodeDist) node  [label=below:{$\sum_{h=1}^k \beta_{h3}\psi_h(x,z)$}]  (E) {};

      \draw (A) -- (B) node [left,pos=0.25] {}(A);
      \draw (A) -- (C) node [right,pos=0.25] {}(A);
      \draw (C) -- (D) node [left,pos=0.25] {}(A);
      \draw (C) -- (E) node [right,pos=0.25] {}(A);
  \end{tikzpicture}
  
  \caption{\label{fig:regtree} (Left) A single tree with multivariate parameters. (Right) The same tree defining leaf-wise regression predictions on a k-dimensional basis expansion $\Psi$ which can use variables defining potential splitting rules (x) or other auxiliary variables (z). }
\end{figure}
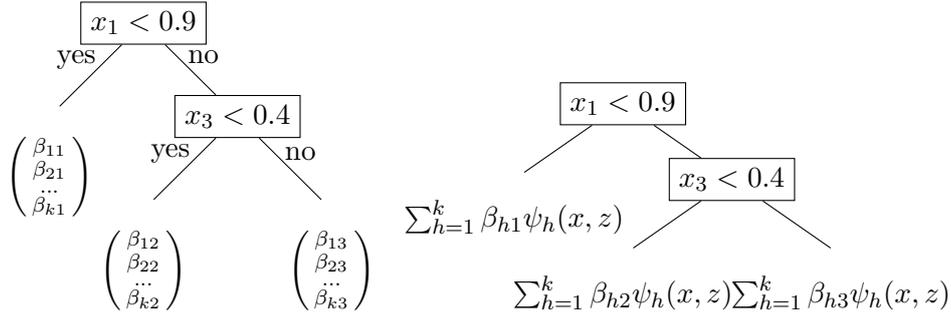

By modifying the way regression trees map to outcomes (equation \ref{eq:g-function}), more elaborate tree models (and ensembles thereof) are possible, which have vector-valued parameters associated with their leaves rather than scalars\footnote{In the future this will allow general shared-tree models; see ~\citep{linero2020shared} for examples and discussion.} (Figure~\ref{fig:regtree}, left). In particular, this allows linear regression at the leaves, as described in \cite{chipman2002bayesian}; see Figure~\ref{fig:regtree} (right panel) for an illustration. Such models specify that
\begin{equation}\label{eq:g-function-linear}
\begin{aligned}
g_s(x,\Psi) &= \sum_{j = 1}^{k_s} \vec{\beta}_{s,j}^t\Psi(x) \mathbb{I}\left(x \in \mathcal{A}_{s,j}\right),\quad f(x) = \sum_{s=1}^m g_s(x, \Psi)\\
\end{aligned}
\end{equation}
where $\Psi$ is a prespecified set of basis functions, and the leaf parameter vectors, $\vec{\beta}_{s,j}$, are given independent multivariate normal priors:
\begin{equation*}
\begin{aligned}
\vec{\beta}_{s,j} &\iid \mathrm{N}\left(\vec{0}, \Sigma_0\right).
\end{aligned}
\end{equation*}

Internally, such a linear regression underlies \pkg{stochtree}'s BCF implementation, taking $\Psi(x,z) := z$. The same formulation works to fit the heterogeneous-linear BCF model in \cite{woody2020estimating} and more esoteric causal models (e.g. multi-arm treatment assignments as in \cite{yeager_synergistic_2022} and the regression discontinuity model below). Other applications of leafwise-regression BART priors include VC-BART models \citep{deshpande2020vcbart} and BART with ``targeted smoothing'' \citep{starling2020bart}. The custom model functionality described in Section~\ref{sec:user-guide-customization} allows multiple forests with distinct bases and updates to the basis specifications during MCMC sampling, making it significantly easier to fit new, flexible model specifications.

\subsubsection{Academic Probation Data RDD analysis} \label{sec:bart-rdd}

\cite{alcantara2025learning} utilize leafwise linear regression in \pkg{stochtree} to perform heterogeneous treatment effect estimation in regression discontinuity designs (RDD). An RDD is characterized by a running variable $X$, and a treatment indicator $Z$ which is 1 whenever  $X > 0$. \cite{alcantara2025learning} show that the treatment effect of $Z$ on outcome $Y$ can be estimated by a leafwise regression by specifying the basis vector
\begin{equation}
\Psi(x,z) = [1, zx, (1-z)x, z].
\end{equation}

This example uses the academic probation data from \cite{lindo2010ability}, which consists of data on college students enrolled in a large Canadian university where students were placed on academic probation based on a sharp cutoff. The treatment $Z$ indicates students that have been placed on academic probation. The running variable, $X$, is the distance away from the probation threshold. Potential moderators, $W$, are:
\begin{itemize}
\itemsep0em 
\item gender
\item age at enrollment
\item North American
\item the number of freshman-year credits 
\item campus (1-3)
\item incoming class rank of high school GPA
\end{itemize}

Fitting this model in \pkg{stochtree} is straightfoward:
\begin{CodeChunk}
\begin{CodeInput}
R> Psi <- cbind(rep(1, n), z * x, (1 - z) * x, z)
R> bart_model <- stochtree::bart(
R>     X_train = cbind(x, w), leaf_basis_train = Psi,
R>     y_train = y, num_gfr = 10, num_mcmc = 500
R> )
\end{CodeInput}
\end{CodeChunk}
We specify \code{Psi0} and \code{Psi1} and the corresponding covariates to extract the leaf-specific regression coefficient on $z$ as follows:
\begin{CodeChunk}
\begin{CodeInput}
R> h <- 0.1 ## window for prediction sample
R> test <- -h < x & x < h
R> Psi0 <- cbind(rep(1, n), rep(0, n), rep(0, n), rep(0, n))[test, ]
R> Psi1 <- cbind(rep(1, n), rep(0, n), rep(0, n), rep(1, n))[test, ]
R> covariates_test <- cbind(x = rep(0, n), w)[test, ]
\end{CodeInput}
\end{CodeChunk}
From here, the CATE estimates are obtained from the fitted model using the predict function:
\begin{CodeChunk}
\begin{CodeInput}
R> cate_posterior <- compute_contrast_bart_model(
R>     bart_model, X_0 = covariates_test,
R>     X_1 = covariates_test, leaf_basis_0 = Psi0,
R>     leaf_basis_1 = Psi1, type = "posterior",
R>     scale = "linear"
R> )
\end{CodeInput}
\end{CodeChunk}

\begin{figure}[t!]
\centering
\includegraphics[width=0.6\linewidth]{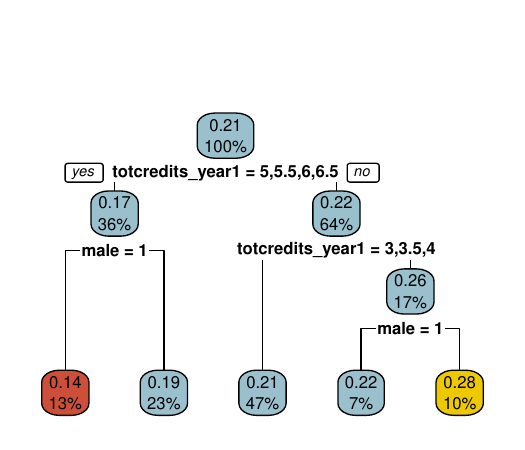}
\caption{\label{fig:rdd-cate-tree} Surrogate CART fit of the treatment effect function estimated using BART in \pkg{stochtree} \proglang{R} package.}
\end{figure}

With this vector of local ($x = 0$) conditional average treatment effects in hand, we may attempt to summarize which variables in $W$ predict distinct treatment effects. Specifically, we visualize a CART tree fit to the posterior point estimates in Figure~\ref{fig:rdd-cate-tree}.

\begin{CodeChunk}
\begin{CodeInput}
R> # Fit regression tree
R> summary_df <- data.frame(y = rowMeans(cate_posterior), w[test, ])
R> cate <- rpart(y ~ ., summary_df, control = rpart.control(cp = 0.015))
R> 
R> # Define separate colors for left and rightmost nodes
R> plot.cart <- function(rpart.obj) {
R>   rpart.frame <- rpart.obj$frame
R>   left <- which.min(rpart.frame$yval)
R>   right <- which.max(rpart.frame$yval)
R>   nodes <- rep(NA, nrow(rpart.frame))
R>   for (i in 1:length(nodes)) {
R>     if (rpart.frame$yval[i] == rpart.frame$yval[right]) {
R>       nodes[i] <- "gold2"
R>     } else if (rpart.frame$yval[i] == rpart.frame$yval[left]) {
R>       nodes[i] <- "tomato3"
R>     } else {
R>       nodes[i] <- "lightblue3"
R>     }
R>   }
R>   return(nodes)
R> }
R> 
R> # Plot regression tree
R> rpart.plot(cate, main = "", box.col = plot.cart(cate))
\end{CodeInput}
\end{CodeChunk}

It appears that age at entry and first year credits predict differences in how students respond to probation. For more details, consult \cite{alcantara2025learning}.

\section{Sampling Custom Models} \label{sec:user-guide-customization}

As illustrated in the previous sections \pkg{stochtree} provides a range of flexible models out-of-the-box. However, for particular applications there is often a need to include bespoke terms in a regression function, alternative error distributions, or other model extensions. \pkg{stochtree} exposes low-level interfaces to forests and data classes at the \proglang{R} and \proglang{Python} levels to make these kinds of model extensions straightforward, as illustrated in the following examples.

\subsection{Overview of the Custom Sampler Interface}

The underlying \proglang{C++} codebase centers around a handful of objects and their interactions. We provide \proglang{R} wrappers for these objects to enable greater customization of stochastic tree samplers than can be furnished by flexible \code{stochtree::bart()} and \code{stochtree::bcf()} functions.The \proglang{C++} class wrappers are managed as R6 classes in \proglang{R} \citep{chang2025r6} and we introduce each object and its purpose below. 

The \code{ForestDataset} class manages covariates, bases, and variance weights used in a forest model, and contains methods for updating the underlying data as well as querying numeric attributes of the data (i.e. \code{num_observations}, \code{num_covariates}, \code{has_basis}, etc...). The \code{RandomEffectsDataset} class manages all of the data used to sample a random effects model, including group indices and regression bases. The \code{Outcome} class wraps the model outcome, which is updated in-place during sampling to reflect the full, or partial, residual net of mean forest or random effects predictions. The \code{ForestSamples} class is a container of sampled tree ensembles, essentially a very thin wrapper around a \proglang{C++} \code{std::vector} of \code{std::unique_ptr} to \code{Ensemble} objects. The \code{Forest} class is a thin wrapper around \code{Ensemble} \proglang{C++} objects, which is used as the ``active forest'' or ``state'' of the forest model during sampling. The \code{ForestModel} class maintains all of the ``temporary'' data structures used to sample a forest, and its \code{ForestModel$sample_one_iteration()} method performs one iteration of the requested forest sampler (i.e. Metropolis-Hastings or Grow-From-Root). 

Writing a custom Gibbs sampler with one or more stochastic forest terms requires initializing each of these objects and then deploying them in a sampling loop. We illustrate two straightforward examples in the following vignettes, with some of the verbose parameter lists converted to ellipses for brevity's sake. Interested readers are referred to our online vignettes for more detailed demonstrations of the ``low-level'' \pkg{stochtree} interface \footnote{\url{https://stochtree.ai/R_docs/pkgdown/articles/CustomSamplingRoutine.html}}.

\subsection{Additive Linear Model} \label{sec:user-guide-additive-linear-model}

In Section~\ref{sec:acic-bcf-rfx} we fit a BCF model with random intercepts and slopes, assuming particular grouping structure and priors on the random effect variances. We might instead want to allow correlated random effects, or fixed effects/linear terms with fixed priors on coefficient variances. The low-level interface exposed by stochtree makes this straightforward. In the general case we have a model specification like:

\begin{equation} \label{eq:linear-term}
\begin{aligned}
Y_i \mid X_i = x_i, W_i = w_i &\iid \mathrm{N}\left(w_i\gamma + f(x_i), \sigma^2\right),\\
f &\sim \mathrm{BART}(\alpha,\beta,m),\\
\sigma^2 &\sim \mathrm{IG}\left(a, b \right),\\
\gamma &\sim \mathrm{N}\left(0, \sigma_{\gamma}\right)
\end{aligned}
\end{equation}
where $X$ and $W$ may contain common features (or not). 

To illustrate, we generate a modified Friedman DGP, where $W \sim \text{U}\left(0,1\right)$ and $\gamma = 5$. To fit the model in \eqref{eq:linear-term}:

\begin{enumerate}

\item We initialize the data and random number generator objects in \proglang{R}:
\begin{CodeChunk}
\begin{CodeInput}
R> forest_dataset <- createForestDataset(X)
R> outcome <- createOutcome(y_standardized)
R> rng <- createCppRNG()
\end{CodeInput}
\end{CodeChunk}

\item Then we initialize the ``configuration'' objects. The full code is in the supplementary materials; the relevant function is \code{createForestModelConfig} and key arguments are:

\begin{enumerate}
    \item \code{feature_types}: a vector of integer-coded feature types (0 denotes numeric and 1 denotes ordered categorical)
    \item \code{variable_weights}: a vector of selection probabilities for variables in generating split rules
    \item \code{leaf_dimension}: the dimensionality of the leaf node parameters
    \item \code{leaf_model_type}: integer code for the leaf model expressed by a given forest (0 denotes constant Gaussian, 1 denotes univariate Gaussian regression, 2 denotes multivariate Gaussian regression, 3 denotes a log-linear Inverse Gamma variance forest model)
    \item \code{num_trees}: the number of trees in a forest
    \item \code{num_features}: the dimensionality of the covariates used to define a forest's split rules
    \item \code{num_observations}: the number of samples in a model's training dataset
\end{enumerate}

\begin{CodeChunk}
\begin{CodeInput}
R> outcome_model_type <- 0
R> leaf_dimension <- 1
R> sigma2_init <- 1.
R> num_trees <- 200
R> feature_types <- as.integer(rep(0, p)) # 0 = numeric
R> variable_weights <- rep(1/p, p)
R> forest_model_config <- createForestModelConfig(
R>     feature_types = feature_types, num_trees = num_trees, 
R>     num_features = p, num_observations = n, 
R>     variable_weights = variable_weights, 
R>     leaf_dimension = leaf_dimension, 
R>     leaf_model_type = outcome_model_type
R> )
R> global_model_config <- createGlobalModelConfig(
R>     global_error_variance = sigma2_init
R> )
\end{CodeInput}
\end{CodeChunk}

\item Then we initialize the forests and temporary tracking objects
\begin{CodeChunk}
\begin{CodeInput}
R> forest_model <- createForestModel(forest_dataset, forest_model_config, 
R>                                   global_model_config)
R> forest_samples <- createForestSamples(num_trees, 1, T)
R> active_forest <- createForest(num_trees, 1, T)
\end{CodeInput}
\end{CodeChunk}

\item To prepare the objects for sampling, we set the root of every tree in the \code{active_forest} to a constant value and propagate these values to the partial residual and temporary tracking data structures
\begin{CodeChunk}
\begin{CodeInput}
R> leaf_init <- mean(y_standardized)
R> active_forest$prepare_for_sampler(forest_dataset, outcome, forest_model, 
R>                                   outcome_model_type, leaf_init)
\end{CodeInput}
\end{CodeChunk}

\item Initialize containers to store the results of our custom sampler (this code is left to the supplement for brevity of presentation).

\item And then we run the MCMC sampler
\begin{CodeChunk}
\begin{CodeInput}
R> for (i in 1:num_mcmc) {
R>     # Add linear regression predictions back to outcome
R>     outcome$add_vector(W %*% current_gamma)
R>     
R>     # Compute a partial residual for regression sampling
R>     # by subtracting out forest predictions
R>     partial_res <- linreg_partial_residual(
R>         y_standardized, forest_dataset, active_forest
R>     )
R>   
R>     # Sample gamma from bayesian linear model with gaussian prior
R>     current_gamma <- sample_linreg_gamma_gibbs(
R>             partial_res, W[, 1],
R>             current_sigma2, gamma_tau
R>     )
R>     
R>     # Subtract updated linear regression predictions from outcome
R>     outcome$subtract_vector(W %*% current_gamma)
R>     
R>     # Sample forest
R>     forest_model$sample_one_iteration(
R>         forest_dataset, outcome, forest_samples, active_forest, rng, 
R>         forest_model_config, global_model_config, 
R>         keep_forest = T, gfr = F
R>     )
R>     
R>     # Sample global variance parameter
R>     current_sigma2 <- sampleGlobalErrorVarianceOneIteration(
R>         outcome, forest_dataset, rng, 1, 1
R>     )
R>     global_model_config$update_global_error_variance(current_sigma2)
R> }
\end{CodeInput}
\end{CodeChunk}

\end{enumerate}

\begin{figure}[t!]
\centering
\includegraphics{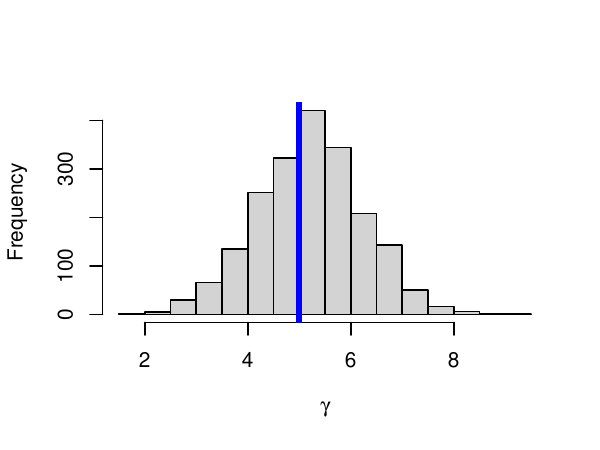}
\caption{\label{fig:custom-interface-bart-reg-gamma-histogram} Histogram of posterior samples of linear regression parameter $\gamma$, estimated using the \pkg{stochtree} \proglang{R} package.}
\end{figure}

We can see the posterior of $\gamma$ in Figure \ref{fig:custom-interface-bart-reg-gamma-histogram}.

\subsection{Robust Errors} \label{sec:user-guide-additive-robust-errors}

A common question about BART is how sensitive it is to deviations from the normal errors assumption. In general, a conditional mean function can be estimated effectively even if the true DGP has non-Gaussian errors and in practice we observe that BART works well for supervised learning in those cases. One notable exception is data which exhibit extremely heavy-tailed errors. In that case, individual observations with extremely large error terms may be mistaken for signal, leading the mean function to be mis-estimated at that point in order accommodate the error.

If this possibility is a concern, \pkg{stochtree} can be used to build a custom posterior sampler for a model with t-distributed errors. To illustrate this, consider a modified Friedman DGP with heavy-tailed errors
\begin{equation*}
\begin{aligned}
Y_i \mid X_i = x_i &\iid t_{2}\left(f(x_i), \sigma^2\right),\\
f(x) &= 10 \sin \left(\pi x_1 x_2\right) + 20 (x_3 - 1/2)^2 + 10 x_4 + 5 x_5,\\
X_1, \dots, X_p &\iid \text{U}\left(0,1\right),
\end{aligned}
\end{equation*}
where $t_{\nu}(\mu,\sigma^2)$ represented a generalized $t$ distribution with location $\mu$, scale $\sigma^2$ and $\nu$ degrees of freedom. Code to sample from this DGP is in Appendix \ref{app:friedman-dataset-robust-error}.

We can obtain $t$-distributed errors by augmenting the basic BART model in~\eqref{eq:bart-model} with a further prior on the individual variances:
\begin{equation} \label{eqn:bart-robust}
\begin{aligned}
Y_i \mid (X_i = x_i) &\iid \mathrm{N}(f(x_i), \phi_i),\\
\phi_i &\iid \text{IG}\left(\frac{\nu}{2}, \frac{\nu\sigma^2}{2}\right),\\
f &\sim \mathrm{BART}(\alpha,\beta,m).
\end{aligned}
\end{equation}
Any Gamma prior on $\sigma^2$ ensures conditional conjugacy, though for simplicity's sake we use a log-uniform prior $\sigma^2\propto 1 / \sigma^2$. In the implementation below, we sample from a ``parameter-expanded'' variant of this model discussed in Section 12.1 of \cite{gelman2013bayesian}, which possesses favorable convergence properties.
\begin{equation} \label{eqn:bart-robust-expanded}
\begin{aligned}
Y_i \mid (X_i = x_i) &\iid \mathrm{N}(f(x_i), a^2\phi_i),\\
\phi_i &\iid \text{IG}\left(\frac{\nu}{2}, \frac{\nu\tau^2}{2}\right),\\
a^2 &\propto 1/a^2,\\
\tau^2 &\propto 1/\tau^2,\\
f &\sim \mathrm{BART}(\alpha,\beta,m).
\end{aligned}
\end{equation}

We can connect \eqref{eqn:bart-robust-expanded} to \eqref{eqn:bart-robust} by noting that $\sigma^2$ is equivalent to $a^2\tau^2$. In order to sample the model in \eqref{eqn:bart-robust-expanded}:

\begin{enumerate}

\item We initialize the data and random number generator objects in R:
\begin{CodeChunk}
\begin{CodeInput}
R> forest_dataset <- createForestDataset(
R>     X, variance_weights = 1 / (var_weights_init)
R> )
R> outcome <- createOutcome(y)
R> rng <- createCppRNG()
\end{CodeInput}
\end{CodeChunk}

\item Then we initialize the ``configuration'' objects. The full code is in the supplement; the relevant functions are \code{createGlobalModelConfig} and \code{createForestModelConfig} and their key arguments are:

\begin{enumerate}
    \item \code{feature_types}: a vector of integer-coded feature types (0 denotes numeric and 1 denotes ordered categorical)
    \item \code{variable_weights}: a vector of selection probabilities for variables in generating split rules
    \item \code{leaf_dimension}: the dimensionality of the leaf node parameters
    \item \code{leaf_model_type}: integer code for the leaf model expressed by a given forest (0 denotes constant Gaussian, 1 denotes univariate Gaussian regression, 2 denotes multivariate Gaussian regression, 3 denotes a log-linear Inverse Gamma variance forest model)
    \item \code{num_trees}: the number of trees in a forest
    \item \code{num_features}: the dimensionality of the covariates used to define a forest's split rules
    \item \code{num_observations}: the number of samples in a model's training dataset
\end{enumerate}

\begin{CodeChunk}
\begin{CodeInput}
R> outcome_model_type <- 0
R> leaf_dimension <- 1
R> num_trees <- 200
R> feature_types <- as.integer(rep(0, p)) # 0 = numeric
R> variable_weights <- rep(1/p, p)
R> forest_model_config <- createForestModelConfig(
R>     feature_types = feature_types, num_trees = num_trees, 
R>     num_features = p, num_observations = n, 
R>     variable_weights = variable_weights, 
R>     leaf_dimension = leaf_dimension, 
R>     leaf_model_type = outcome_model_type
R> )
R> global_model_config <- createGlobalModelConfig(
R>     global_error_variance = sigma2_init
R> )
\end{CodeInput}
\end{CodeChunk}

\item Then we initialize the forests and temporary tracking objects
\begin{CodeChunk}
\begin{CodeInput}
R> forest_model <- createForestModel(forest_dataset, forest_model_config, 
R>                                   global_model_config)
R> forest_samples <- createForestSamples(num_trees, 1, T)
R> active_forest <- createForest(num_trees, 1, T)
\end{CodeInput}
\end{CodeChunk}

\item To prepare the objects for sampling, we set the root of every tree in the \code{active_forest} to a constant value and propagate these values to the partial residual and temporary tracking data structures
\begin{CodeChunk}
\begin{CodeInput}
R> leaf_init <- mean(y)
R> active_forest$prepare_for_sampler(forest_dataset, outcome, forest_model, 
R>                                   outcome_model_type, leaf_init)
\end{CodeInput}
\end{CodeChunk}

\item Initialize containers to store the results of our custom sampler (as in Section \ref{sec:user-guide-additive-linear-model}, this code is deferred to the supplement for brevity's sake).

\item And then we run the MCMC sampler
\begin{CodeChunk}
\begin{CodeInput}
R> for (i in 1:num_mcmc) {
R>     # Sample forest
R>     forest_model$sample_one_iteration(
R>         forest_dataset, outcome, forest_samples, active_forest, rng, 
R>         forest_model_config, global_model_config, keep_forest = T, gfr = F
R>     )
R>     
R>     # Sample local variance parameters
R>     current_phi_i <- sample_phi_i(
R>         y_standardized, forest_dataset, active_forest, 
R>         current_a2, current_tau2, nu
R>     )
R>     
R>     # Sample a2
R>     current_a2 <- sample_a2(
R>         y_standardized, forest_dataset, active_forest, current_phi_i
R>     )
R>     
R>     # Sample tau2
R>     current_tau2 <- sample_tau2(current_phi_i, nu)
R>     
R>     # Update observation-specific variance weights
R>     forest_dataset$update_variance_weights(current_phi_i * current_a2)
R> }
\end{CodeInput}
\end{CodeChunk}

\end{enumerate}

\begin{figure}[t!]
\centering
\includegraphics[width=0.7\linewidth]{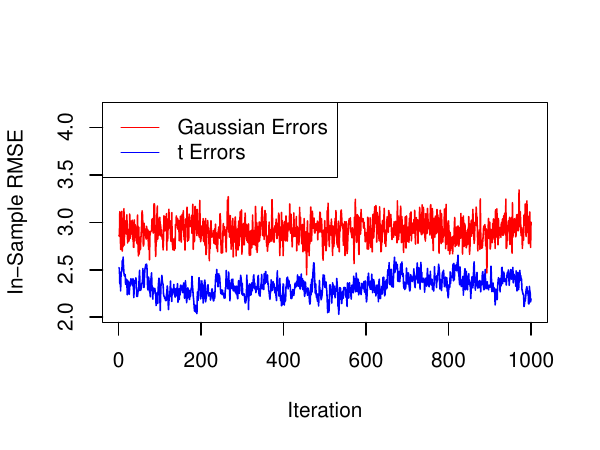}
\caption{\label{fig:custom-interface-bart-robust-rmse-comparison-r} Comparison of in-sample sum of squared error (SSE) between a BART model with robust errors and a BART model with normal errors, both estimated using the \pkg{stochtree} \proglang{R} package. Here, the true DGP has t-distributed errors with 2 degrees of freedom; the BART model with t-distributed errors estimates a smaller error standard deviation compared to a BART model with Gaussian errors.}
\end{figure}

\begin{figure}[t!]
\centering
\includegraphics{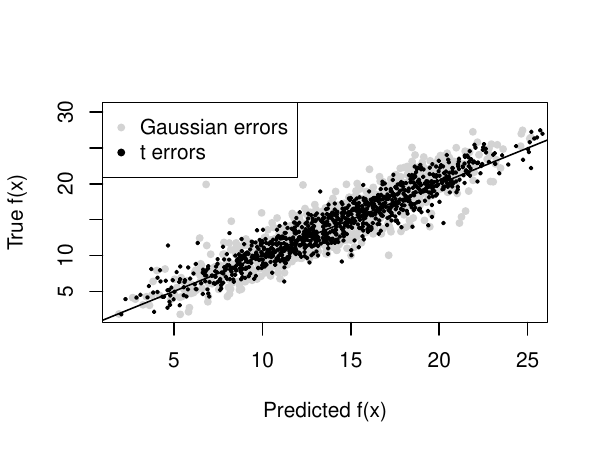}
\caption{\label{fig:custom-interface-bart-robust-pred-actual-comparison-r} Comparison of actual conditional mean values and their posterior mean point estimates for BART models with robust errors and with Gaussian errors, both estimated using the \pkg{stochtree} \proglang{R} package. Here, the true DGP has t-distributed errors with 2 degrees of freedom; the BART model with t-distributed errors more accurately estimates the conditional mean compared to a BART model with Gaussian errors, as would be expected.}
\end{figure}

Figures \ref{fig:custom-interface-bart-robust-rmse-comparison-r} and \ref{fig:custom-interface-bart-robust-pred-actual-comparison-r} show the results of fitting a custom BART model with t-distributed errors compared to a default BART model using default Gaussian errors, looking at (in-sample) root mean squared estimation error (as a traceplot) and an actual-versus-predicted scatterplot. Both plots  show that the custom BART model with t-distributed errors estimates the mean function more accurately when the data come from a (scaled) t-distribution with $\nu = 2$ degrees of freedom.

\section{Discussion} \label{sec:summary}
The \pkg{stochtree} package has been built for both primary data analyses and also methods development. This dual-functionality is the package's core strength -- rather than attempting to be a comprehensive BART supervised learning tool {\em today}, \pkg{stochtree} was designed with extensibility in mind. By having core sampling functionality written modularly in \proglang{C++} and accessible from the \proglang{R} and \proglang{Python} interfaces, users are encouraged to expand on the package. Models and algorithms developed in this way can potentially be incorporated into the core package functions; reliance on identical underlying algorithms makes the merging process less onerous for developers. 

For example, at present, the \pkg{BART} package can handle two models that \pkg{stochtree} does not: monotone function estimation \citep{chipman2022mbart} and survival analysis \citep{sparapani2016nonparametric, sparapani2023nonparametric}. However, in the near future, \pkg{stochtree} will incorporate these features {\em without a complete rebuild/rewrite of the underlying code base}. Likewise, we foresee that multi-class classification \citep{wang2024computational} will be incorporated in time. 

Conversely, \pkg{stochtree} has narrower ambitions than \pkg{pymc-bart}; we believe that tree ensembles posses enough special structure that computational efficiency and also statistical accuracy (MCMC approximation) benefit from dedicated algorithms. 
Similarly, although the particle sampling approach nominally provides extreme flexibility so that BART can be incorporated into essentially any model, the particle Gibbs sampler is less well-studied than the (generalized) Bayesian back-fitting algorithms around which \pkg{stochtree} is built.


\section*{Computational details}

The results in this paper were obtained using:
\begin{itemize}
    \item \proglang{R}~4.4.2
    \item \proglang{Python}~3.12.9
\end{itemize}
with the \pkg{stochtree}~0.2.1 package. \proglang{Python} package dependencies are detailed in the \code{requirements.txt} file included in the replication materials for this paper. \proglang{R} itself and all packages used are available from the Comprehensive \proglang{R} Archive Network (CRAN) at \url{https://CRAN.R-project.org/}. \proglang{Python} is available via numerous popular distributions. For this paper, we used the Anaconda \proglang{Python} distribution \url{https://www.anaconda.com/download} and installed all packages from the \proglang{Python} Package Index (PyPI) at \url{https://pypi.org/}.

\section*{Acknowledgments}

\begin{leftbar}
The authors gratefully acknowledge \dots
\end{leftbar}


\bibliography{refs}


\newpage

\begin{appendix}

\section{Dataset Details} \label{app:dataset-details}

\subsection{Generating the Friedman Dataset} \label{app:friedman-dataset}

This classic data generating process is implemented in \proglang{R} as follows:
\begin{CodeChunk}
\begin{CodeInput}
R> friedman_mean <- function(x) {
R>    10*sin(pi*x[,1]*x[,2]) + 20*(x[,3] - 0.5)^2 + 10*x[,4] + 5*x[,5]
R> }
R> n <- 500
R> p <- 100
R> X <- matrix(runif(n*p),ncol=p)
R> m_x <- friedman_mean(X)
\end{CodeInput}
\end{CodeChunk}

The homoskedastic Friedman vignettes generate $\sigma^2$ according to a user-specified signal-to-noise ratio
\begin{CodeChunk}
\begin{CodeInput}
R> snr <- 3
R> eps <- rnorm(n, 0, 1) * (sd(m_x) / snr)
R> y <- m_x + eps
\end{CodeInput}
\end{CodeChunk}

\subsubsection{Causal Friedman Dataset} \label{app:causal-friedman-dataset}

The ``causal Friedman dataset'' uses the \code{friedman_mean} function defined above to determine the prognostic and propensity functions
\begin{CodeChunk}
\begin{CodeInput}
R> prog_fn <- function(x) {
R>    10*sin(pi*x[,1]*x[,2]) + 20*(x[,3] - 0.5)^2 + 10*x[,4] + 5*x[,5]
R> }
R> propensity_fn <- function(x) {
R>   pnorm(0.05 * (prog_fn(x) - mean(prog_fn(x))))
R> }
\end{CodeInput}
\end{CodeChunk}

The CATE function is linear in $X_1$: 
\begin{CodeChunk}
\begin{CodeInput}
R> cate_fn <- function(x) {
R>   5 * x[, 1]
R> }
\end{CodeInput}
\end{CodeChunk}

And the data is generated as
\begin{CodeChunk}
\begin{CodeInput}
R> X <- matrix(runif(n * p), nrow = n, ncol = p)
R> mu_x <- prog_fn(X)
R> pi_x <- propensity_fn(X)
R> tau_x <- cate_fn(X)
R> Z <- rbinom(n, 1, pi_x)
R> E_Y_ZX <- mu_x + tau_x * Z
R> y <- E_Y_ZX + rnorm(n, 0, 1)
\end{CodeInput}
\end{CodeChunk}

\subsubsection{Additive Regression Model} \label{app:friedman-dataset-additive-reg}

For the ``custom interface'' demos of Section \ref{sec:user-guide-customization}, we use the \code{friedman_mean} function with two modifications. The first demo (Section \ref{sec:user-guide-additive-linear-model}) adds a linear regression mean term, $W\beta$, with univariate uniform $W$ and $\beta = 5$.
\begin{CodeChunk}
\begin{CodeInput}
R> p_W <- 1
R> W <- matrix(runif(n*p_W), ncol = p_W)
R> beta_W <- c(5)
R> lm_term <- W %*% beta_W
R> y <- lm_term + m_x + eps
\end{CodeInput}
\end{CodeChunk}

\subsubsection{Robust Errors} \label{app:friedman-dataset-robust-error}

The second demo (Section \ref{sec:user-guide-additive-robust-errors}) replaces the homoskedastic Gaussian error with a scaled $\sigma t_{\nu}$ error term with $\nu = 2$ and $\sigma^2 = 9$.
\begin{CodeChunk}
\begin{CodeInput}
R> sigma2 <- 9
R> nu <- 2
R> eps <- rt(n, df = nu) * sqrt(sigma2)
\end{CodeInput}
\end{CodeChunk}

\subsection{ACIC Dataset} \label{app:acic-dataset}

\cite{carvalho2019assessing} introduce a semi-synthetic dataset, in which covariates are informed by a randomized controlled trial, but outcome and treatment are simulated to ensure confounding. The synthetic outcome and treatment are referred to as $Y$ and $Z$, respectively, while the covariates are defined in the data dictionary in Table \ref{tab:acic-data-dictionary} (drawn from \cite{carvalho2019assessing}).

\begin{table}[t!]
\centering
\begin{tabular}{l|p{32em}}
Covariate & Description \\ \hline\hline
\code{S3} & Student’s self-reported expectations for success in the future, a proxy for prior achievement, measured prior to random assignment \\ \hline  
\code{C1} & Categorical variable for student race/ethnicity \\ \hline
\code{C2} & Categorical variable for student identified gender \\ \hline
\code{C3} & Categorical variable for student firstgeneration status (i.e., first in family to go to college) \\ \hline
\code{XC} & School-level categorical variable for urbanicity of the school (i.e., rural, suburban, etc.) \\ \hline
\code{X1} & School-level mean of students’ fixed mindsets, reported prior to random assignment \\ \hline
\code{X2} & School achievement level, as measured by test scores and college preparation for the previous four cohorts of students \\ \hline 
\code{X3} & School racial/ethnic minority composition – i.e., percentage black, Latino, or Native American \\ \hline
\code{X4} & School poverty concentration – i.e., percentage of students who are from families whose incomes fall below the federal poverty line \\ \hline
\code{X5} & School size – Total number of students in all four grade levels in the school \\ 
\end{tabular}
\caption{\label{tab:acic-data-dictionary} ACIC Covariate Data Dictionary}
\end{table}

We load this dataset from Github as follows
\begin{CodeChunk}
\begin{CodeInput}
R> url_string <- paste0("https://raw.githubusercontent.com/andrewherren/acic2024/",
R>                      "refs/heads/main/data/acic2018/synthetic_data.csv")
R> df <- read.csv(url_string)
\end{CodeInput}
\end{CodeChunk}
and we unpack the data into a format needed for \code{stochtree::bcf()} as follows
\begin{CodeChunk}
\begin{CodeInput}
R> y <- df$Y
R> Z <- df$Z
R> covariate_df <- df[,!(colnames(df) %in% c("schoolid","Z","Y"))]
R> unordered_categorical_cols <- c("C1","XC")
R> ordered_categorical_cols <- c("S3","C2","C3")
R> for (col in unordered_categorical_cols) {
R>     covariate_df[,col] <- factor(covariate_df[,col], ordered = F)
R> }
R> for (col in ordered_categorical_cols) {
R>     covariate_df[,col] <- factor(covariate_df[,col], ordered = T)
R> }
\end{CodeInput}
\end{CodeChunk}

\subsection{Academic Probation Dataset} \label{app:academic-probation-dataset}

In studying the Academic Probation dataset \citep{lindo2010ability} for RDD, our outcome of interest is each student's GPA at the end of the current academic term (\code{nextGPA} in the data) and the running variable (\code{X}) is the negative difference between a student's previous-term GPA and the threshold for being placed on academic probation. From this running variable, we define a binary treatment variable \code{Z} as \code{X > 0}. Covariates are defined in the data dictionary in Table \ref{tab:probation-data-dictionary}.

\begin{table}[t!]
\centering
\begin{tabular}{l|p{30em}}
Variable & Description \\ \hline\hline
\code{totcredits_year1} & Number of academic credits enrolled in first year \\ \hline
\code{hsgrade_pct} & Percentile of student's high school GPA relative to university cohort \\ \hline
\code{age_at_entry} & Student's age when beginning university \\ \hline
\code{male} & Whether or not student's gender is male \\ \hline 
\code{bpl_north_america} & Whether or not student was born in North America \\ \hline
\code{loc_campus1} & Whether or not student is located on campus 1 \\ \hline
\code{loc_campus2} & Whether or not student is located on campus 2 \\ \hline
\code{loc_campus3} & Whether or not student is located on campus 3 \\ 
\end{tabular}
\caption{\label{tab:probation-data-dictionary} Academic Probation Data Dictionary}
\end{table}

We load the dataset as follows
\begin{CodeChunk}
\begin{CodeInput}
R> url_string <- paste0("https://raw.githubusercontent.com/rdpackages-replication/",
R>                      "CIT_2024_CUP/refs/heads/main/CIT_2024_CUP_discrete.csv")
R> data <- read.csv(url_string)
\end{CodeInput}
\end{CodeChunk}
and extract the individual data components needed for the BART regression discontinuity model 
\begin{CodeChunk}
\begin{CodeInput}
R> y <- data$nextGPA
R> x <- data$X
R> x <- x/sd(x) ## we always standardize X
R> w <- data[,4:11]
R> # Define categorical features as ordered/unordered factors
R> w$totcredits_year1 <- factor(w$totcredits_year1,ordered=TRUE)
R> w$male <- factor(w$male,ordered=FALSE)
R> w$bpl_north_america <- factor(w$bpl_north_america,ordered=FALSE)
R> w$loc_campus1 <- factor(w$loc_campus1,ordered=FALSE)
R> w$loc_campus2 <- factor(w$loc_campus2,ordered=FALSE)
R> w$loc_campus3 <- factor(w$loc_campus3,ordered=FALSE)
R> # Define the cutoff
R> c <- 0
R> z <- as.numeric(x>c)
\end{CodeInput}
\end{CodeChunk}

\section{Basic Python stochtree Workflow} \label{sec:python-user-guide-basic-function-call}

The main paper introduces \pkg{stochtree} and demonstrates many of its features in \proglang{R}. This appendix provides the same demon for the \proglang{Python} interface. We use the same five datasets for the \proglang{Python} vignettes; code for loading and preprocessing datasets is in Appendix \ref{app:python-dataset-details}. 

\subsection{Overview}

The \code{BARTModel} class in \pkg{stochtree} defines a \code{sample()} method which takes as input response and predictor data and produces posterior samples from a BART model. After sampling, \code{BARTModel} objects include model metadata, parameter draws, and pointers to \proglang{C++} representations of tree ensemble samples. Out-of-sample prediction is accomplished through a \code{predict()} method which accepts an array of predictors dictating the points at which predictions are desired and returns predictions as \pkg{numpy} arrays.

In its simplest form (using default priors and default MCMC settings) this process consists of one instantiation of a \code{BARTModel} object:
\begin{CodeChunk}
\begin{CodeInput}
>>> bart_model = BARTModel()
\end{CodeInput}
\end{CodeChunk}
one call to \code{sample()}:
\begin{CodeChunk}
\begin{CodeInput}
>>> bart_model.sample(X_train = X_train, y_train = y_train)
\end{CodeInput}
\end{CodeChunk}
and one call to \code{predict()}:
\begin{CodeChunk}
\begin{CodeInput}
>>> bart_preds = bart_model.predict(X = X_test)
\end{CodeInput}
\end{CodeChunk}

However, the \pkg{stochtree} workflow consists of a number of choices prior to calling these core functions and a number of choices after as well. In the following subsections we will examine three preparatory steps: data preprocessing (Section \ref{sec:python-preprocessing}), prior specification (Section \ref{sec:python-prior-parameters}), and algorithm settings (Section \ref{sec:python-mcmc-settings}); as well as
 three post-fitting steps: convergence diagnostic plots, extracting model fit information, and extracting (point and interval) predictions.

We present this basic workflow in terms of the \code{BARTModel} object, but the other core \code{BCFModel} class for causal inference, is closely analagous and will be covered in one of the illustrations. 

\subsubsection{Data Preprocessing} \label{sec:python-preprocessing}

\pkg{stochtree} can accommodate both categorical and numerical predictor variables, but some preprocessing is necessary for the algorithm to utilize them properly. We demonstrate this below on the ACIC dataset, assuming that it has been loaded into a dataframe called \code{df} (see Appendix \ref{app:python-acic-dataset} for details on loading this dataset).

We begin by extracting the outcome (and treatment variable, in the case of BCF) from our data frame and storing them as vectors:
\begin{CodeChunk}
\begin{CodeInput}
>>> y = df.loc[:, "Y"].to_numpy()
>>> Z = df.loc[:, "Z"].to_numpy()
\end{CodeInput}
\end{CodeChunk}

Categorical variables should be coded in \pkg{pandas} as ordered or unordered categoricals; by default the response variable is standardized internally but results are reported on the original scale.
\begin{CodeChunk}
\begin{CodeInput}
>>> unordered_categorical_cols = ["C1", "XC"]
>>> ordered_categorical_cols = ["S3", "C2", "C3"]
>>> for col in unordered_categorical_cols:
>>>     covariate_df.loc[:, col] = pd.Categorical(
>>>         covariate_df.loc[:, col], ordered=False
>>>     )
>>> for col in ordered_categorical_cols:
>>>     covariate_df.loc[:, col] = pd.Categorical(
>>>         covariate_df.loc[:, col], ordered=True
>>>     )
\end{CodeInput}
\end{CodeChunk}

\subsubsection{Specifying Prior Parameters} \label{sec:python-prior-parameters}

While BART models are known to work remarkably well using default prior distributions, users can depart freely from these defaults in various ways. \pkg{stochtree}'s prior parameters are organized according to user-passed \code{dict}s. In the basic case of a BART fit, the relevant dictionaries are  \code{mean_forest_params}, which governs the prior over the trees, and \code{general_params}, which governs global model parameters as well as MCMC parameters, including model initialization. 

If we wish to change model parameters, such as the number of trees or node split hyperparameters $\alpha$ and $\beta$, we do so via parameter dictionaries
\begin{CodeChunk}
\begin{CodeInput}
>>> mean_forest_params = {"alpha" : 0.25, "beta" : 3, "num_trees" : 20}
>>> bart_model = BARTModel()
>>> bart_model.sample(X_train = X_train, y_train = y_train, 
>>>                   mean_forest_params = mean_forest_params, 
>>>                   num_gfr = 0, num_burnin = 1000, num_mcmc = 1000)
\end{CodeInput}
\end{CodeChunk}

\subsubsection{Algorithm Settings} \label{sec:python-mcmc-settings}

Users may specify the length of the desired Markov chain, number of burn-in steps, and certain Markov chain initialization choices. 

In particular, initialization can be determined by the XBART algorithm, a greedy approximation to the BART MCMC procedure that rapidly converges to high likelihood tree ensembles \citep{he2023stochastic}, by regrowing each tree from root at each iteration of the algorithm. Because XBART involves a number of additional innovations, here we will refer to its core algorithm separately as grow-from-root (GFR). On its own, grow-from-root represents a fast approximate BART fit for prediction problems, particularly those with large training sample sizes; but grow-from-root can also be used to initialize the BART MCMC algorithm at favorable parameter values, thereby substantially reducing the necessary burn-in period. Following \cite{he2023stochastic} we refer to this strategy as ``warm start''.

By default, \code{stochtree::bart()} runs 10 iterations of the grow-from-root (GFR) algorithm \citep{he2023stochastic} and 100 MCMC iterations initialized by the final GFR forest. We can modify these values directly in the \code{BARTModel.sample()} method call
\begin{CodeChunk}
\begin{CodeInput}
>>> bart_model = BARTModel()
>>> bart_model.sample(X_train = X_train, y_train = y_train, 
>>>                   num_gfr = 0, num_burnin = 1000, num_mcmc = 1000)
\end{CodeInput}
\end{CodeChunk}

\subsubsection{Prediction}
Finally, predictions are obtained as pointwise posterior means via the \code{predict()} method. 
\begin{CodeChunk}
\begin{CodeInput}
>>> y_hat_test = bart_model.predict(X = X_test, terms = "y_hat", type = "mean")
\end{CodeInput}
\end{CodeChunk}

\subsubsection{Serialization and Advanced Sampling Options} \label{sec:python-user-guide-mcmc-niceties}

Should one need to revisit and modify a previously-fit BART model, \pkg{stochtree} model objects can be saved to disk via the \code{to_json()} method and reloaded via the \code{from_json()} method. This can be useful when further MCMC iterations are needed or when a BART model is being used as one step of a more elaborate model. This functionality will be demonstrated in Section \ref{sec:python-friedman-bart-supervised}. Note that these objects can be ported back and forth between \proglang{R} and \proglang{Python}, facilitating convenient cross-platform interoperability. 

\subsection{Example: Supervised Learning with BART (Friedman dataset)} \label{sec:python-friedman-bart-supervised}

As in Section \ref{sec:friedman-bart-supervised}, we use the Friedman dataset with additive Gaussian errors, setting $p = 100$ and $n = 500$. Our Bayesian model will simply be $f \sim \mbox{BART}(\alpha = 0.25,\beta = 2, m = 200)$. We initialize the model by running 20 GFR iterations using default settings ($\alpha = 0.95,\beta = 2, m = 200$).

\begin{CodeChunk}
\begin{CodeInput}
>>> xbart_model = BARTModel()
>>> xbart_model.sample(X_train = X_train, y_train = y_train, 
>>>                    num_gfr = 20, num_mcmc = 0)
\end{CodeInput}
\end{CodeChunk}

We then save the model and restart sampling according to the Markov chain, obtaining 10,000 samples under a more conservative prior ($\alpha = 0.25$ and a maximum tree depth of eight). 

\begin{CodeChunk}
\begin{CodeInput}
>>> xbart_json = xbart_model.to_json()
>>> mean_forest_params = {"alpha" : 0.25, "beta" : 2, 
>>>                       "min_samples_leaf" : 10, "max_depth" : 8}
>>> bart_model = BARTModel()
>>> bart_model.sample(X_train = X_train, y_train = y_train, 
>>>                   mean_forest_params = mean_forest_params, 
>>>                   num_gfr = 0, num_burnin = 0, num_mcmc = 1000, 
>>>                   previous_model_json = xbart_json, 
>>>                   previous_model_warmstart_sample_num = 19
>>> )
\end{CodeInput}
\end{CodeChunk}

\begin{figure}[t!]
\centering
\includegraphics{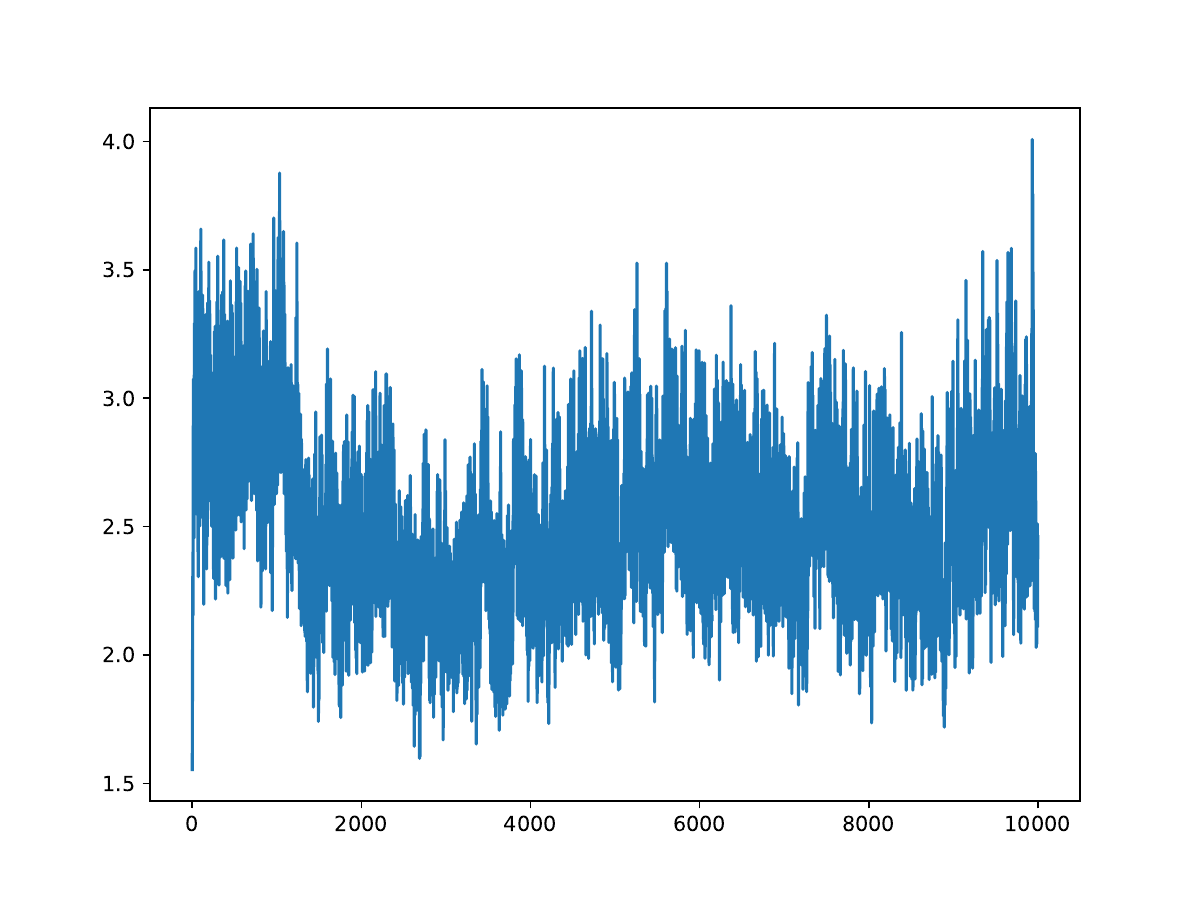}
\caption{\label{fig:python-friedman-bart-python-traceplot-warm-start} Traceplot of the global error variance parameter, $\sigma^2$, sampled as part of a homoskedastic BART model fit with the \pkg{stochtree} \proglang{Python} package.}
\end{figure}

Figure \ref{fig:python-friedman-bart-python-traceplot-warm-start} shows the traceplot for the residual variance paramter, $\sigma^2$, produced with the following code:
\begin{CodeChunk}
\begin{CodeInput}
>>> plt.plot(bart_model.global_var_samples, linestyle="-")
\end{CodeInput}
\end{CodeChunk}

\begin{figure}[t!]
\centering
\includegraphics{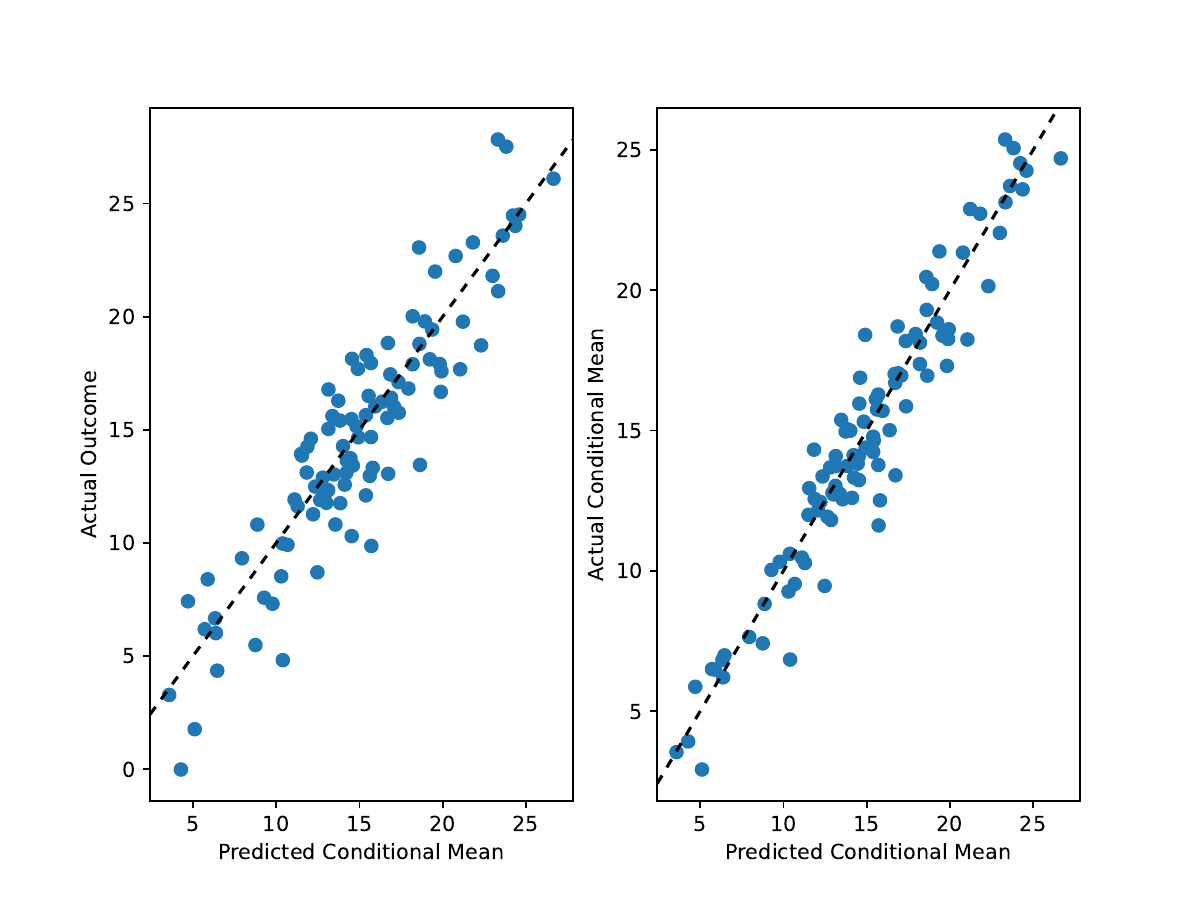}
\caption{\label{fig:python-friedman-bart-python-pred-actual-warm-start} Predictions (posterior means) plotted against actual outcomes on a hold-out set, for a homoskedastic BART model fit with the \pkg{stochtree} \proglang{Python} package.}
\end{figure}

Figure \ref{fig:python-friedman-bart-python-pred-actual-warm-start} plots the predicted conditional mean estimates to observed outcomes for a holdout set, produced with the following code:
\begin{CodeChunk}
\begin{CodeInput}
>>> y_hat_test = bart_model.predict(
>>>     X=X_test, terms="y_hat", type="mean"
>>> )
>>> plt.scatter(y_hat_test, y_test)
>>> y_bar = np.mean(y_test)
>>> plt.axline((y_bar, y_bar), slope=1, color="black", 
>>>     linestyle=(0, (3, 3)))
\end{CodeInput}
\end{CodeChunk}

\subsection{Example: Causal Inference with BCF (Causal Friedman dataset)}
\label{sec:python-causal-friedman-data}

\pkg{stochtree}'s \proglang{Python} interface can also be used to fit the BCF model defined in~\eqref{eq:bcf-model}. Basic function calls to BCF mirror those of BART. Models are fitted via the \code{sample()} method of the \code{BCFModel} class. The \code{predict()} method computes response predictions, as well as prognostic ($f_0$) and treatment effect ($\tau$) functions. Likewise, models are serialized via \code{to_json()} methods and reloaded via \code{from_json()}.

A number of features make BCF more suitable for causal inference problems than regular BART. One, the BART priors on $f_0$ and $\tau$ can be specified separately, including involving distinct subsets of features. Two, a propensity function is automatically fit and used as a control covariate (or one may be supplied by the user), mitigating ``regularization-induced confounding", cf. \cite{hahn2020bayesian}. 

The following code fits a BCF model to the causal Friedman data set. The model in~\eqref{eq:bart-model-probit} can be fit in \pkg{stochtree} simply by specifying a probit model in the general parameters dictionary and suppressing sampling of the residual error variance ($\sigma^2$), which is unidentified for probit models.

\begin{CodeChunk}
\begin{CodeInput}
>>> general_params_propensity = {
>>>     "probit_outcome_model": True,
>>>     "sample_sigma2_global": False,
>>> }
>>> propensity_model = BARTModel()
>>> propensity_model.sample(
>>>     X_train=covariate_df, y_train=Z, general_params=general_params_propensity
>>> )
>>> propensity = propensity_model.predict(
>>>     X=covariate_df, type="mean", terms="y_hat"
>>> )
\end{CodeInput}
\end{CodeChunk}

We then pass this estimated propensity score array as an argument to the \code{BCFModel.sample()} method.
\begin{CodeChunk}
\begin{CodeInput}
>>> bcf_model = BCFModel()
>>> bcf_model.sample(
>>>     X_train=covariate_df,
>>>     Z_train=Z,
>>>     y_train=y,
>>>     propensity_train=propensity,
>>>     num_gfr=10,
>>>     num_burnin=2000,
>>>     num_mcmc=1000,
>>>     treatment_effect_forest_params={"keep_vars": ["X1", "X2"]},
>>> )
\end{CodeInput}
\end{CodeChunk}

If propensities are not provided to the \code{BCFModel.sample()} method, \pkg{stochtree} will fit a propensity model automatically unless the user specifies otherwise by specifying \code{propensity_covariate = "none"} in the \code{general_params} dictionary.
Users retain more control\footnote{It could be fit with a method other than \pkg{stochtree}, as \code{BCFModel.sample()} merely requires an array of estimated propensity scores.} over the propensity model specification by pre-fitting it, as shown here. 
By default, this option is set to \code{propensity_covariate = "mu"}, indicating that the propensity score is included in the $f_0$ function.

\begin{figure}[t!]
\centering
\includegraphics{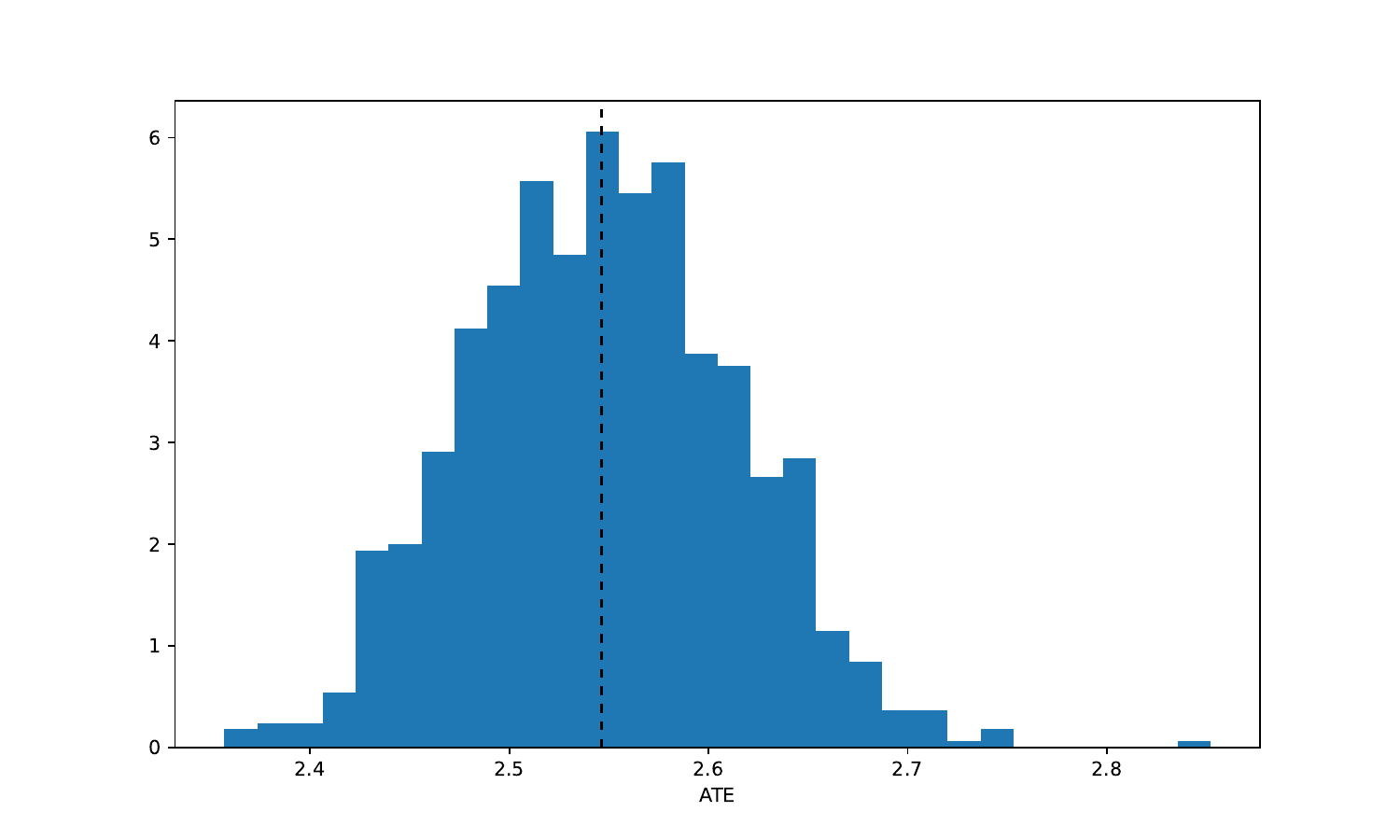}
\caption{\label{fig:python-simulated-ate-python} Posterior distribution of average treatment effect (ATE) estimated by homoskedastic BCF model using the \pkg{stochtree} \proglang{Python} package.}
\end{figure}

Figure \ref{fig:python-simulated-ate-python} plots a histogram of posterior samples of the average treatment effect (ATE) as estimated by BCF: 
\begin{CodeChunk}
\begin{CodeInput}
>>> tau_hat_posterior = bcf_model.predict(
>>>     X=covariate_df, Z=Z, propensity=propensity, type="posterior", terms="cate"
>>> )
>>> ate_posterior = np.mean(tau_hat_posterior, axis=0)
>>> plt.hist(ate_posterior, bins=30)
>>> plt.axvline(x = np.mean(tau_x), color = "black", linestyle = (0, (3, 3)))
\end{CodeInput}
\end{CodeChunk}

\begin{figure}[t!]
\centering
\includegraphics{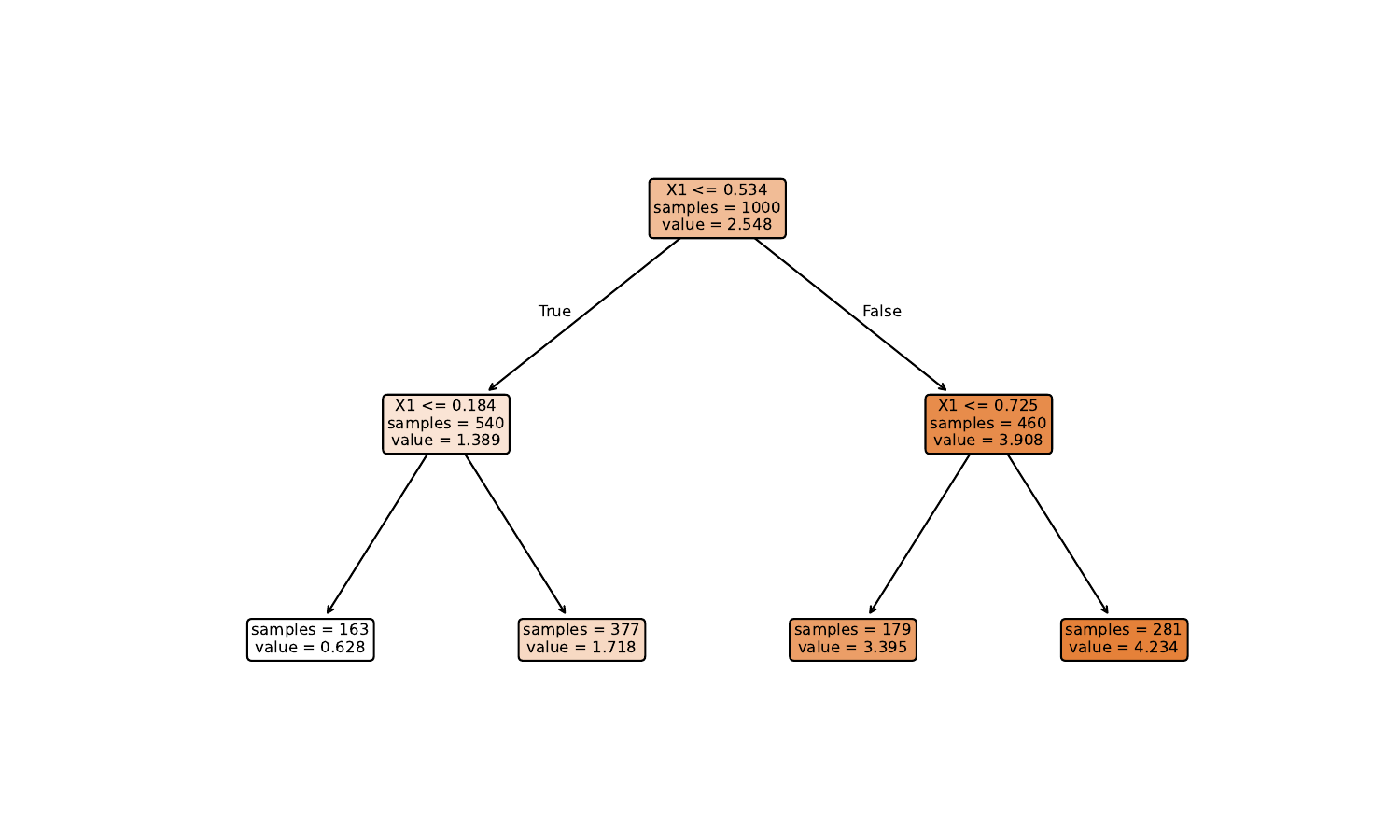}
\caption{\label{fig:python-simulated-cate-tree-python} Decision tree summary of the CATE posterior mean, as estimated by a homoskedastic BCF model using the \pkg{stochtree} \proglang{Python} package.}
\end{figure}

Now, we investigate the nature of the observed treatment effect moderation by fitting a CART tree to the posterior mean of $\tau(X)$. The tree presented in Figure \ref{fig:python-simulated-cate-tree-python} shows evidence of strong moderation by $X_1$, matching the simulated $\tau(X) = 5 X_1$. 

\begin{figure}[t!]
\centering
\includegraphics{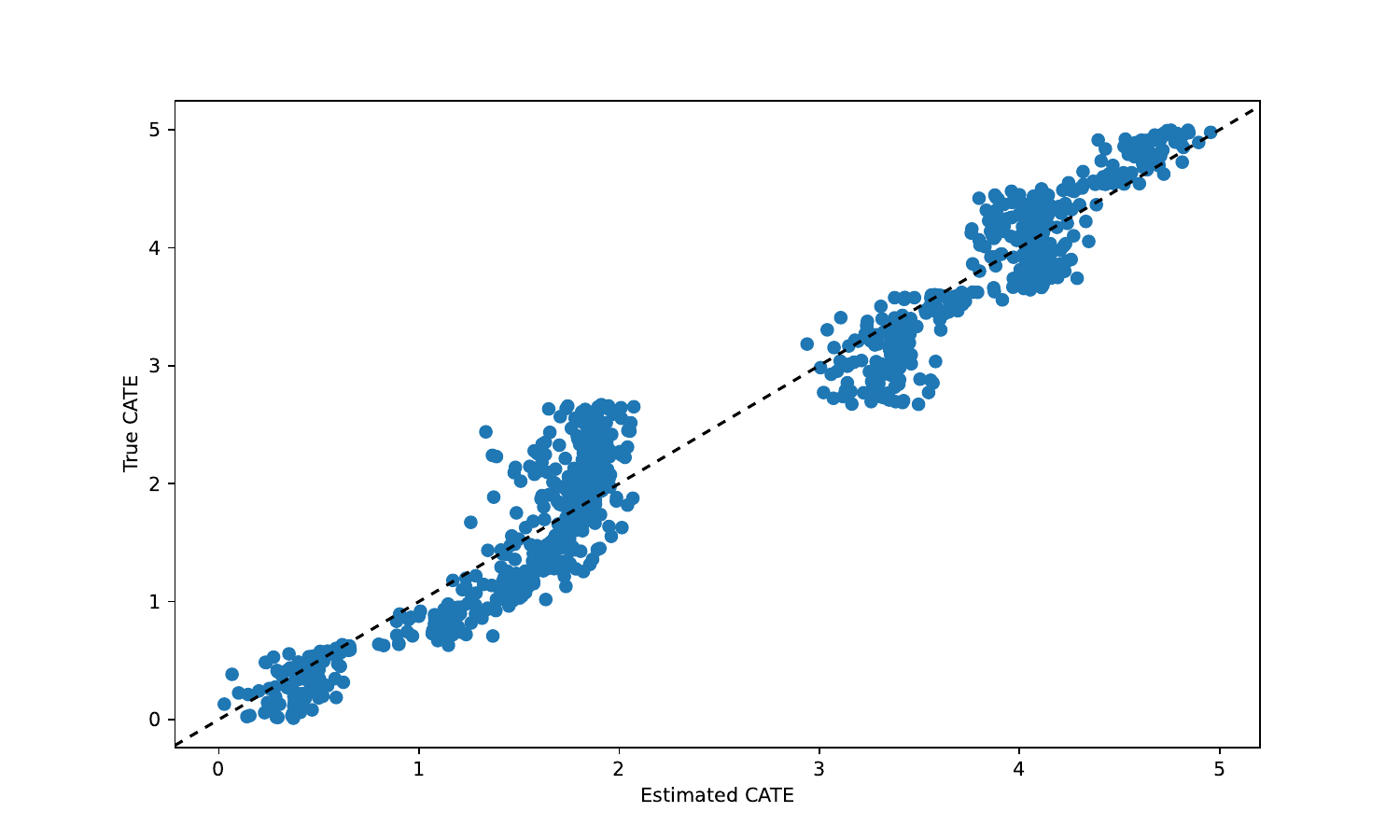}
\caption{\label{fig:python-simulated-cate-true-fitted-python} True conditional average treatment effect (CATE) compared to the CATE estimated by a homoskedastic BCF model using the \pkg{stochtree} \proglang{R} package.}
\end{figure}

Figure \ref{fig:python-simulated-cate-true-fitted-python} plots the estimated conditional average treatment effect (CATE) against the true function. 

\section{Advanced Models} \label{sec:python-user-guide-advanced}

\subsection{Additive Random Effects Model}\label{sec:python-random-effects}

\subsubsection{A Random Effects Model} \label{sec:python-rfx-model}

BART and BCF models in \pkg{stochtree} include specialized algorithms for fitting one-way random intercepts and random slopes with independent variance components (using the redundant parameterizations of \citep{gelman2008redundant}). These models are appropriate for observations with group-wise structure and dependence, such as patients nested within providers or students within schools. As noted in Section \ref{sec:rfx-model}, the most common use-case for BCF in our applied work is given by ~\eqref{eq:bcf-model-rfx}.

\subsubsection{A random effects analysis on the ACIC 2019 Data} \label{sec:python-acic-bcf-rfx}

The data in \cite{carvalho2019assessing} exhibit a pattern of school-level clustering, which is naturally accommodated by the model in~\eqref{eq:bcf-model-rfx}. In \pkg{stochtree}, this is supported directly in both the \code{BARTModel} and \code{BCFModel} classes. We fit school-level random intercept and treatment effects by specifying:
\begin{itemize}
    \item \code{rfx_group_ids_train}: labels that define groups or clusters in a random effects model, and 
    \item \code{rfx_basis_train}: bases on which random slopes are fitted.
\end{itemize}
For the ACIC data, group IDs are in the \code{schoolid} column of the dataframe and the basis for the leaf regression model -- the array $w_{ij}$ in the generalized formulation~\eqref{eq:bcf-model-rfx-gen} -- is simply \code{as.matrix(1,Z)}.
\begin{CodeChunk}
\begin{CodeInput}
>>> group_ids = df.loc[:, "schoolid"].to_numpy() - 1
>>> rfx_basis = np.concatenate(
>>>     (np.ones((n, 1)), np.expand_dims(Z, 1)), axis=1
>>> )
\end{CodeInput}
\end{CodeChunk}

The model specified in \eqref{eq:bcf-model-rfx} is common enough that it is supported in \code{BCFModel} without users having to construct and specify \code{rfx_basis} manually. If \code{model_spec = "intercept_plus_treatment"} is specified in the \code{random_effects_params} parameter dictionary, \pkg{stochtree} will handle basis construction at both sampling and prediction time. In particular, the model in \eqref{eq:bcf-model-rfx} distinguishes between the conditional average treatment effect (CATE) and $\tau(x)$, where the CATE is the sum of the $\tau(x)$ forest predictions and the relevant group regression slope on $Z$. Users who fit a model with random slopes on $Z$ without specifying \code{model_spec = "intercept_plus_treatment"} (i.e. by passing \code{np.c_[np.ones(n),Z]} manually as a basis) will not recover the correct CATE from the \code{predict} method, though this can still be computed via a contrast of \code{yhat} predictions (which is implemented as a \code{compute_contrast()} method). \pkg{stochtree} also supports a \code{model_spec = "intercept_only"} random effects specification for both BART and BCF.

\cite{hahn2020bayesian} note that, in many cases, researchers only consider a subset of their covariates to be plausible treatment effect moderators. \pkg{stochtree} offers a simple way to specify this expectation without providing separate covariate data for each forest. Parameter dictionaries for the prognostic forest (\code{prognostic_forest_params}) and the treatment effect forest (\code{treatment_effect_forest_params}) both allow users to specify a list of variables to retain (\code{keep_vars}) or drop from a forest (\code{drop_vars}). In the analytical challenge that accompanied this dataset, \cite{carvalho2019assessing} asked researchers to assess whether $X_1$ and $X_2$ moderate the treatment effect, so we fit a BCF model with only these two variables included as moderators.

The BCF model with an additive random effects term and $\tau(X)$ specified as a function of $X_1$ and $X_2$ is fit as follows:
\begin{CodeChunk}
\begin{CodeInput}
>>> treatment_forest_params = {"keep_vars": ["X1", "X2"]}
>>> bcf_model_rfx = BCFModel()
>>> bcf_model_rfx.sample(
>>>     X_train=covariate_df,
>>>     Z_train=Z,
>>>     y_train=y,
>>>     propensity_train=propensity,
>>>     rfx_group_ids_train=group_ids,
>>>     rfx_basis_train=rfx_basis,
>>>     num_gfr=10,
>>>     num_burnin=2000,
>>>     num_mcmc=1000,
>>>     treatment_effect_forest_params=treatment_forest_params,
>>>     random_effects_params={"model_spec": "intercept_plus_treatment"},
>>> )
\end{CodeInput}
\end{CodeChunk}

\begin{figure}[t!]
\centering
\includegraphics{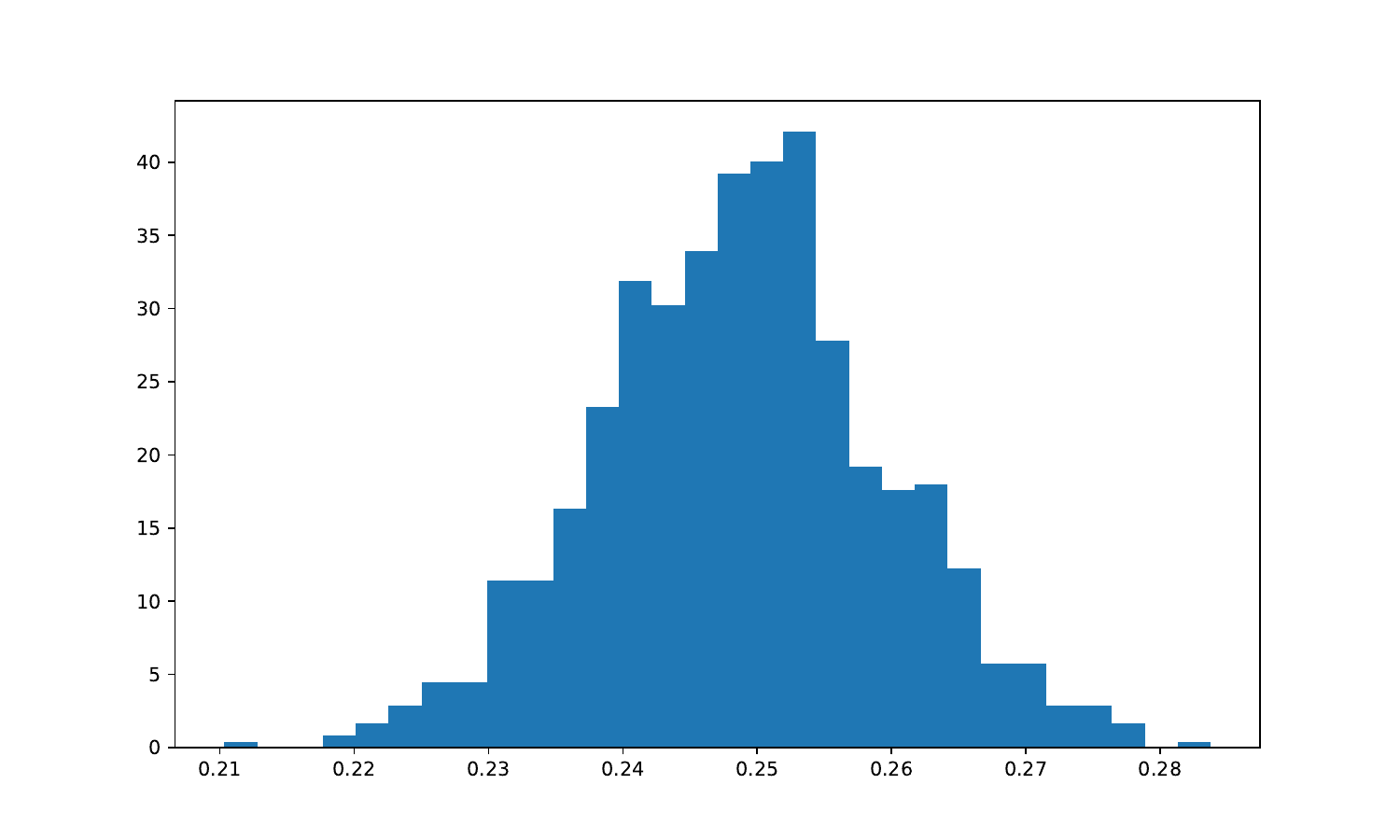}
\caption{\label{fig:python-acic-bcf-ate-posterior-rfx} Posterior distribution of average treatment effect (ATE) estimated by homoskedastic BCF model using the \pkg{stochtree} \proglang{R} package.}
\end{figure}

\begin{figure}[t!]
\centering
\includegraphics{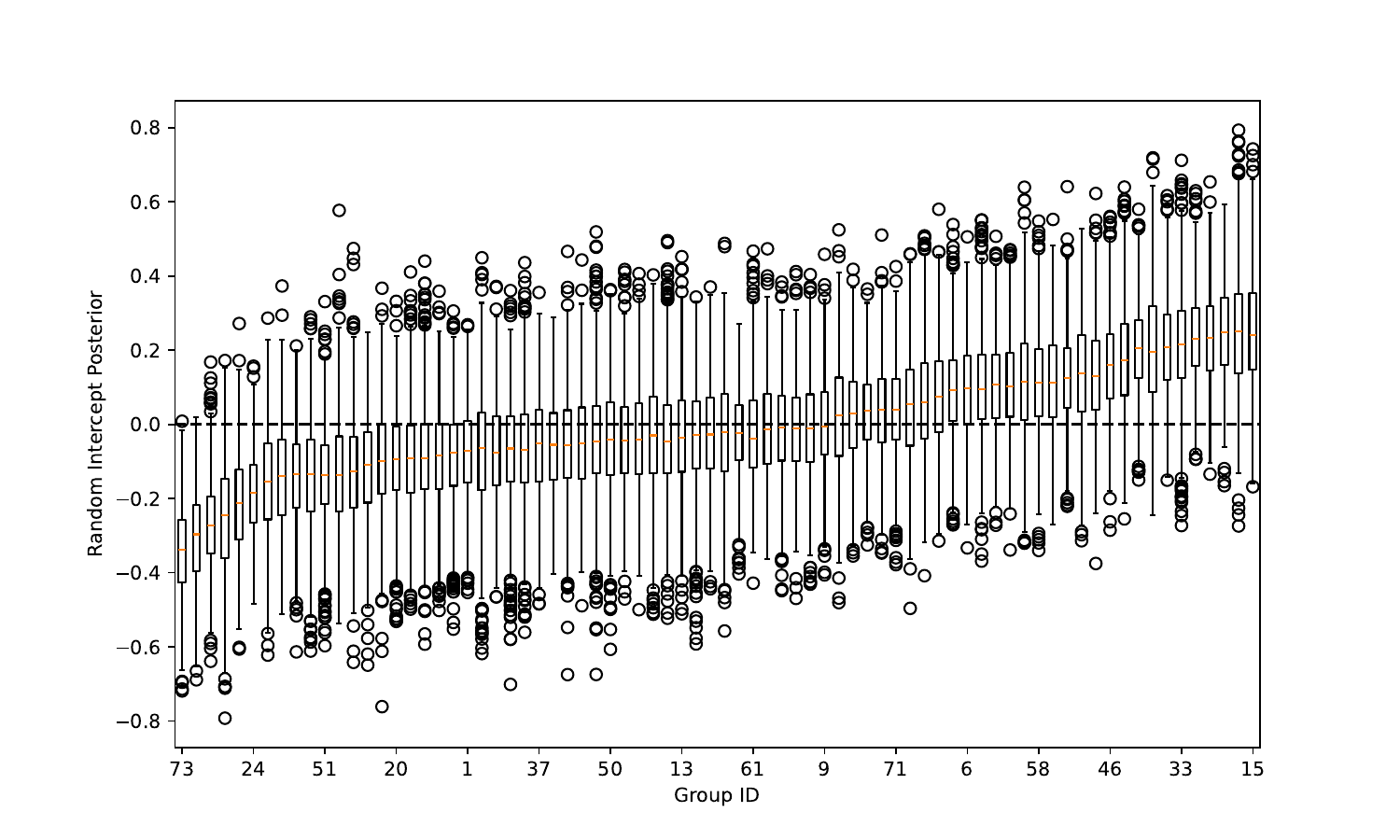}
\caption{\label{fig:python-random-intercept-boxplot} Boxplot of random intercept posterior samples for all 76 schools, estimated using the \pkg{stochtree} \proglang{R} package.}
\end{figure}

After fitting this BCF model, we see in Figure \ref{fig:python-acic-bcf-ate-posterior-rfx} that the posterior of the ATE is largely unchanged, but that the school-level intercepts are in some cases quite pronounced (Figure \ref{fig:python-random-intercept-boxplot}).

Below we demonstrate how to extract posterior samples from the random effects model to produce Figure \ref{fig:python-random-intercept-boxplot}. 
The \code{extract_parameter_samples} method of the \code{BCFModel} class's internal \code{rfx_container} object returns a dictionary of arrays that correspond to terms in the \cite{gelman2008redundant} parameterization. The \code{beta_samples} entry of this dictionary corresponds to $\beta_j$ in \eqref{eq:bcf-model-rfx-gen} and it is stored as an array of dimension $(k,l,m)$ where $k$ is the dimension of $\beta_j$, $l$ is the number of groups in the study, and $m$ is the number of posterior samples. As we are interested in the random intercepts, we restrict our attention to \code{rfx_betas[0, :, :]}.
\begin{CodeChunk}
\begin{CodeInput}
>>> rfx_samples = bcf_model_rfx.rfx_container.extract_parameter_samples()
>>> rfx_betas = rfx_samples["beta_samples"]
>>> rfx_intercept_group_means = np.array(
>>>     [np.mean(rfx_betas[0, i, :]) for i in range(rfx_betas.shape[1])]
>>> )
>>> rfx_intercept_sort_inds = np.argsort(rfx_intercept_group_means).tolist()
>>> rfx_per_group_intercept = [
>>>     rfx_betas[0, i, :] for i in rfx_intercept_sort_inds
>>> ]
>>> plt.boxplot(rfx_per_group_intercept)
>>> plt.xticks([y + 1 for y in range(len(rfx_per_group_intercept))],
>>>            labels=rfx_intercept_sort_inds)
>>> plt.xlabel('Group ID')
>>> plt.ylabel('Random Intercept Posterior')
\end{CodeInput}
\end{CodeChunk}

\begin{figure}[t!]
\centering
\includegraphics[width=0.8\linewidth]{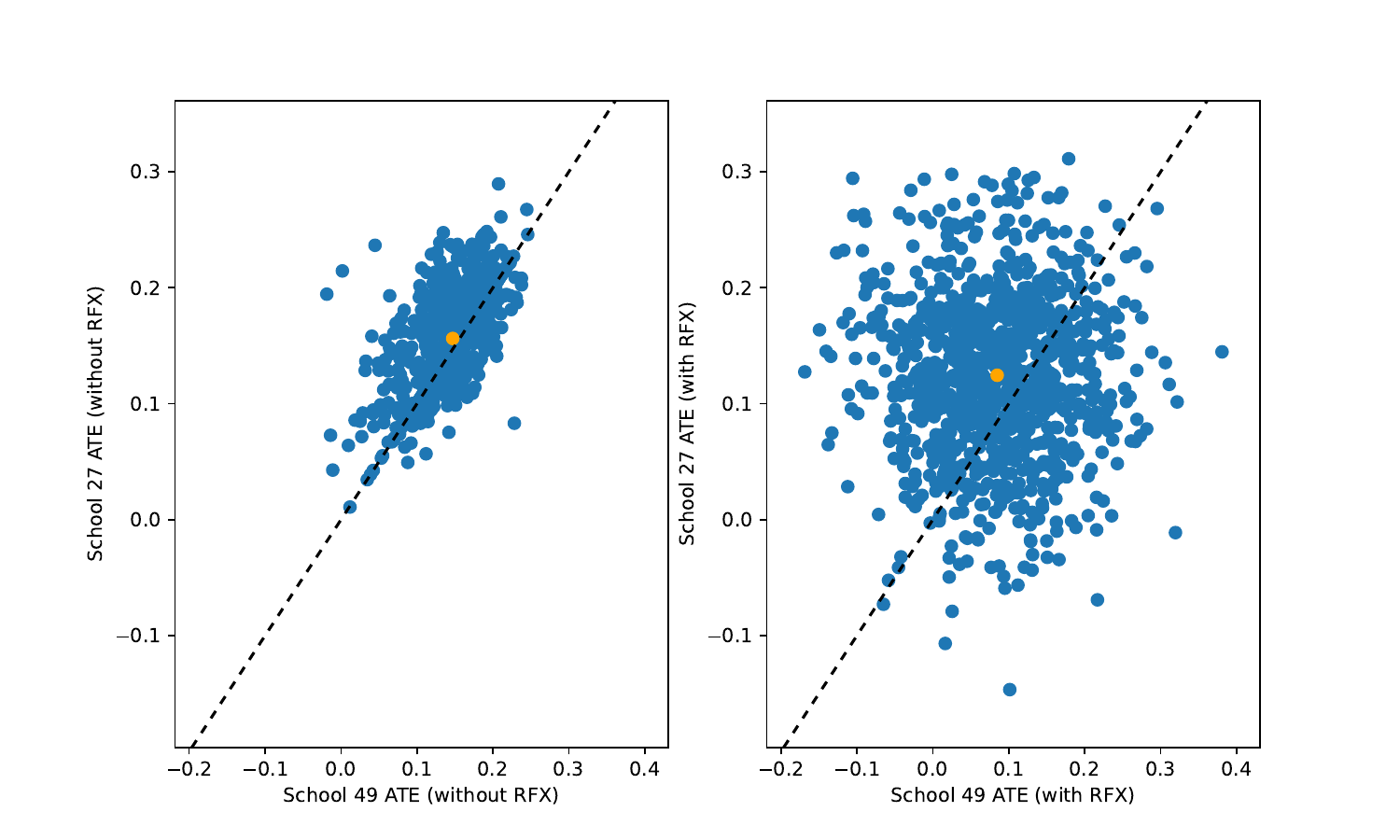}
\caption{\label{fig:python-acic-bcf-rfx-comparison} Comparison of subgroup ATE for two schools in BCF models fit with random effects and without random effects, both estimated using the \pkg{stochtree} \proglang{Python} package.}
\end{figure}

To visualize the impact of fitting a random effects BCF model, compared to a BCF model with no random effects, Figure \ref{fig:python-acic-bcf-rfx-comparison} plots posterior samples of the school-average treatment effect for two schools, given by:
\begin{equation}
\frac{1}{n_j}\sum_{i\in \text{group } j}\E[Y_i(1) - Y_i(0) \mid X=x_{ij}, i\in \text{group } j] = \xi_j + \frac{1}{n_j}\sum_{i\in \text{group } j}\tau(x_{ij})
\end{equation}
Observe that the school-level treatment effects show stronger correlation in the no-random-effects model, due to shared group level covariates, while the random effects specification allows the observed differences in the data to be attributed to idiosyncratic random effects, leading to weaker posterior dependence. 

\subsection{Heteroskedastic errors} \label{sec:python-user-guide-heteroskedasticity}

\subsubsection{BART with Heteroskedastic Errors}

\cite{pratola2020heteroscedastic} and \cite{murray2021log} extend the classic BART model with a log-linear forest modeling the conditional variance function. 
\begin{equation*}
\begin{aligned}
Y_i \mid x_i &\sim \mathrm{N}\left(f(x_i), \sigma^2_0 \exp{(h(x_i)})\right)\\
f \sim &\mathrm{BART}(\alpha_f, \beta_f, m_f) \\
h \sim &\mathrm{logBART}(\alpha_h, \beta_h, m_h)
\end{aligned}
\end{equation*}
where $\mathrm{logBART}$ denotes the same tree prior as BART, but with leaf parameters $\lambda_{s,j}$ assigned independent (log) inverse-gamma priors: 
\begin{equation*}
\begin{aligned}
\exp(\lambda_{s,j}) &\iid \text{IG}\left(a, b\right).
\end{aligned}
\end{equation*}

\pkg{stochtree} enables heteroskedastic forests for both supervised learning (BART) and causal inference (BCF); the next example demonstrates a simple heteroskedastic supervised learning model.

\subsubsection{Heteroskedastic Motorcycle Data} \label{sec:python-bart-motorcycle}

We can direct \pkg{stochtree} to model the conditional variance of the motorcycle dataset by simply specifying \code{num_trees > 0} in the \code{variance_forest_params} parameter dictionary.
\begin{CodeChunk}
\begin{CodeInput}
>>> num_gfr = 10
>>> num_burnin = 0
>>> num_mcmc = 100
>>> general_params = {'sample_sigma2_global': True}
>>> variance_forest_params = {'num_trees' : 20, 'alpha' : 0.5, 
>>>                           'beta' : 3.0, 'min_samples_leaf' : 20}
>>> bart_model_het = BARTModel()
>>> bart_model_het.sample(
>>>     X_train = mcycle[:,0], y_train = mcycle[:,1], 
>>>     num_gfr = num_gfr, num_burnin = num_burnin, num_mcmc = num_mcmc, 
>>>     general_params = general_params, 
>>>     variance_forest_params = variance_forest_params
>>> )
\end{CodeInput}
\end{CodeChunk}

We also construct a homoskedastic model for comparison.
\begin{CodeChunk}
\begin{CodeInput}
>>> general_params = {'sample_sigma2_global': True}
>>> bart_model = BARTModel()
>>> bart_model.sample(
>>>     X_train = mcycle[:,0], y_train = mcycle[:,1], 
>>>     num_gfr = num_gfr, num_burnin = num_burnin, num_mcmc = num_mcmc, 
>>>     general_params = general_params
>>> )
\end{CodeInput}
\end{CodeChunk}

\begin{figure}[t!]
\centering
\includegraphics{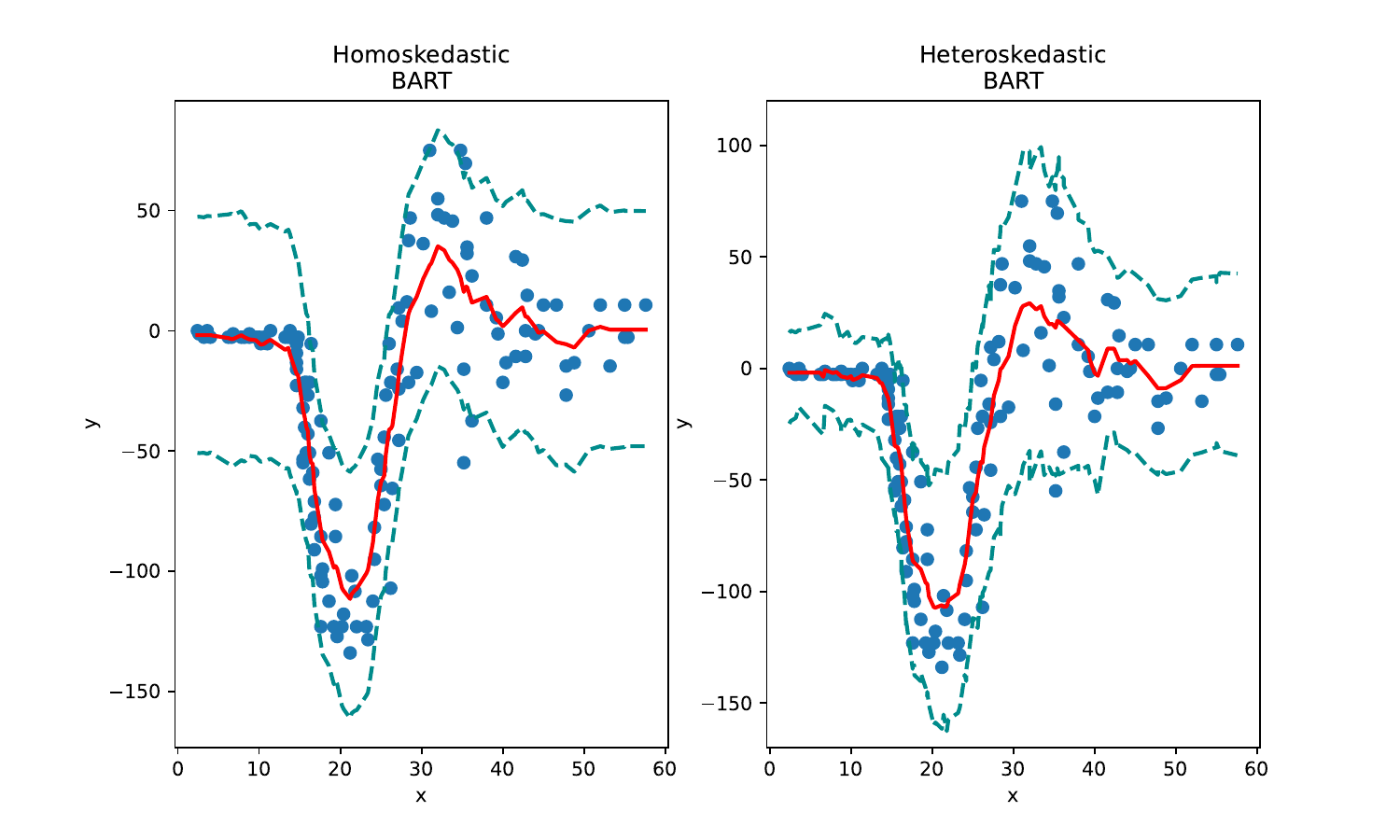}
\caption{\label{fig:python-motorcycle-model-comparison} Comparison of mean predictions and intervals on the motorcycle dataset from heteroskedastic and homoskedastic BART models using the \pkg{stochtree} \proglang{R} package.}
\end{figure}

Posterior mean and interval estimates of the two models are shown for comparison in Figure~\ref{fig:python-motorcycle-model-comparison}.

\newpage
\subsection{Leafwise-Regression Models} \label{sec:python-user-guide-linear-leaf-model}

By modifying the way regression trees map to outcomes ~\eqref{eq:g-function}, more elaborate tree models (and ensembles thereof) are possible, which have vector-valued parameters associated with their leaves rather than scalars\footnote{In the future this will allow general shared-tree models; see ~\citep{linero2020shared} for examples and discussion.} In particular, this allows linear regression at the leaves, as in ~\eqref{eq:g-function-linear}.

Internally, such a linear regression underlies \pkg{stochtree}'s BCF implementation, taking $\Psi(x,z) := z$. The same formulation works to fit the heterogeneous-linear BCF model in \cite{woody2020estimating} and more esoteric causal models (e.g. multi-arm treatment assignments as in \cite{yeager_synergistic_2022} and the regression discontinuity model below). Other applications include VC-BART \citep{deshpande2020vcbart} and BART with ``targeted smoothing'' \citep{starling2020bart}.

\subsubsection{Academic Probation Data RDD analysis} \label{sec:python-bart-rdd}

\cite{alcantara2025learning} utilize leafwise linear regression in \pkg{stochtree} to perform heterogeneous treatment effect estimation in regression discontinuity designs (RDD). An RDD is characterized by a running variable $X$, and a treatment indicator $Z_i$ which is one whenever  $X_i > 0$. \cite{alcantara2025learning} show that the treatment effect of $Z$ on outcome $Y$ can be estimated by a leafwise regression by specifying the basis vector
\begin{equation}
\Psi(x,z) = [1, zx, (1-z)x, z].
\end{equation}

This example uses the academic probation data from \cite{lindo2010ability}, which consists of data on college students enrolled in a large Canadian university where students were placed on academic probation based on a sharp cutoff. The treatment $Z$ indicates students that have been placed on academic probation. The running variable, $X$, is the distance away from the probation threshold. Potential moderators, $W$, are:
\begin{itemize}
\itemsep0em 
\item gender
\item age at enrollment
\item North American
\item the number of freshman-year credits 
\item campus (1-3)
\item incoming class rank of high school GPA
\end{itemize}

Fitting this model in \pkg{stochtree} is straightfoward:
\begin{CodeChunk}
\begin{CodeInput}
>>> covariates_train = w
>>> covariates_train.loc[:,'x'] = x
>>> bart_model = BARTModel()
>>> bart_model.sample(
>>>   X_train = covariates_train,
>>>   leaf_basis_train = Psi,
>>>   y_train = y,
>>>   num_gfr = 10,
>>>   num_burnin = 0,
>>>   num_mcmc = 500,
>>>   general_params = global_params,
>>>   mean_forest_params = forest_params
>>> )
\end{CodeInput}
\end{CodeChunk}
We specify \code{Psi0} and \code{Psi1} and the corresponding covariates to extract the leaf-specific regression coefficient on $z$ as follows:
\begin{CodeChunk}
\begin{CodeInput}
>>> h = 0.1 ## window for prediction sample
>>> test = (-h < x) & (x < h)
>>> Psi0 = np.c_[np.ones(n), np.zeros(n), np.zeros(n), np.zeros(n)][test, :]
>>> Psi1 = np.c_[np.ones(n), np.zeros(n), np.zeros(n), np.ones(n)][test, :]
>>> covariates_test = w.iloc[test,:]
>>> covariates_test.loc[:,'x'] = np.zeros(ntest)
\end{CodeInput}
\end{CodeChunk}
From here, the CATE estimates are obtained from the fitted model using the predict function:
\begin{CodeChunk}
\begin{CodeInput}
>>> cate_posterior = bart_model.compute_contrast(
>>>     X_0 = covariates_test,
>>>     X_1 = covariates_test,
>>>     leaf_basis_0 = Psi0,
>>>     leaf_basis_1 = Psi1,
>>>     type = "posterior",
>>>     scale = "linear"
>>> )
\end{CodeInput}
\end{CodeChunk}

\begin{figure}[t!]
\centering
\includegraphics{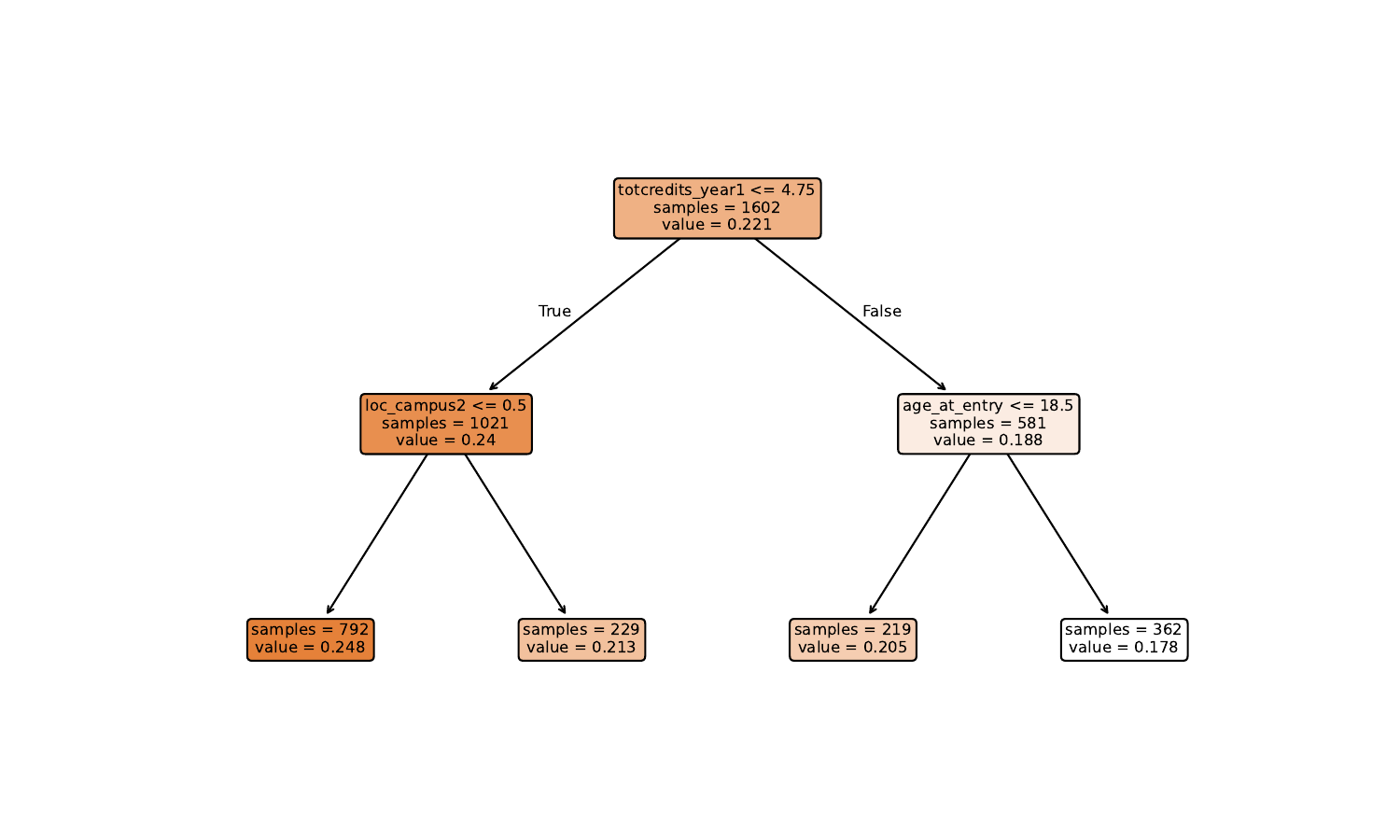}
\caption{\label{fig:python-rdd-cate-tree} Surrogate CART fit of the treatment effect function estimated using BART in \pkg{stochtree} \proglang{Python} package.}
\end{figure}

With this vector of local ($x = 0$) conditional average treatment effects in hand, we may attempt to summarize which variables in $W$ predict distinct treatment effects. Specifically, we visualize a CART tree fit to the posterior point estimates in Figure~\ref{fig:python-rdd-cate-tree}.

\begin{CodeChunk}
\begin{CodeInput}
>>> # Fit regression tree
>>> tau_hat_bar = np.mean(cate_posterior, axis=1)
>>> surrogate_tree = DecisionTreeRegressor(max_depth=2)
>>> surrogate_tree.fit(w.iloc[test,:], tau_hat_bar)
>>> 
>>> # Plot regression tree
>>> plot_tree(
>>>     surrogate_tree,
>>>     feature_names=[f"W{i+1}" for i in range(w.shape[1])],
>>>     filled=True,
>>>     rounded=True,
>>>     impurity=False,
>>>     fontsize=8,
>>> )
\end{CodeInput}
\end{CodeChunk}

It appears that gender, high-school performance, and first year credits predict differences in how students respond to probation. For more details, consult \cite{alcantara2025learning}.

\section{Sampling Custom Models} \label{sec:python-user-guide-customization}

As illustrated in the previous sections \pkg{stochtree} provides a range of flexible models out-of-the-box. However, for particular applications there is often a need to include bespoke terms in a regression function, alternative error distributions, or other model extensions. \pkg{stochtree} exposes low-level interfaces to forests and data classes at the \proglang{R} and \proglang{Python} levels to make these kinds of model extensions straightforward, as illustrated in the following examples.

\subsection{Overview of the Custom Sampler Interface}

The underlying \proglang{C++} codebase centers around a handful of objects and their interactions. We provide \proglang{Python} wrappers for these objects to enable greater customization of stochastic tree samplers than can be furnished by flexible \code{BARTModel} and \code{BCFModel} classes. The \proglang{C++} class wrappers are managed via \proglang{Python} classes and we introduce each object and its purpose below. 

The \code{Dataset} class manages covariates, bases, and variance weights used in a forest model, and contains methods for updating the underlying data as well as querying numeric attributes of the data (i.e. \code{num_observations}, \code{num_covariates}, \code{has_basis}, etc...). The \code{RandomEffectsDataset} class manages all of the data used to sample a random effects model, including group indices and regression bases. The \code{Residual} class wraps the model outcome, which is updated in-place during sampling to reflect the full, or partial, residual net of mean forest or random effects predictions. The \code{ForestContainer} class is a container of sampled tree ensembles, essentially a very thin wrapper around a \proglang{C++} \code{std::vector} of \code{std::unique_ptr} to \code{Ensemble} objects. The \code{Forest} class is a thin wrapper around \code{Ensemble} \proglang{C++} objects, which is used as the ``active forest'' or ``state'' of the forest model during sampling. The \code{ForestSampler} class maintains all of the ``temporary'' data structures used to sample a forest, and its \code{ForestSampler.sample_one_iteration()} method performs one iteration of the requested forest sampler (i.e. Metropolis-Hastings or Grow-From-Root). 

Writing a custom Gibbs sampler with one or more stochastic forest terms requires initializing each of these objects and then deploying them in a sampling loop. We illustrate two straightforward examples in the following vignettes, with some of the verbose parameter dictionaries converted to ellipses for brevity's sake. Interested readers are referred to our online vignettes for more detailed demonstrations of the ``low-level'' \pkg{stochtree} interface \footnote{\url{https://stochtree.ai/python_docs/demo/prototype_interface.html}}.

\subsection{Additive Linear Model} \label{sec:python-user-guide-additive-linear-model}

In Section~\ref{sec:python-acic-bcf-rfx} we fit a BCF model with random intercepts and slopes, assuming particular grouping structure and priors on the random effect variances. We might instead want to allow correlated random effects, or fixed effects/linear terms with fixed priors on coefficient variances. The low-level interface exposed by \pkg{stochtree} makes this straightforward. In the general case, we have a model specification like ~\eqref{eq:linear-term}.

To illustrate, we generate a modified Friedman dataset, where $W \sim \text{U}\left(0,1\right)$ and $\gamma = 5$. To fit the model in \eqref{eq:linear-term}:

\begin{enumerate}

\item We initialize the data and random number generator objects in \proglang{R}:
\begin{CodeChunk}
\begin{CodeInput}
>>> forest_dataset = Dataset()
>>> forest_dataset.add_covariates(X)
>>> residual = Residual(y_standardized)
>>> cpp_rng = RNG()
\end{CodeInput}
\end{CodeChunk}

\item Then we initialize the ``configuration'' objects. The full code is in the supplementary materials; the relevant class is \code{ForestModelConfig} and key initialization arguments are:

\begin{enumerate}
    \item \code{feature_types}: a numpy array of integer-coded feature types (0 denotes numeric and 1 denotes ordered categorical)
    \item \code{variable_weights}: a numpy array of selection probabilities for variables in generating split rules
    \item \code{leaf_dimension}: the dimensionality of the leaf node parameters
    \item \code{leaf_model_type}: integer code for the leaf model expressed by a given forest (0 denotes constant Gaussian, 1 denotes univariate Gaussian regression, 2 denotes multivariate Gaussian regression, 3 denotes a log-linear Inverse Gamma variance forest model)
    \item \code{num_trees}: the number of trees in a forest
    \item \code{num_features}: the dimensionality of the covariates used to define a forest's split rules
    \item \code{num_observations}: the number of samples in a model's training dataset
\end{enumerate}

\begin{CodeChunk}
\begin{CodeInput}
>>> outcome_model_type = 0
>>> leaf_dimension = 1
>>> sigma2_init = 1.
>>> num_trees = 200
>>> feature_types = np.repeat(0, p).astype(int)  # 0 = numeric
>>> var_weights = np.repeat(1 / p, p)
>>> global_model_config = GlobalModelConfig(global_error_variance=sigma2_init)
>>> forest_model_config = ForestModelConfig(
>>>     feature_types=feature_types,
>>>     num_trees=num_trees,
>>>     num_features=p,
>>>     num_observations=n,
>>>     variable_weights=var_weights,
>>>     leaf_dimension=leaf_dimension,
>>>     leaf_model_type=outcome_model_type,
>>> )
\end{CodeInput}
\end{CodeChunk}

\item Then we initialize the forests and temporary tracking objects
\begin{CodeChunk}
\begin{CodeInput}
>>> forest_sampler = ForestSampler(
>>>     forest_dataset, global_model_config, forest_model_config
>>> )
>>> forest_container = ForestContainer(num_trees, leaf_dimension, True, False)
>>> active_forest = Forest(num_trees, leaf_dimension, True, False)
\end{CodeInput}
\end{CodeChunk}

\item To prepare the objects for sampling, we set the root of every tree in the \code{active_forest} to a constant value and reflect these values through to the partial residual and temporary tracking data structures
\begin{CodeChunk}
\begin{CodeInput}
>>> leaf_init = np.mean(y_standardized, keepdims=True)
>>> forest_sampler.prepare_for_sampler(
>>>     forest_dataset,
>>>     residual,
>>>     active_forest,
>>>     outcome_model_type,
>>>     leaf_init,
>>> )
\end{CodeInput}
\end{CodeChunk}

\item Initialize containers to store the results of our custom sampler (this code is left to the supplement for brevity of presentation).

\item And then we run the MCMC sampler
\begin{CodeChunk}
\begin{CodeInput}
>>> for i in range(num_mcmc):
>>>     # Update partial residual, both in the C++ object and the R vector
>>>     # used for coefficient sampling
>>>     partial_res = linreg_partial_residual(
>>>         y_standardized,
>>>         forest_dataset,
>>>         active_forest
>>>     )
>>>     residual.add_vector(lm_term_estimate)
>>> 
>>>     # Sample gamma from bayesian linear model with gaussian prior
>>>     current_gamma = sample_linreg_gamma_gibbs(
>>>         partial_res,
>>>         W[:, 0],
>>>         current_sigma2,
>>>         gamma_tau,
>>>         rng,
>>>     )
>>> 
>>>     # Update partial residual before sampling forest
>>>     lm_term_estimate = (W * current_gamma).squeeze()
>>>     residual.subtract_vector(lm_term_estimate)
>>> 
>>>     # Sample from the forest
>>>     forest_sampler.sample_one_iteration(
>>>         forest_container,
>>>         active_forest,
>>>         forest_dataset,
>>>         residual,
>>>         cpp_rng,
>>>         global_model_config,
>>>         forest_model_config,
>>>         keep_sample,
>>>         False,
>>>     )
>>> 
>>>     # Sample global variance parameter
>>>     current_sigma2 = global_var_model.sample_one_iteration(
>>>         residual, cpp_rng, 1.0, 1.0
>>>     )
>>>     global_model_config.update_global_error_variance(current_sigma2)
\end{CodeInput}
\end{CodeChunk}

\end{enumerate}

\begin{figure}[t!]
\centering
\includegraphics{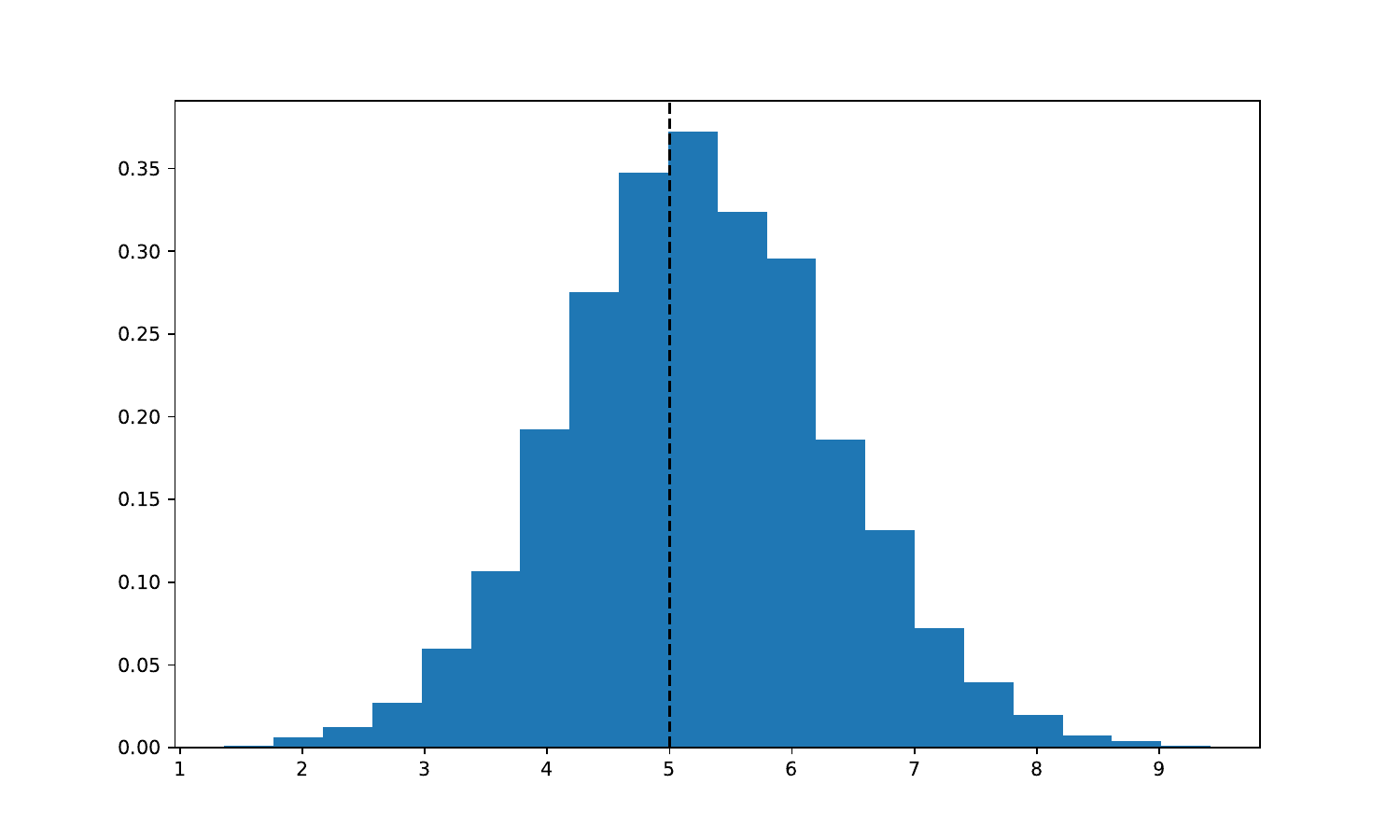}
\caption{\label{fig:python-custom-interface-bart-reg-gamma-histogram} Histogram of posterior samples of linear regression parameter $\gamma$, estimated using the \pkg{stochtree} \proglang{Python} package.}
\end{figure}

We can see the posterior of $\gamma$ in Figure \ref{fig:python-custom-interface-bart-reg-gamma-histogram}.

\subsection{Robust Errors} \label{sec:python-user-guide-additive-robust-errors}

Consider a modified Friedman DGP with heavy-tailed errors
\begin{equation*}
\begin{aligned}
Y_i \mid X_i = x_i &\iid t_{\nu}\left(f(x_i), \sigma^2\right),\\
f(x) &= 10 \sin \left(\pi x_1 x_2\right) + 20 (x_3 - 1/2)^2 + 10 x_4 + 5 x_5,\\
X_1, \dots, X_p &\iid \text{U}\left(0,1\right),
\end{aligned}
\end{equation*}
where $t_{\nu}(\mu,\sigma^2)$ represented a generalized $t$ distribution with location $\mu$, scale $\sigma^2$ and $\nu$ degrees of freedom.

For this dataset, we wish to fit a BART model with a robust error term, rather than the homogeneous Gaussian errors of ``classic'' BART and BCF. We can obtain $t$-distributed errors via a heteroskedastic version of the basic BART model in~\eqref{eq:bart-model} with a further prior on the individual variances ~\eqref{eqn:bart-robust}. Any Gamma prior on $\sigma^2$ ensures conditional conjugacy, though for simplicity's sake we use a log-uniform prior $\sigma^2\propto 1 / \sigma^2$. In the implementation below, we sample from a ``parameter-expanded'' variant of the model given in ~\eqref{eqn:bart-robust-expanded} , which can have favorable convergence properties.

We can connect \eqref{eqn:bart-robust-expanded} to \eqref{eqn:bart-robust} by noting that $\sigma^2$ s equivalent to $a^2\tau^2$. In order to sample the model in \eqref{eqn:bart-robust-expanded}:

\begin{enumerate}

\item We initialize the data and random number generator objects in \proglang{R}:
\begin{CodeChunk}
\begin{CodeInput}
>>> forest_dataset = Dataset()
>>> forest_dataset.add_covariates(X)
>>> forest_dataset.add_variance_weights(1.0 / phi_i_init)
>>> residual = Residual(y_standardized)
>>> cpp_rng = RNG()
\end{CodeInput}
\end{CodeChunk}

\item Then we initialize the ``configuration'' objects. The full code is in the supplement; the relevant functions are \code{createGlobalModelConfig} and \code{createForestModelConfig} and their key arguments are:

\begin{enumerate}
    \item \code{feature_types}: a vector of integer-coded feature types (0 denotes numeric and 1 denotes ordered categorical)
    \item \code{variable_weights}: a vector of selection probabilities for variables in generating split rules
    \item \code{leaf_dimension}: the dimensionality of the leaf node parameters
    \item \code{leaf_model_type}: integer code for the leaf model expressed by a given forest (0 denotes constant Gaussian, 1 denotes univariate Gaussian regression, 2 denotes multivariate Gaussian regression, 3 denotes a log-linear Inverse Gamma variance forest model)
    \item \code{num_trees}: the number of trees in a forest
    \item \code{num_features}: the dimensionality of the covariates used to define a forest's split rules
    \item \code{num_observations}: the number of samples in a model's training dataset
\end{enumerate}

\begin{CodeChunk}
\begin{CodeInput}
>>> outcome_model_type = 0
>>> leaf_dimension = 1
>>> num_trees = 200
>>> feature_types = np.repeat(0, p).astype(int)  # 0 = numeric
>>> var_weights = np.repeat(1 / p, p)
>>> global_model_config = GlobalModelConfig(global_error_variance=sigma2_init)
>>> forest_model_config = ForestModelConfig(
>>>     feature_types=feature_types,
>>>     num_trees=num_trees,
>>>     num_features=p,
>>>     num_observations=n,
>>>     variable_weights=var_weights,
>>>     leaf_dimension=leaf_dimension,
>>>     leaf_model_type=outcome_model_type,
>>> )
\end{CodeInput}
\end{CodeChunk}

\item Then we initialize the forests and temporary tracking objects
\begin{CodeChunk}
\begin{CodeInput}
>>> forest_sampler = ForestSampler(
>>>     forest_dataset, global_model_config, forest_model_config
>>> )
>>> forest_container = ForestContainer(num_trees, leaf_dimension, True, False)
>>> active_forest = Forest(num_trees, leaf_dimension, True, False)
\end{CodeInput}
\end{CodeChunk}

\item To prepare the objects for sampling, we set the root of every tree in the \code{active_forest} to a constant value and reflect these values through to the partial residual and temporary tracking data structures
\begin{CodeChunk}
\begin{CodeInput}
>>> leaf_init = np.mean(y_standardized, keepdims=True)
>>> forest_sampler.prepare_for_sampler(
>>>     forest_dataset,
>>>     residual,
>>>     active_forest,
>>>     outcome_model_type,
>>>     leaf_init,
>>> )
\end{CodeInput}
\end{CodeChunk}

\item Initialize containers to store the results of our custom sampler (as in Section \ref{sec:python-user-guide-additive-linear-model}, this code is deferred to the supplement for brevity's sake).

\item And then we run the MCMC sampler
\begin{CodeChunk}
\begin{CodeInput}
>>> for i in range(num_burnin + num_mcmc):
>>>     # Sample from the forest
>>>     forest_sampler.sample_one_iteration(
>>>         forest_container, active_forest,
>>>         forest_dataset, residual,
>>>         cpp_rng, global_model_config,
>>>         forest_model_config, keep_sample,
>>>         False,
>>>     )
>>> 
>>>     # Sample local variance parameters
>>>     current_phi_i = sample_phi_i(
>>>         y_standardized, forest_dataset,
>>>         active_forest, current_a2,
>>>         current_tau2, nu, rng,
>>>     )
>>> 
>>>     # Sample a2
>>>     current_a2 = sample_a2(
>>>         y_standardized, forest_dataset,
>>>         active_forest, current_phi_i, rng,
>>>     )
>>> 
>>>     # Sample tau2
>>>     current_tau2 <- sample_tau2(current_phi_i, nu, rng)
>>>     
>>>     # Update observation-specific variance weights
>>>     forest_dataset.update_variance_weights(current_phi_i * current_a2)
\end{CodeInput}
\end{CodeChunk}

\end{enumerate}

\begin{figure}[t!]
\centering
\includegraphics{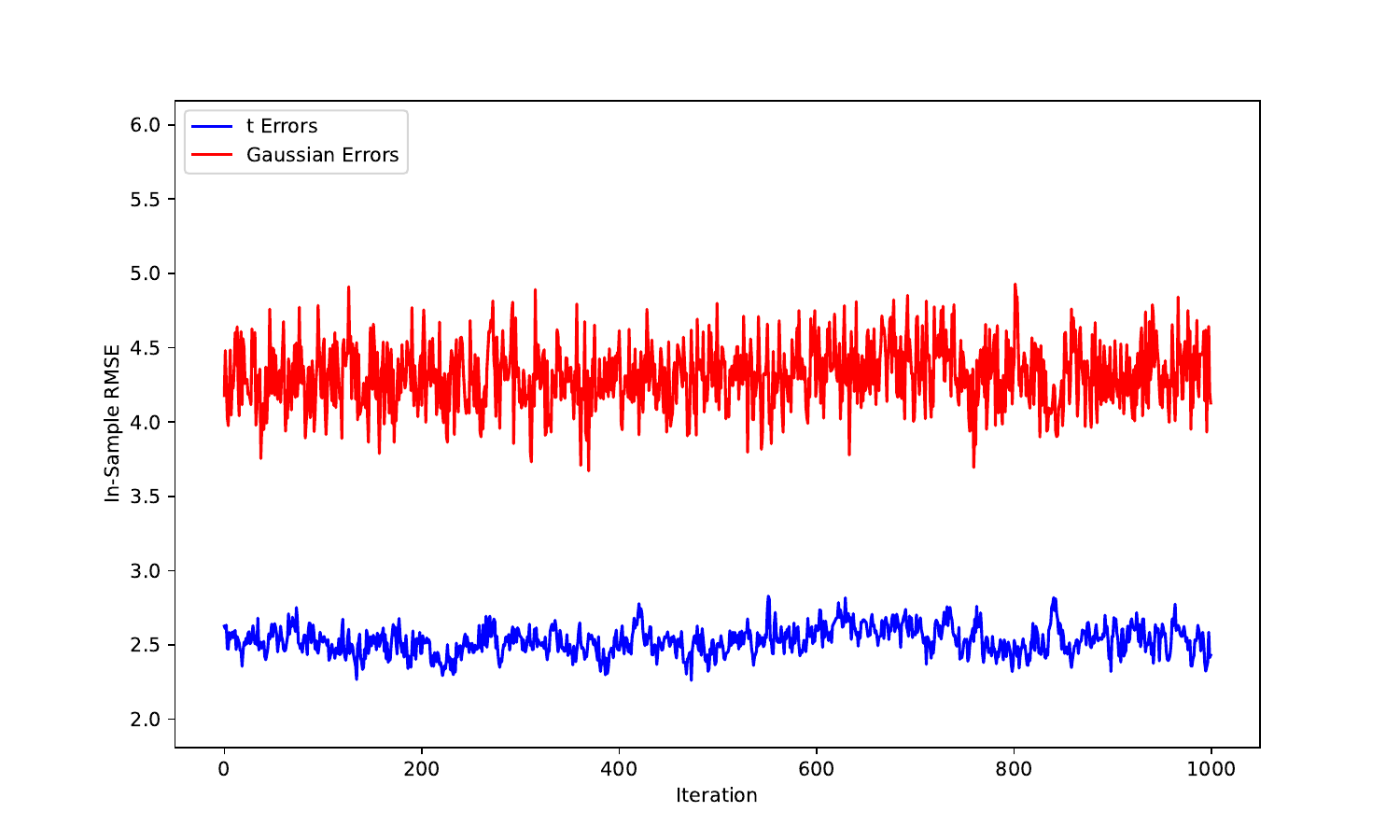}
\caption{\label{fig:python-custom-interface-bart-robust-rmse-comparison-r} Comparison of in-sample sum of squared error (SSE) between a BART model with robust errors and a BART model with normal errors, both estimated using the \pkg{stochtree} \proglang{Python} package. Here, the true DGP has t-distributed errors with 2 degrees of freedom; the BART model with t-distributed errors estimates a smaller error standard deviation compared to a BART model with Gaussian errors.}
\end{figure}

\begin{figure}[t!]
\centering
\includegraphics{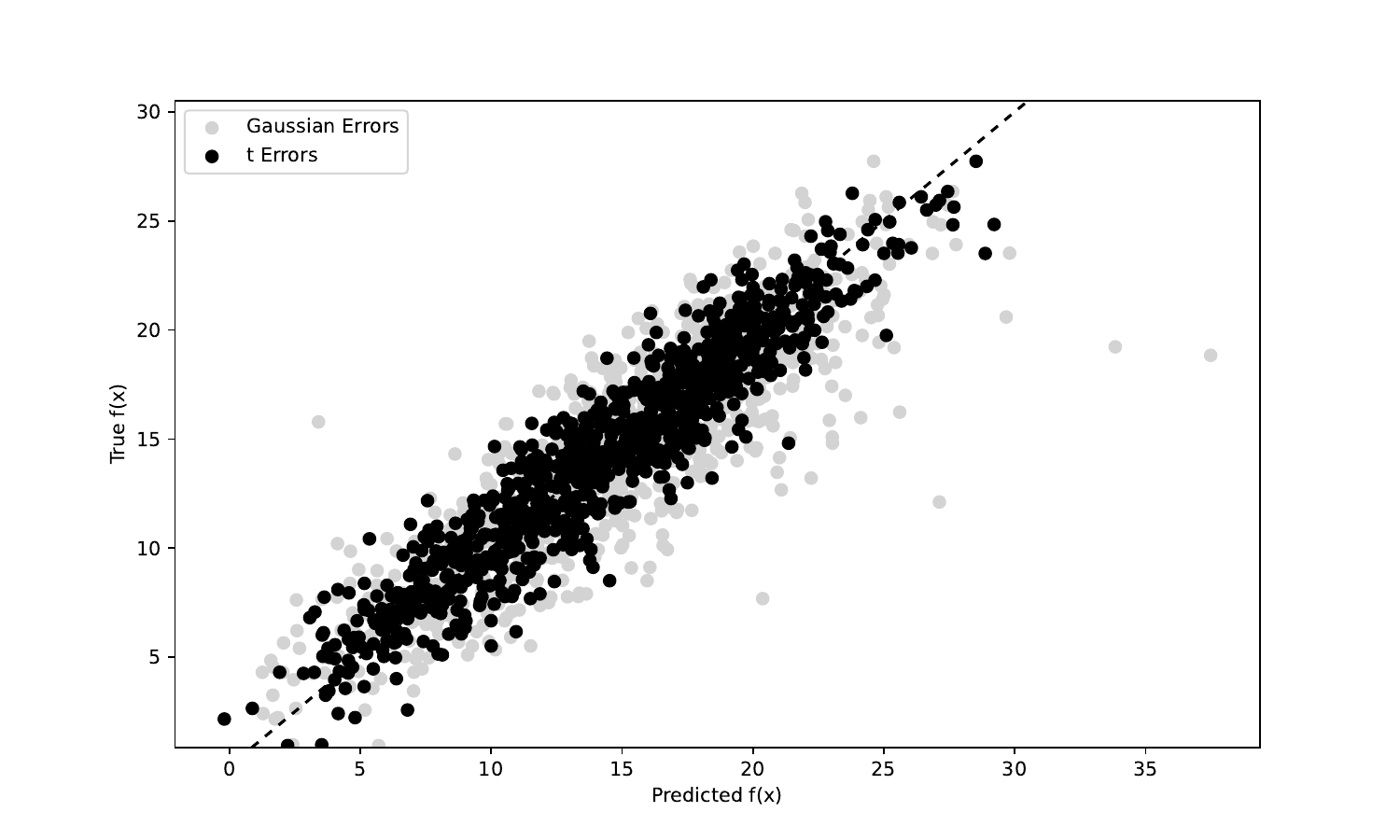}
\caption{\label{fig:python-custom-interface-bart-robust-pred-actual-comparison-r} Comparison of actual conditional mean values and their posterior mean point estimates for BART models with robust errors and with Gaussian errors, both estimated using the \pkg{stochtree} \proglang{Python} package. Here, the true DGP has t-distributed errors with 2 degrees of freedom; the BART model with t-distributed errors more accurately estimates the conditional mean compared to a BART model with Gaussian errors, as would be expected.}
\end{figure}

Figures \ref{fig:python-custom-interface-bart-robust-rmse-comparison-r} and \ref{fig:python-custom-interface-bart-robust-pred-actual-comparison-r} show the results of fitting a custom BART model with t-distributed errors compared to a default BART model using default Gaussian errors, looking at (in-sample) root mean squared estimation error (as a traceplot) and an actual-versus-predicted scatterplot. Both plots  show that the custom BART model with t-distributed errors estimates the mean function more accurately when the data come from a (scaled) t-distribution with $\nu = 2$ degrees of freedom.

\newpage

\section{Scikit-Learn Interface} \label{sec:python-sklearn-interface}

While the primary interface for supervised learning in \pkg{stochtree} is the \code{BARTModel} class, we have further simplified this use case via compatibility with \pkg{scikit-learn} \citep{pedrogosa2011scikit}. The primary simplification is that instead of centering the workflow around specifying, sampling, and analyzing a posterior distribution, we treat the posterior mean of a BART model as a single model, or ``estimator'' in the notation of \pkg{scikit-learn}\footnote{\url{https://scikit-learn.org/stable/developers/develop.html}}.

Below, we demonstrate how to use this API for regression and classification tasks and to tune hyperparameters via \pkg{scikit-learn}'s model selection module.

\subsection{Regression}
\label{app:sklearn-regression}

For continuous outcomes, the \code{StochTreeBARTRegressor} class has \code{fit}, \code{predict} and \code{score} methods in keeping with the standard \pkg{scikit-learn} API.

Consider a simulated regression dataset
\begin{CodeChunk}
\begin{CodeInput}
>>> n = 100
>>> p = 10
>>> X = rng.normal(size=(n, p))
>>> y = X[:, 0] * 3 + rng.normal(size=n)
\end{CodeInput}
\end{CodeChunk}

We fit a \code{StochTreeBARTRegressor} as follows
\begin{CodeChunk}
\begin{CodeInput}
>>> reg = StochTreeBARTRegressor(general_params={"random_seed": random_seed})
>>> reg.fit(X, y)
\end{CodeInput}
\end{CodeChunk}

As with the \code{BARTModel} class, \code{StochTreeBARTRegressor} accepts direct arguments for sampling parameters (\code{num_gfr}, \code{num_burnin}, \code{num_mcmc}) as well as parameter dictionaries for the components of a \pkg{stochtree} BART model: 
\begin{itemize}
\item \code{general_params},
\item \code{mean_forest_params},
\item \code{variance_forest_params}, and
\item \code{rfx_params}.
\end{itemize}

While each of these terms have calibrated defaults, we can use the cross-validation machinery of \pkg{scikit-learn} to tune any of the above parameters, with appropriate nesting for dictionaries.
\begin{CodeChunk}
\begin{CodeInput}
>>> param_grid = {
>>>     "num_gfr": [10, 40],
>>>     "num_mcmc": [0, 1000],
>>>     "mean_forest_params": [
>>>         {"num_trees": 50, "alpha": 0.95, "beta": 2.0},
>>>         {"num_trees": 100, "alpha": 0.90, "beta": 1.5},
>>>         {"num_trees": 200, "alpha": 0.85, "beta": 1.0},
>>>     ],
>>> }
>>> grid_search = GridSearchCV(
>>>     estimator=StochTreeBARTRegressor(),
>>>     param_grid=param_grid,
>>>     cv=5,
>>>     scoring="r2",
>>>     n_jobs=-1,
>>> )
>>> grid_search.fit(X, y)
\end{CodeInput}
\end{CodeChunk}

\subsection{Binary Classification}

For binary outcomes, we provide a \code{StochTreeBARTBinaryClassifier} class which has \code{fit}, \code{predict}, \code{predict_proba}, \code{decision_function}, and \code{score} methods. This classifier uses the probit link BART model introduced in Section \ref{sec:acic-data}.

Consider a binary classification dataset provided by \pkg{scikit-learn}\footnote{\url{https://scikit-learn.org/stable/modules/generated/sklearn.datasets.load_breast_cancer.html}}:
\begin{CodeChunk}
\begin{CodeInput}
>>> dataset = load_breast_cancer()
>>> X = dataset.data
>>> y = dataset.target
\end{CodeInput}
\end{CodeChunk}

We can fit a classifier to this dataset via the same \code{fit} method introduced in Section \ref{app:sklearn-regression}:
\begin{CodeChunk}
\begin{CodeInput}
>>> clf = StochTreeBARTBinaryClassifier(
>>>     general_params={"random_seed": random_seed}
>>> )
>>> clf.fit(X=X, y=y)
\end{CodeInput}
\end{CodeChunk}

We similarly query predicted probabilities with \code{predict_proba}
\begin{CodeChunk}
\begin{CodeInput}
>>> probs = clf.predict_proba(X)
\end{CodeInput}
\end{CodeChunk}

\subsection{Multiclass Classification}

For categorical outcomes, the \code{OneVsRestClassifier} estimator from \pkg{scikit-learn} can be wrapped around a \code{StochTreeBARTBinaryClassifier} to convert it to a generic multiclass classifier.

Consider a classification dataset from scikit-learn \footnote{\url{https://scikit-learn.org/stable/modules/generated/sklearn.datasets.load_wine.html}}
\begin{CodeChunk}
\begin{CodeInput}
>>> dataset = load_wine()
>>> X = dataset.data
>>> y = dataset.target
\end{CodeInput}
\end{CodeChunk}

We can fit a multi-class model to this data via:
\begin{CodeChunk}
\begin{CodeInput}
>>> clf = OneVsRestClassifier(
>>>     StochTreeBARTBinaryClassifier(
>>>         general_params={"random_seed": random_seed}
>>>     )
>>> )
>>> clf.fit(X=X, y=y)
\end{CodeInput}
\end{CodeChunk}

And compute predicted probabilities exactly as above
\begin{CodeChunk}
\begin{CodeInput}
>>> probs = clf.predict_proba(X)
\end{CodeInput}
\end{CodeChunk}

\section{Python Dataset Details} \label{app:python-dataset-details}

\subsection{Generating the Friedman Dataset} \label{app:python-friedman-dataset}

This classic data generating process is implemented in \proglang{R} as follows:
\begin{CodeChunk}
\begin{CodeInput}
>>> def friedman_mean(x: np.array) -> np.array:
>>>     return (
>>>         10 * np.sin(np.pi * x[:, 0] * x[:, 1])
>>>         + 20 * np.power(x[:, 2] - 0.5, 2.0)
>>>         + 10 * x[:, 3]
>>>         + 5 * x[:, 4]
>>>     )
>>> n = 500
>>> p = 100
>>> X = rng.uniform(low=0.0, high=1.0, size=(n, p))
>>> m_x = friedman_mean(X)
\end{CodeInput}
\end{CodeChunk}

The homoskedastic Friedman vignettes generate $\sigma^2$ according to a user-specified signal-to-noise ratio
\begin{CodeChunk}
\begin{CodeInput}
>>> snr = 3.0
>>> eps = rng.normal(0, 1, n) * np.std(cond_mean) / snr
>>> y = m_x + eps
\end{CodeInput}
\end{CodeChunk}

\subsubsection{Causal Friedman Dataset} \label{app:python-causal-friedman-dataset}

The ``causal Friedman dataset'' uses the \code{friedman_mean} function defined above to determine the prognostic and propensity functions
\begin{CodeChunk}
\begin{CodeInput}
>>> def prog_fn(x: np.array) -> np.array:
>>>     return (
>>>         10 * np.sin(np.pi * x[:, 0] * x[:, 1])
>>>         + 20 * np.power(x[:, 2] - 0.5, 2.0)
>>>         + 10 * x[:, 3]
>>>         + 5 * x[:, 4]
>>>     )
>>> def propensity_fn(x: np.array) -> np.array:
>>>     return norm.cdf(0.05 * (prog_fn(x) - np.mean(prog_fn(x))))
\end{CodeInput}
\end{CodeChunk}

The CATE function is linear in $X_1$: 
\begin{CodeChunk}
\begin{CodeInput}
>>> def cate_fn(x: np.array) -> np.array:
>>>     return 5 * x[:, 0]
\end{CodeInput}
\end{CodeChunk}

And the data is generated as
\begin{CodeChunk}
\begin{CodeInput}
>>> X = rng.uniform(0, 1, size=(n, p))
>>> mu_x = prog_fn(X)
>>> pi_x = propensity_fn(X)
>>> tau_x = cate_fn(X)
>>> Z = rng.binomial(1, pi_x, size=n)
>>> E_Y_ZX = mu_x + tau_x * Z
>>> y = E_Y_ZX + rng.normal(0, 1, size=n)
\end{CodeInput}
\end{CodeChunk}

\subsubsection{Additive Regression Model} \label{app:python-friedman-dataset-additive-reg}

For the ``custom interface'' demos of Section \ref{sec:python-user-guide-customization}, we use the \code{friedman_mean} function with two modifications.
The first demo (Section \ref{sec:python-user-guide-additive-linear-model}) adds a linear regression mean term, $W\beta$, with univariate uniform $W$ and $\beta = 5$.
\begin{CodeChunk}
\begin{CodeInput}
>>> p_W = 1
>>> W = rng.uniform(low=0.0, high=1.0, size=(n, p_W))
>>> gamma_W = 5.0
>>> lm_term = (W * gamma_W).squeeze()
>>> y = lm_term + m_x + eps
\end{CodeInput}
\end{CodeChunk}

\subsubsection{Robust Errors} \label{app:python-friedman-dataset-robust-errors}

The second demo (Section \ref{sec:python-user-guide-additive-robust-errors}) replaces the homoskedastic Gaussian error with a scaled $\sigma t_{\nu}$ error term with $\nu = 5$ and $\sigma^2 = 2$.
\begin{CodeChunk}
\begin{CodeInput}
>>> sigma2 = 9
>>> nu = 2
>>> eps = rng.standard_t(df=nu, size=n) * np.sqrt(sigma2)
>>> y = m_x + eps
\end{CodeInput}
\end{CodeChunk}

\subsection{ACIC Dataset} \label{app:python-acic-dataset}

\cite{carvalho2019assessing} introduce a semi-synthetic dataset, in which covariates are informed by a randomized controlled trial, but outcome and treatment are simulated to ensure confounding. The synthetic outcome and treatment are referred to as $Y$ and $Z$, respectively, while the covariates are defined in the data dictionary in Table \ref{tab:acic-data-dictionary} (drawn from \cite{carvalho2019assessing}).

We load this dataset from Github as follows
\begin{CodeChunk}
\begin{CodeInput}
>>> url_string = "https://raw.githubusercontent.com/andrewherren/acic2024/" \
>>>              "refs/heads/main/data/acic2018/synthetic_data.csv"
>>> df = pd.read_csv(url_string)
\end{CodeInput}
\end{CodeChunk}
and we unpack the data into a format needed for \code{BCFModel()} as follows
\begin{CodeChunk}
\begin{CodeInput}
>>> y = df.loc[:, "Y"].to_numpy()
>>> Z = df.loc[:, "Z"].to_numpy()
>>> covariate_df = df.loc[:, ~np.isin(df.columns, ["schoolid", "Z", "Y"])]
>>> unordered_categorical_cols = ["C1", "XC"]
>>> ordered_categorical_cols = ["S3", "C2", "C3"]
>>> for col in unordered_categorical_cols:
>>>     covariate_df.loc[:, col] = pd.Categorical(
>>>         covariate_df.loc[:, col], ordered=False
>>>     )
>>> for col in ordered_categorical_cols:
>>>     covariate_df.loc[:, col] = pd.Categorical(
>>>         covariate_df.loc[:, col], ordered=True
>>>     )
\end{CodeInput}
\end{CodeChunk}

\subsection{Academic Probation Dataset} \label{app:python-academic-probation-dataset}

In studying the Academic Probation dataset \citep{lindo2010ability} for RDD, our outcome of interest is each student's GPA at the end of the current academic term (\code{nextGPA} in the data) and the running variable (\code{X}) is the negative difference between a student's previous-term GPA and the threshold for being placed on academic probation. From this running variable, we define a binary treatment variable \code{Z} as \code{X > 0}. Covariates are defined in the data dictionary in Table \ref{tab:probation-data-dictionary}.

We load the dataset as follows
\begin{CodeChunk}
\begin{CodeInput}
>>> url_string = "https://raw.githubusercontent.com/rdpackages-replication/" \
>>>              "CIT_2024_CUP/refs/heads/main/CIT_2024_CUP_discrete.csv"
>>> data = read.csv(url_string)
\end{CodeInput}
\end{CodeChunk}
and extract the individual data components needed for the BART regression discontinuity model 
\begin{CodeChunk}
\begin{CodeInput}
>>> y = data.loc[:,'nextGPA'].to_numpy().squeeze()
>>> x = data.loc[:,'X'].to_numpy().squeeze()
>>> x = x / np.std(x) ## we always standardize X
>>> w = data.iloc[:, 3:11]
>>> # Define categorical features as ordered/unordered factors
>>> w.loc[:,'totcredits_year1'] = pd.Categorical(
>>>     w.loc[:,'totcredits_year1'], ordered=True
>>> )
>>> unordered_categorical_cols = [
>>>     'male', 'bpl_north_america', 'loc_campus1', 'loc_campus2', 'loc_campus3'
>>> ]
>>> for col in unordered_categorical_cols:
>>>     w.loc[:, col] = pd.Categorical(
>>>         w.loc[:, col], ordered=False
>>>     )
>>> # Define the cutoff
>>> c = 0
>>> z = (x > c).astype(float)
\end{CodeInput}
\end{CodeChunk}

\end{appendix}


\end{document}